\documentclass[aps,prl,showpacs,onecolumn,floatfix]{revtex4}
\usepackage{amsfonts}
\usepackage{color}
\usepackage{epic}
\usepackage{graphics}
\usepackage{times}

\usepackage{changebar}

\renewcommand{\@}{\partial}

\renewcommand{\d}{\mathrm{d}}
\newcommand{\df}[2]{\frac{\partial #1}{\partial #2}}
\newcommand{\ddf}[2]{\frac{\partial^2 #1}{\partial #2^2}}
\renewcommand{\d}{\mathrm{d}}
\newcommand{\Df}[2]{\frac{\d #1}{\d #2}}
\newcommand{\DDf}[2]{\frac{\d^2 #1}{\d #2^2}}
\newcommand{\eq}[1]{(\ref{#1})}
\def\eqtwo(#1,#2){(\ref{#1},\ref{#2})}
\newcommand{\fig}[1]{fig.~\ref{#1}}
\newcommand{\Fig}[1]{Fig.~\ref{#1}}

\newlength{\panelwidth}

\renewcommand{\L}{\mathcal{L}}

\begin{document}

\title{
Soliton-like phenomena in one-dimensional cross-diffusion systems: \\
a predator-prey pursuit and evasion example
}

\author{M. A. Tsyganov}
\affiliation{Institute of Theoretical and Experimental Biophysics,
Pushchino, Moscow Region 142290, Russia}
\author{J. Brindley}
\affiliation{Department of Applied Mathematics, University of Leeds,
Leeds LS2 9JT, UK}
\author{A. V. Holden}
\affiliation{School of Biomedical Sciences, University of Leeds, Leeds
LS2 9JT, UK}
\author{V. N. Biktashev}
\affiliation{Department of Mathematical Sciences, University of
Liverpool, Liverpool L69 7ZL, UK}
\date{\today}

\begin{abstract}
We have studied properties of nonlinear waves in a mathematical
model of a predator-prey system with pursuit and evasion. We
demonstrate a new type of propagating wave in this system. The
mechanism of propagation of these waves essentially depends on the
``taxis'',
represented by nonlinear ``cross-diffusion'' terms in the mathematical formulation.
We have shown that the dependence of the velocity of wave
propagation on the taxis has two distinct forms, ``parabolic'' and
``linear''. Transition from one form to the other correlates with
changes in the shape of the wave profile. Dependence of the
propagation velocity on diffusion in this system differs from the
square-root dependence typical of reaction-diffusion waves.  We
demonstrate also that, for systems with negative and positive taxis,
for example, pursuit and evasion,
there typically exists a large region in the parameter space,
where the waves demonstrate quasisoliton interaction: colliding waves can
penetrate through each other, and waves can also reflect from impermeable
boundaries.
\end{abstract}

\maketitle


\section{Introduction.}

Population waves demonstrate an example of the reconciliation of
random movement of individual organisms with ordered, deterministic behaviour of
their communities. Study of the mechanisms of such reconciliation of
disorder and order is important.  To explain
specific phenomena characteristic of various population communities,
we have performed studies of pursuit-evasion waves in a mathematical
model of a predator-prey system. Our results presented in \cite{PRL03}
and here demonstrate that population taxis waves exhibit
special properties, which provide insight not only into specific mechanisms of
population dynamics, but also broaden the class of nonlinear
waves to include quasi-solitons in reaction-diffusion-taxis
systems.

The spatio-temporal dynamics of a system of two interacting species are often
described in terms of a reaction-diffusion system of the type
\begin{eqnarray}
\df{P}{t} &=& f(P,Z) + D_1\ddf{P}{x} , \nonumber\\
\df{Z}{t} &=& g(P,Z) + D_2\ddf{Z}{x} , \label{RD}
\end{eqnarray}
where $P$ is the density of the prey population, $Z$ is the density of
the predator population, $D_1$ and $D_2$ are their diffusion coefficients,
the nonlinear functions $f(P,Z)$ and $g(P,Z)$ describe local dynamics,
including growth and interaction of the species, whereas the diffusion terms
describe their spread in space, e.g. resulting from individual random motions.
However, one characteristic feature of living systems is their ability to react
to changes of the environment, and to move towards, or away from, an
environmental stimulus, behavior known as taxis. Examples are chemotaxis,
phototaxis, thermotaxis and gyrotaxis \cite{UFN91,Kessler}. Many models of
spatial dynamics of populations take taxis into account, and its importance has
been  recognized in modelling various biological and ecological processes,
including propagation of epidemics, bacterial population waves, aggregation in
the cellular slime mold \textit{Dictyostelium discoideum}, dynamics of
planktonic communities and of insect populations \cite{UFN91, ameba95,
MedvSIAM}. The existence of travelling waves, and also stationary spatially-
inhomogeneous structures, in interacting populations with taxis has been
demonstrated experimentally and theoretically \cite{Adler66, KelSeg71, Proc93,
ameba94,BerezUFN,Sherr00}.

Here we investigate a system of partial differential equations describing two
spatially distributed populations in a ``predator-prey'' relationship with each
other. The spatial evolution is governed by three processes, positive taxis of
predators up the gradient of prey (pursuit) and negative taxis of prey down the
gradient of predators (evasion), yielding nonlinear ``cross-diffusion'' terms,
and random motion of both species (diffusion).  In this paper we consider the
problem in one spatial dimension, $x$, using the equations
\begin{eqnarray}
\df{P}{t} &=& f(P,Z) + D\ddf{P}{x} + h_-\df{ }{x}P\df{Z}{x} , \nonumber\\
\df{Z}{t} &=& g(P,Z) + D\ddf{Z}{x} - h_+\df{ }{x}Z\df{P}{x} . \label{RDT}
\end{eqnarray}
The diffusion coefficients, $D$, for simplicity are considered constant, uniform and equal for both species,
and
$\df{}{x}\left(P\df{Z}{x}\right)$ and
$\df{}{x}\left(Z\df{P}{x}\right)$ are taxis terms \cite{KelSeg71};
$h_-$ is the coefficient of negative taxis of $P$ on the gradient of
$Z$, $h_+$ is the coefficient of positive taxis of $Z$ on the gradient of $P$.
We choose as local kinetics functions $f(P,Z)$ and $g(P,Z)$ the Holling type III
form used by Truscott and Brindley \cite{TrBrind1} to
describe the population dynamics of phytoplankton, $P$, and zooplankton, $Z$:
\begin{eqnarray}
f(P,Z) &=& \beta P(1-P) - Z P^2/(P^2+\nu^2), \nonumber \\
g(P,Z) &=& \gamma Z P^2/(P^2+\nu^2)-w Z .      \label{TB}
\end{eqnarray}

It is known that these kinetics demonstrate ``excitable'' behavior,
and the reaction-diffusion system \eq{RD} has propagating solitary
wave solutions \cite{TrBrind1,MatBrind}.
We now show how inclusion of the taxis terms can alter the properties
of such solutions.
Though predator-prey systems, with one or both populations demonstrating
``intelligent'' taxis have been studied before, by means of individual-based
Monte-Carlo simulations \cite{Rozenf, Monetti} and by using partial differential
equations \cite{Murray83, Murray, Pettet00, Droz01, Arditi1, Cast02}, 
our objective here to isolate and identify the specific role of the taxis 
terms in creating novel behavior.
We demonstrate a new type of propagating waves in the system \eq{RDT}.
The mechanism of propagation of these waves essentially depends on
the taxis of the species and is entirely different from
waves in reaction-diffusion system. Unlike typical waves in reaction-diffusion
systems, the pursuit-evasion taxis waves can penetrate through each other and
reflect from impermeable boundaries.

Behavior in the form of solitary propagating waves is typical for many spatially
extended nonlinear dissipative systems. Solitary waves that remain unchanged
after collision with each other are less typical and are known only for a rather
narrow class of nonlinear dissipative media \cite{S-1}. In this respect, such
waves are analogous to the solitons in conservative systems, whose study, as
stable particle-like features of nonlinear systems, remain a key interdisciplinary
topic of modern mathematical physics.
In the present paper we demonstrate soliton-like behavior and also wave-splitting
in a class of waves which can emerge in population dynamics models
as a consequence of taxis.

\section{Details of the model and numerical methods}

We have performed extensive numerical studies of equations (2,3).
Three finite difference schemes were used, differing in their approximation of
the taxis terms $\L{u}=\df{}{x}u(x,t)\df{S(x,t)}{x}$.
In all cases, we approximated

\begin{eqnarray*}
(\L{u})(x_i,t_j) \approx \left[ a(x_{i+1},t_j)\left(S(x_{i+1},t_j)-S(x_i,t_j)\right)
-a(x_i,t_j)\left(S(x_i,t_j)-S(x_{i-1},t_j)\right) \right]/(\delta x)^2 ,
\end{eqnarray*}
where $a(x_i,t_j)$ was given by one of the following:

Scheme A: The central implicit scheme \cite{SmG}.

$a(x_i,t_j)=0.5(u(x_i,t_{j+1})+u(x_{i-1},t_{j+1}))$

Scheme B: an ``upwind'' explicit scheme.
For the positive taxis (pursuit),

$a(x_i,t_j)=u(x_{i-1},t_j)$ if $S(x_i,t_j)>S(x_{i-1},t_j)$,

$a(x_i,t_j)=u(x_i,t_j)$ if $S(x_i,t_j)<S(x_{i-1},t_j)$.

For the negative taxis (evasion),

$a(x_i,t_j)=u(x_i,t_j)$ if $S(x_i,t_j)>S(x_{i-1},t_j)$,

$a(x_i,t_j)=u(x_{i-1},t_j)$ if $S(x_i,t_j)<S(x_{i-1},t_j)$.

Scheme C: an ``upwind'' implicit scheme.
For the positive taxis,

$a(x_i,t_j)=u(x_{i-1},t_{j+1})$ if $S(x_i,t_{j+1})>S(x_{i-1},t_{j+1})$,

$a(x_i,t_j)=u(x_i,t_{j+1})$ if $S(x_i,t_{j+1})<S(x_{i-1},t_{j+1})$

For the negative taxis,

$a(x_i,t_j)=u(x_i,t_{j+1})$ if $S(x_i,t_{j+1})>S(x_{i-1},t_{j+1})$,

$a(x_i,t_j)=u(x_{i-1},t_{j+1})$  if  $S(x_i,t_{j+1})<S(x_{i-1},t_{j+1})$

So, the ``upwind'' schemes rather than use the mean between values of the
variables subject to taxis at two neighbouring grid nodes as in the
central scheme, select one or the other depending on the direction
of taxis, i.e. sign of the gradient of the attractant (see
e.g. \cite{Morton} for the discussion of upwind schemes).

The majority of calculations were based on scheme C with discretization
steps $\delta x=0.1$, $\delta t=5\times10^{-3}$ for most figures or
scheme B and C with $\delta x=0.5$ and $\delta t=0.01$ for large-scale
parametric studies \fig{hbgw} and scheme \fig{hhU}.
Selected control calculations used scheme B with smaller steps,
down to $\delta x=0.01$, $\delta t=4\times10^{-6}$,
and schemes A and C with $\delta x=0.01$, $\delta t=10^{-3}$.

Unless specified otherwise, we have calculated solutions to equations
\eqtwo(RDT,TB) with the following parameter values: $\nu=0.07$,
$\beta=1$, $w=0.004$ for two different values of $\gamma$:
$\gamma=0.01$, which allows propagation of purely diffusive waves,
i.e. with $h_+=h_-=0$, $D>0$, and $\gamma=0.016$, which prohibits
propagation of such wave.

\section{Case I: the reaction-diffusion-taxis waves for $\gamma=0.01$}
\subsection{The different mechanisms of wave propagation for $\gamma=0.01$, $D=0.04$}

\begin{figure*}[htbp]
\includegraphics{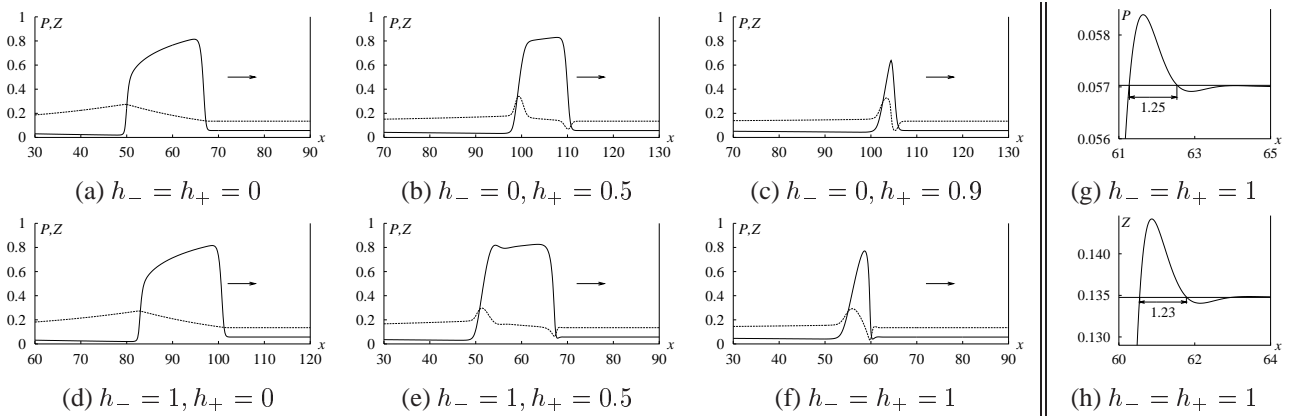}
\caption[]{
(a--f) The profiles of waves with $D=0.04$ and different taxis
coefficients $h_{\pm}$ (at $h_-=0$, $h_+=1$ solitary wave solutions do
not exist).  The different shape of the profiles suggests that
different propagation mechanisms are involved. Note the oscillatory onset
of the pulse front when both $h_+>0$, $h_->0$.
(g,h) The oscillatory onset of the front of the pulse of (f),
magnified. The horizontal lines are at the steady state levels,
$(P_0,Z_0)$. The theoretical value for the oscillation half-length is
$1.256\dots$.
}\label{prof}
\end{figure*}

%-------------------------------------------

\Fig{prof} shows the profiles of population waves in a purely
reaction-diffusion case (a) and with addition of taxis terms (b-f).
The taxis terms significantly change the shape of the profiles.
The value of the pursuit coefficient $h_+$ has much more pronounced effect
than the value of the evasion coefficient $h_-$. If
only evasion ($h_->0$) but no pursuit ($h_+=0$) is added, waves tend to
retain the same shape as purely diffusive waves, with long and smooth plateaus.
The addition of pursuit ($h_+>0$) adds distinctive features, in particular non-monotonic
behavior of predators around the front and/or the back of the wave.

Here we suggest a qualitative explanation of wave shape
change in terms of the pursuit term ($h_+$) only.
Ahead of the wave, the system is at its stable equilibrium.
Consider the effect of a local increase of
the prey density $P$ above the equilibrium.
The resulting flux of predators to the area, described by the
taxis term with the coefficient $h_+$, will deplete the density of predators in surrounding areas,
and the conditions of equilibrium will be violated.
Decreased density of predators will temporarily encourage
growth of prey, followed by influx of predators, and the same sequence of events
occurs progressively at each point in the spatial ($x$) direction,
constituting a travelling wave in the population pattern.
Note that no diffusion of either prey or predators is required;
the phenomenon requires only the presence of taxis terms in equation \eq{RDT}.
The ``excitable'' character of the kinetics in \eq{TB} leads
to a strong magnification of the localized increase of the
prey population, through the prey-escape mechanism
(prey multiply faster than predators).
This feature, of course, is also essential for solitary
waves in purely diffusive systems.
An unusual feature of taxis waves is the oscillatory character
of the front, see \fig{prof}(g,h).
Since these oscillations have a small amplitude, they can be described by linearized theory.
Indeed, in a steadily propagating wave with speed $c$, variables $P$ and $Z$ depend
only on the combination $\xi=x-ct$, and satisfy the ``automodel'' system
\begin{eqnarray}
&& f(P,Z) + D\DDf{P}{\xi} + h_-\Df{}{\xi}P\Df{Z}{\xi}+c\Df{P}{\xi}=0 ,\nonumber\\
&& g(P,Z) + D\DDf{Z}{\xi} - h_+\Df{}{\xi}Z\Df{P}{\xi}+c\Df{Z}{\xi}=0 .\label{automod}
\end{eqnarray}
The speed of the wave in \fig{prof}(f) is $c=0.3535$, and the steady-state values of
the variables are $P_0=0.05703$, $Z_0=0.13480$. With these parameters, a straightforward
calculation gives solutions in the form
$(P,Z)(\xi)\approx(P_0,Z_0)+\mathrm{Re}\left[(P_1,Z_1)e^{\lambda\xi}\right]$,
$|P_1,Z_1|\ll|P_0,Z_0|$, with
$\lambda_{1,2}\approx1.9925+2.5014i$.
This predicts the half-length of oscillations along $\xi$ coordinate of
$\pi/\mathrm{Im}(\lambda_{1,2})\approx1.256$, in a good agreement with
the observed shape, see \fig{prof}(g,h).
This means that these oscillations are not a numerical artifact.
An important feature of solitary taxis waves is the uniformity of their shape,
amplitude and speed: after a transient, these are the same, regardless of the
details of the initial conditions.
In this they are similar to reaction-diffusion
excitation waves and different from solitons in conservative systems.

%****************************************************************

\subsection{Quasisoliton interaction of pulses}
%
%====================================================Interaction-A

\begin{figure*}[htbp]
\setlength{\unitlength}{1mm} \newcommand{\panel}[1]{\begin{picture}(52,65)(0,0) % (0,0)
\put(0,0){\vector(1,0){51}} \put(51,1){$x$}
\put(0,0){\vector(0,1){64}} \put(1,64){$t$}
\put(1,1){\setlength{\fboxsep}{0pt}\fbox{\resizebox{48mm}{62mm}{\includegraphics{#1}}}}
\end{picture}}
\centerline{\begin{tabular}{ccc}
\panel{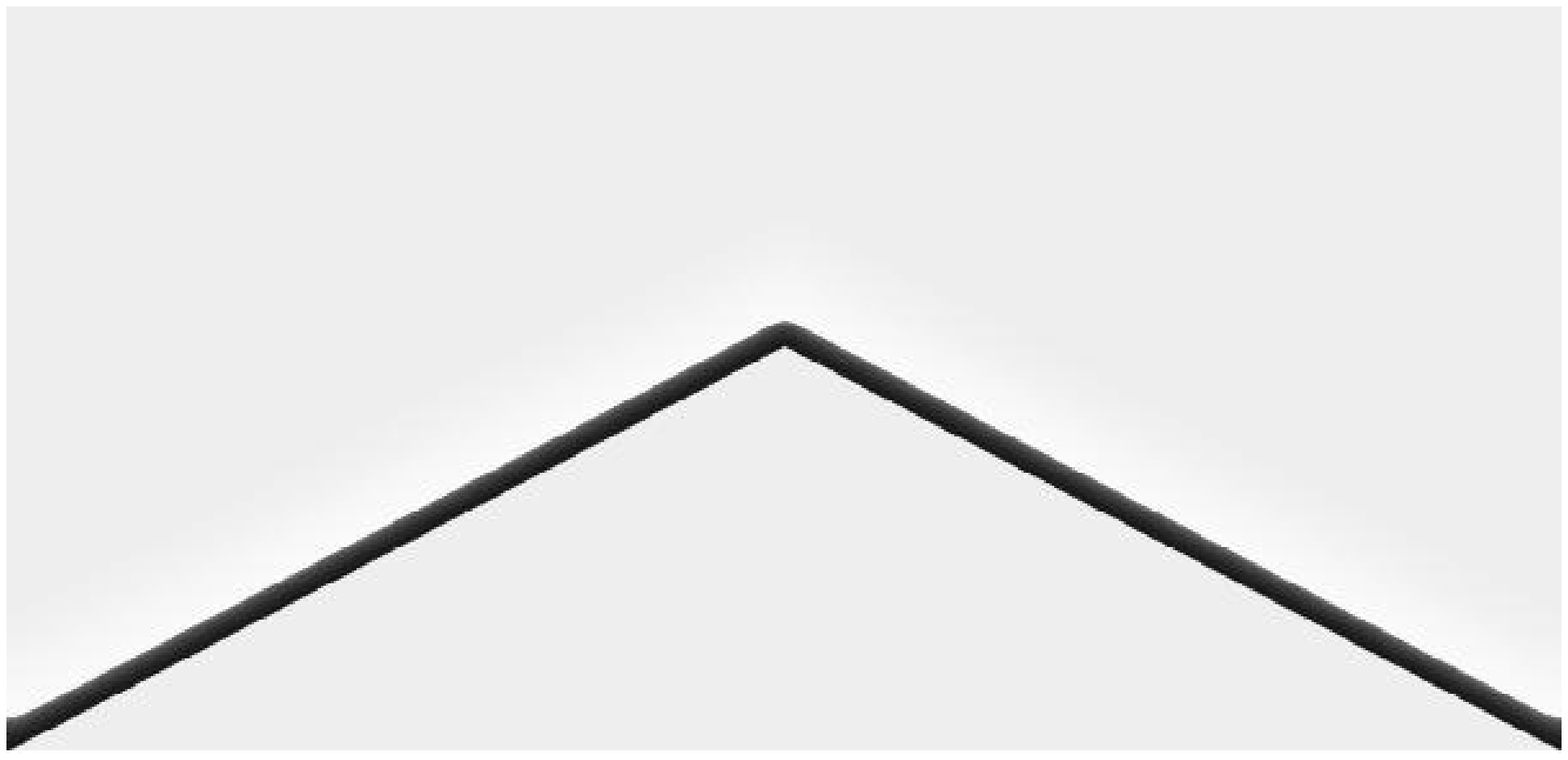} & \panel{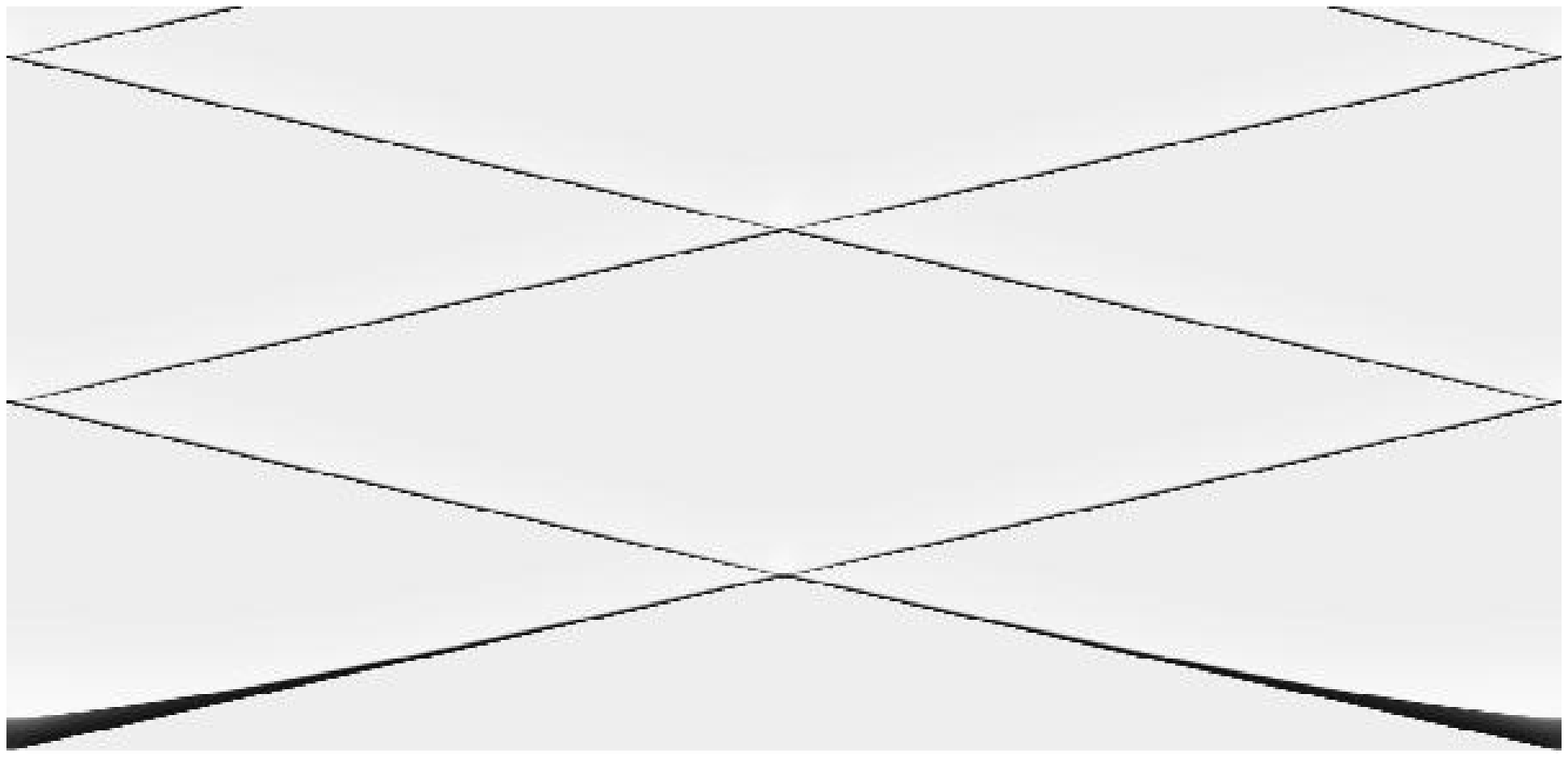} & \panel{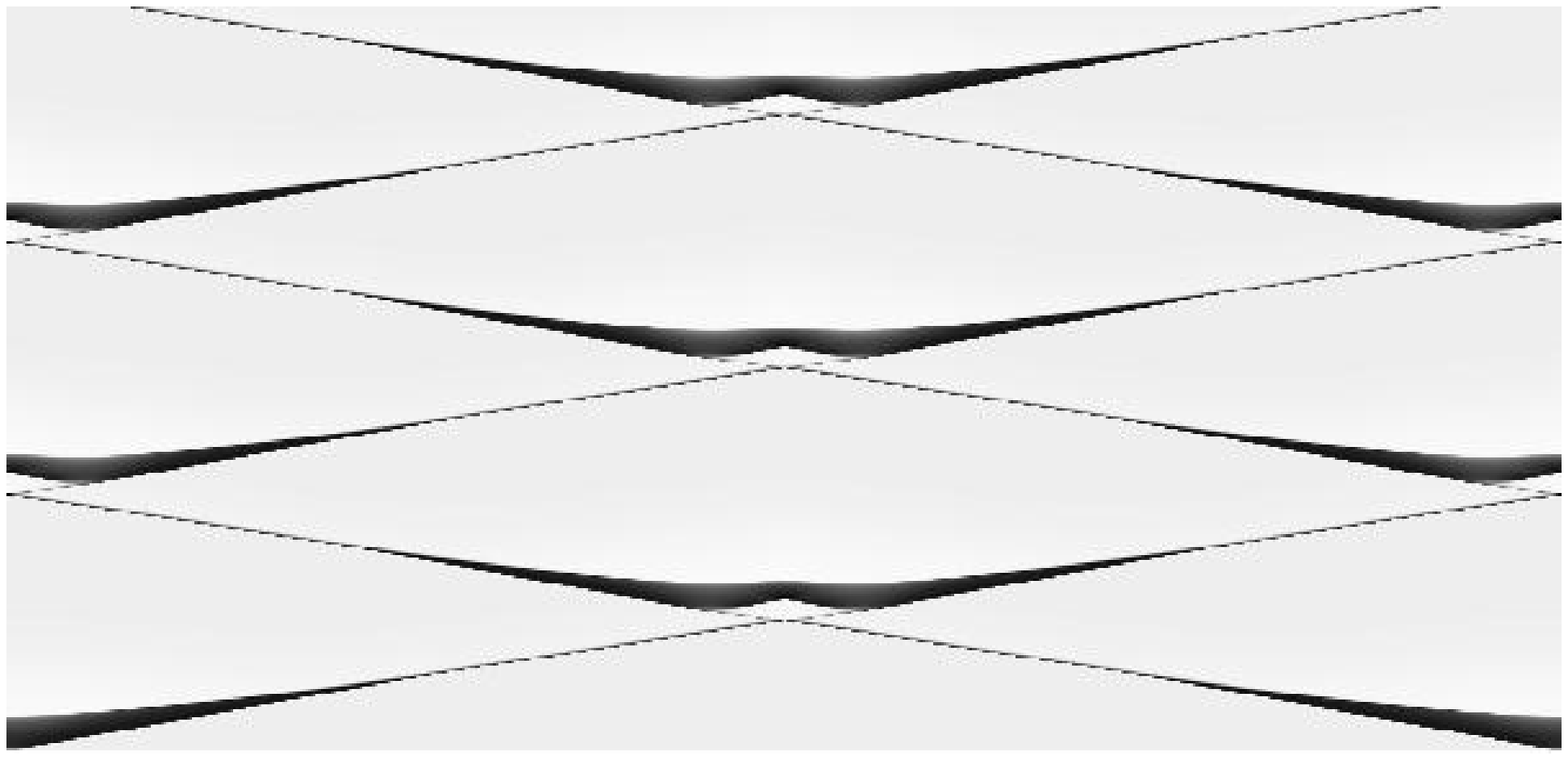} \\
(a) $h_-=h_+=0$
&
(b) $h_-=h_+=1$
&
(c) $h_-=2,  h_+=2.2$
\end{tabular}}
\caption{
Space-time density plots showing interaction of waves in (a)
purely diffusive in (b) and (c) taxis cases. On all panels, the length of the
interval $L=500$, the time scale $t\in[0,3000]$. Black corresponds to
$P=0.9$, white to $P=0$.
}\label{collis}
\end{figure*}

We have found that the system \eqtwo(RDT,TB) has a region of parameter space
where solitary waves interact as if they were solitons.
That is, they do not annihilate, as reaction-diffusion
pulses usually do, but penetrate through, or reflect from,
each other (since the waves are indistinguishable, these two terms describe the same result).
\Fig{collis} shows results of simulations in an interval of finite length $L$ with no-flux
boundary conditions $\df{P}{x}|_{x=0,L}=0$ and $\df{Z}{x}|_{x=0,L}=0$.
Two waves were initiated simultaneously, one at each end of
the interval; the results are shown as density plots.
In the purely diffusive case, panel (a), the waves annihilate at the collision.
With the taxis terms included, panels (b) and (c), two types
of quasisoliton interaction are possible:
(b) - the waves penetrate through each other on collision,
and are then reflected from the boundaries;
(c) - after the waves penetrate through each other on collision
or reflected from the boundaries, they split.

%===========================================Param.Tab.
\begin{figure}[htbp]
\setlength{\unitlength}{1mm} \newcommand{\panel}[1]{\begin{picture}(54,54)(0,0)
\put(0,0){\mbox{\resizebox{53mm}{!}{\includegraphics{#1}}}}
\end{picture}}
\centerline{\begin{tabular}{cccc}
\panel{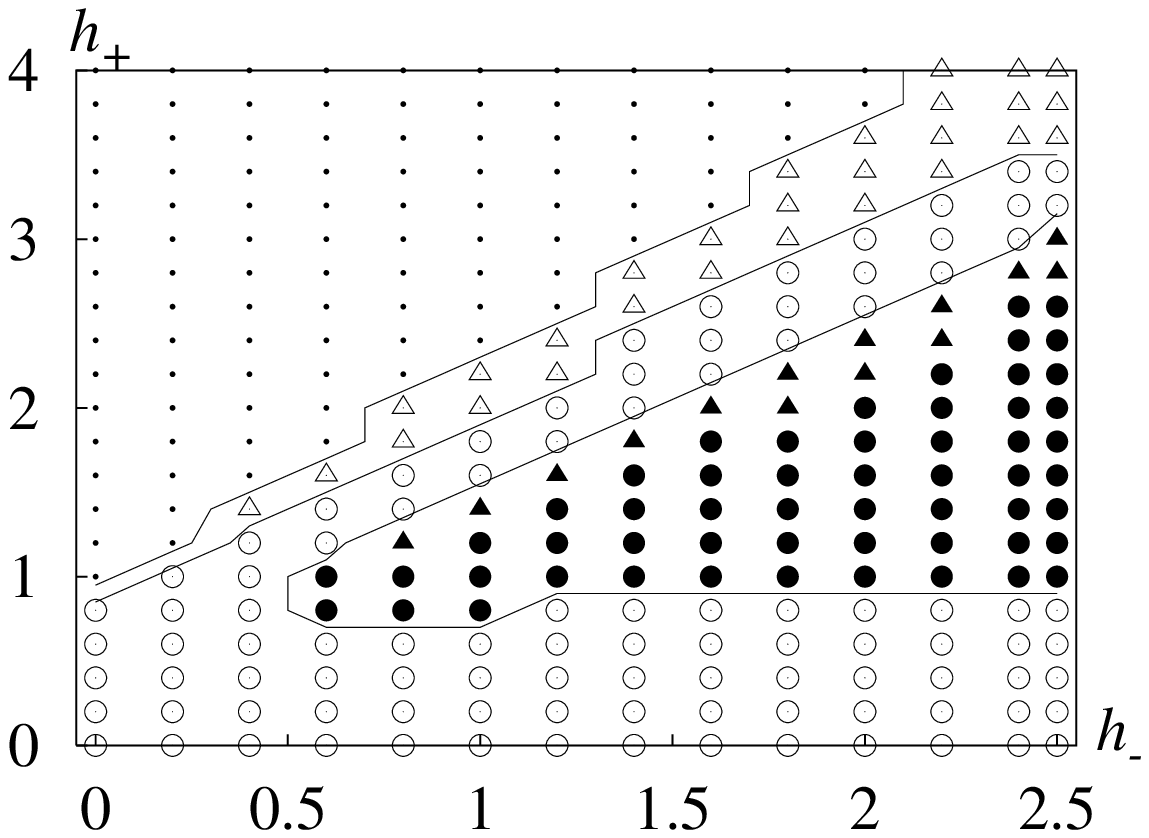}  & \panel{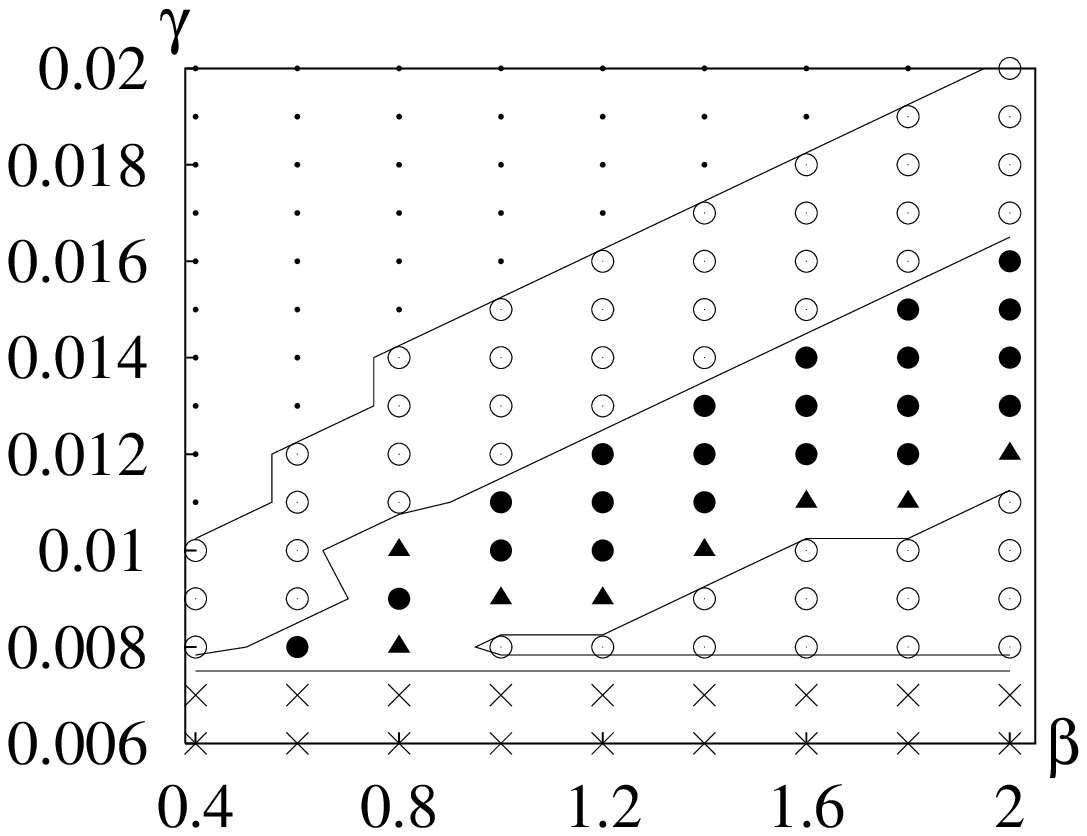}  \\
(a)  & (b) \\
\panel{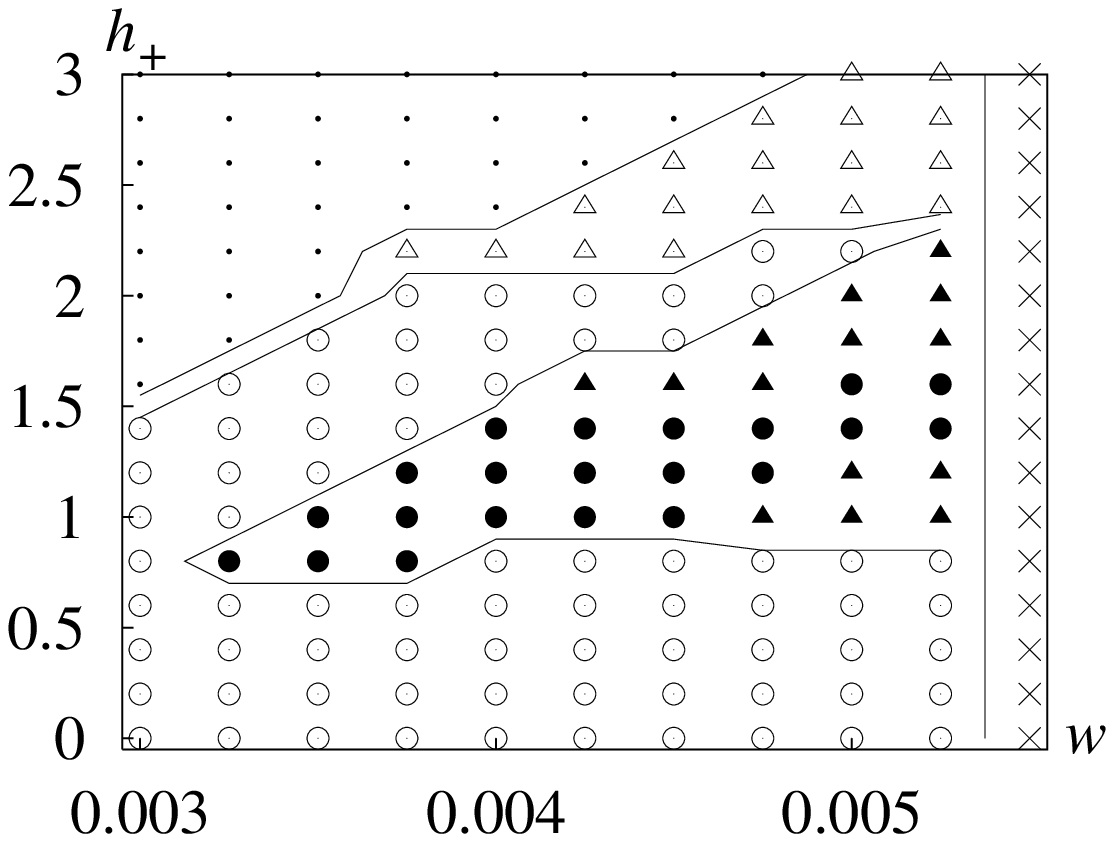} & \panel{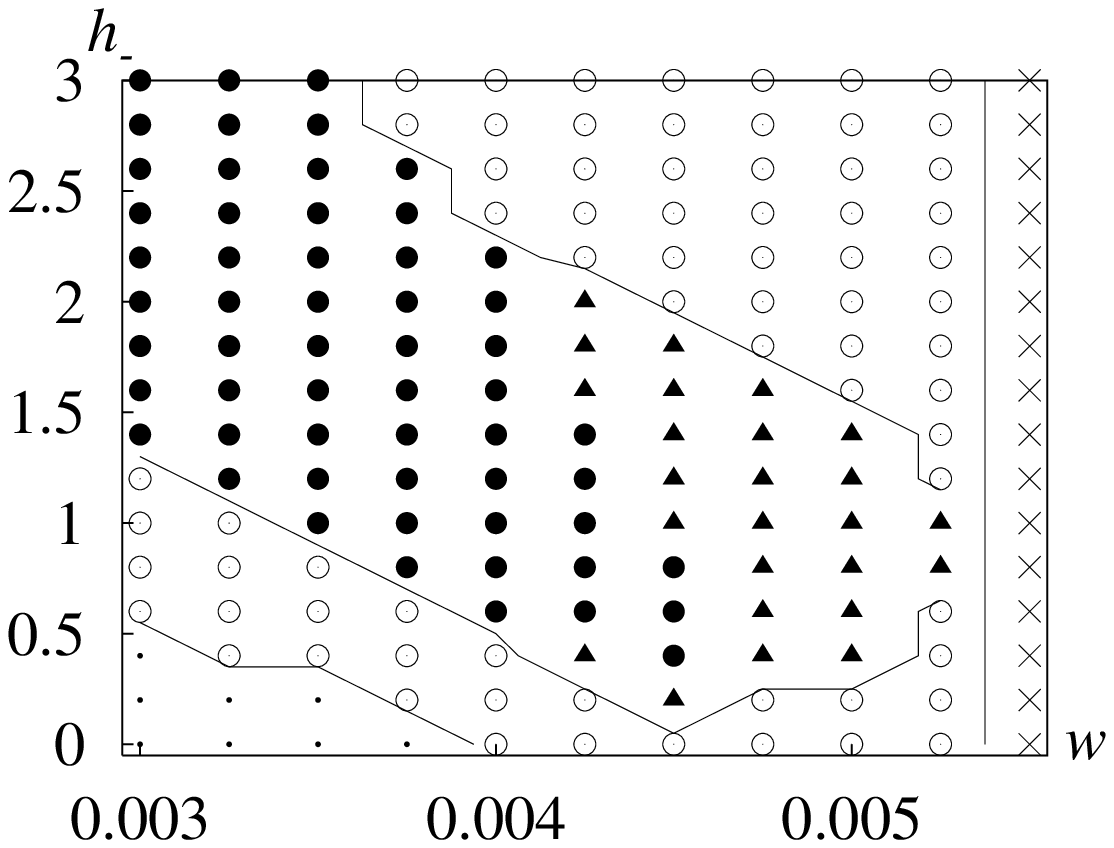} \\
(c)  & (d)
\end{tabular}}
\caption{
Parameter regions corresponding to different regimes of taxis waves ($D=0.04$).
(a) $\beta=1$, $\gamma=0.01$, $w=0.004$,
(b) $h_+=h_-=1$, $w=0.004$,
(c) $\beta=1$, $\gamma=0.01$, $h_-=1$,
(d) $\beta=1$, $\gamma=0.01$, $h_+=1$.
Solid circles: quasisolitons pulses with soliton interaction.
Solid triangles: quasisolitons with the wave splitting.
Hollow circles: stable propagation of pulses with nonsoliton interaction on collision.
Hollow triangles: propagation of pulses with the wave spliting.
Dots: no stable propagation of pulses.
Crosses: oscillatory local kinetics.
Shown are two-dimensional cross-sections in the parameter space, all
through the point corresponding to the standard parameter values.
}\label{hbgw}
\end{figure}

Soliton-like interactions of solitary waves have been
observed in some reaction-diffusion systems with excitable
kinetics, both in numerics
\cite{Koba,Petrov,Kozek,AslMor97,AslMor99,RoterMikh} and in experiments \cite{Rotermund,RoterMikh}.
Such interactions are always limited to narrow parameter
ranges close to the boundaries between excitable and oscillatory (limit cycle)
regimes of the reaction kinetics.
In contrast, \fig{hbgw} shows regions in the parameter space
corresponding to different regimes of interaction and propagation of
taxis waves described by equations \eqtwo(RDT,TB). Both the existence
of steady propagating pulses and their ability to penetrate/reflect
have a complex relationship with the kinetic and propagation
parameters. However, it is quite clear that the ranges of parameters
providing soliton-like behavior are not in any sense narrow, and do
not require proximity to the oscillatory regime in the reaction
kinetics. Although large enough $h_+$ is typically sufficient for
propagation of waves, quasisoliton behavior requires both $h_+$ and
$h_-$ to be non-vanishing.

\begin{figure*}[htbp]
\setlength{\unitlength}{1mm} \newcommand{\panel}[1]{\begin{picture}(52,52)(0,0)
\put(1,1){\mbox{\resizebox{49mm}{!}{\includegraphics{#1}}}}
\end{picture}}
\centerline{\begin{tabular}{ccc}
\panel{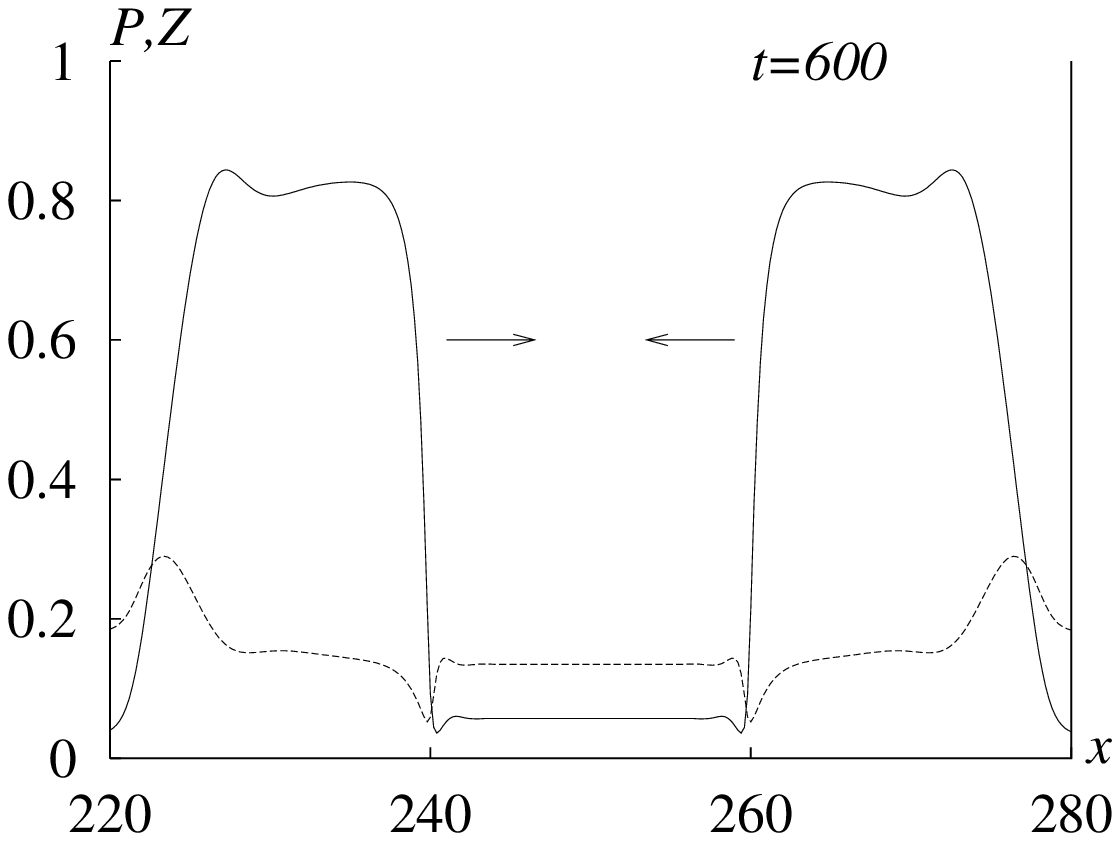} & \panel{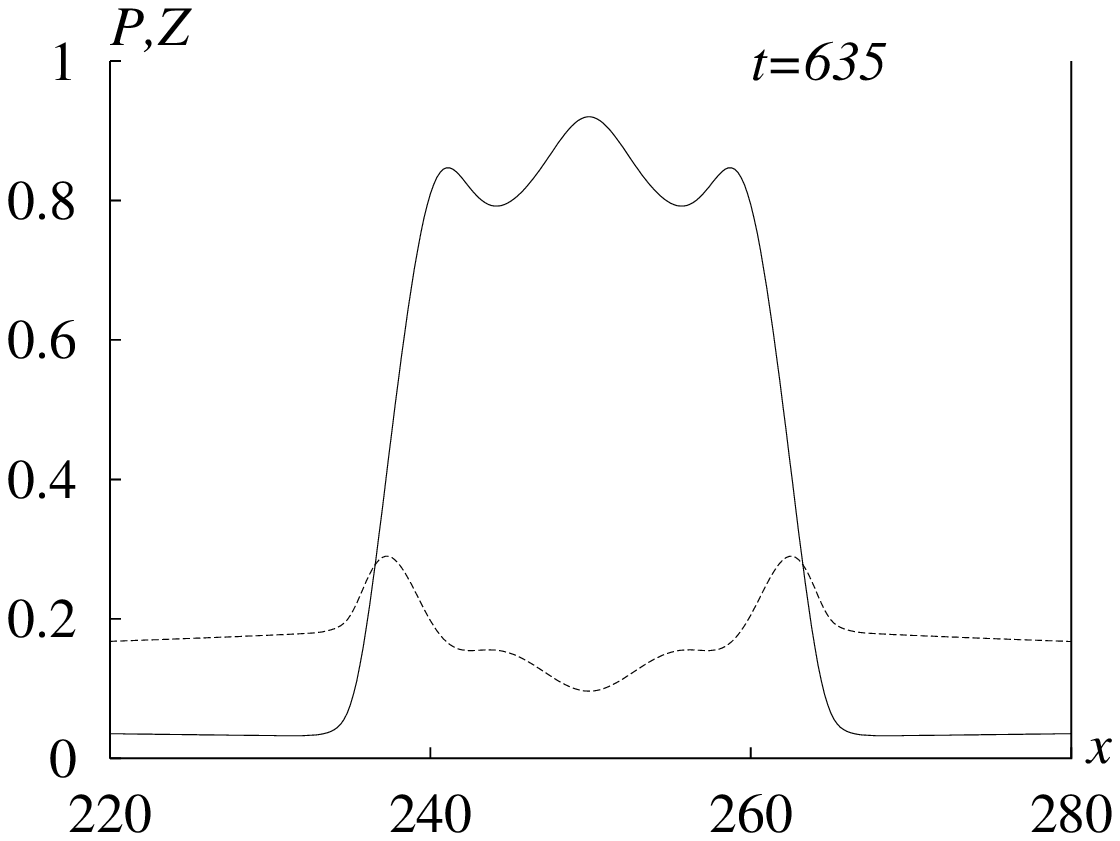} & \panel{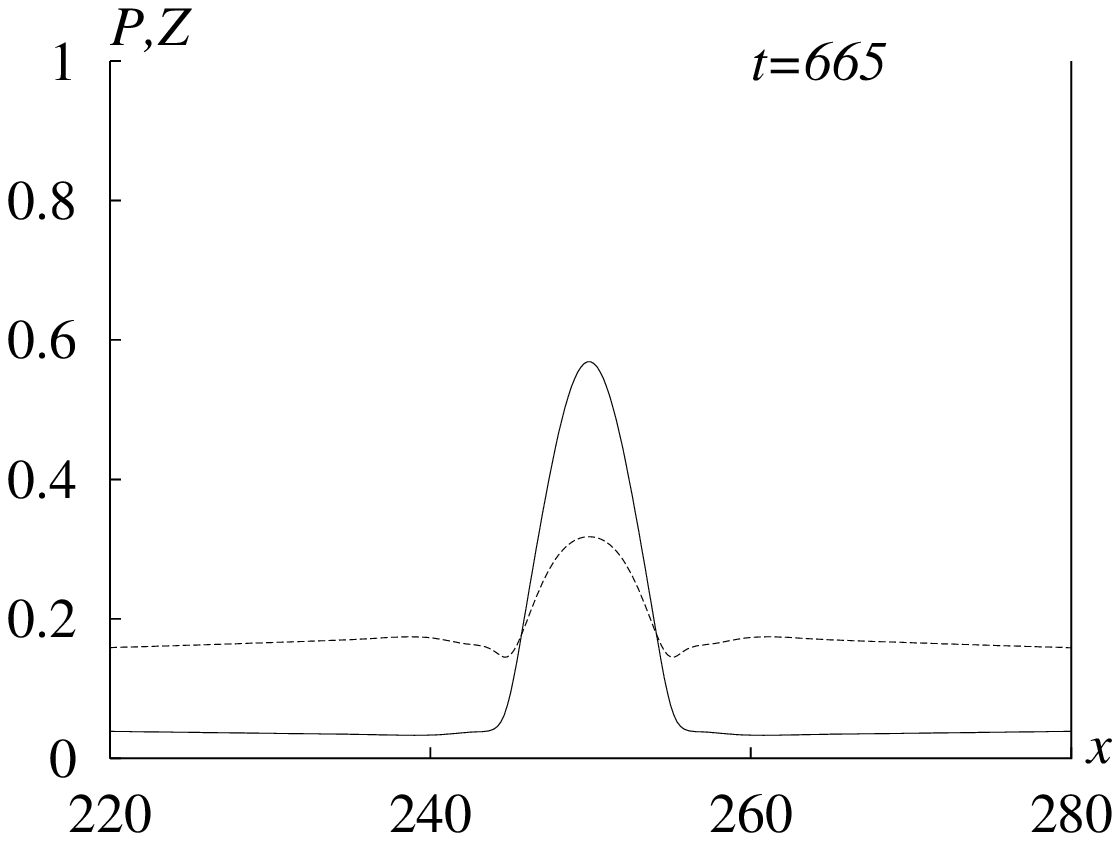} \\
(a)
&
(b)
&
(c)
\end{tabular}}
\centerline{\begin{tabular}{ccc}
\panel{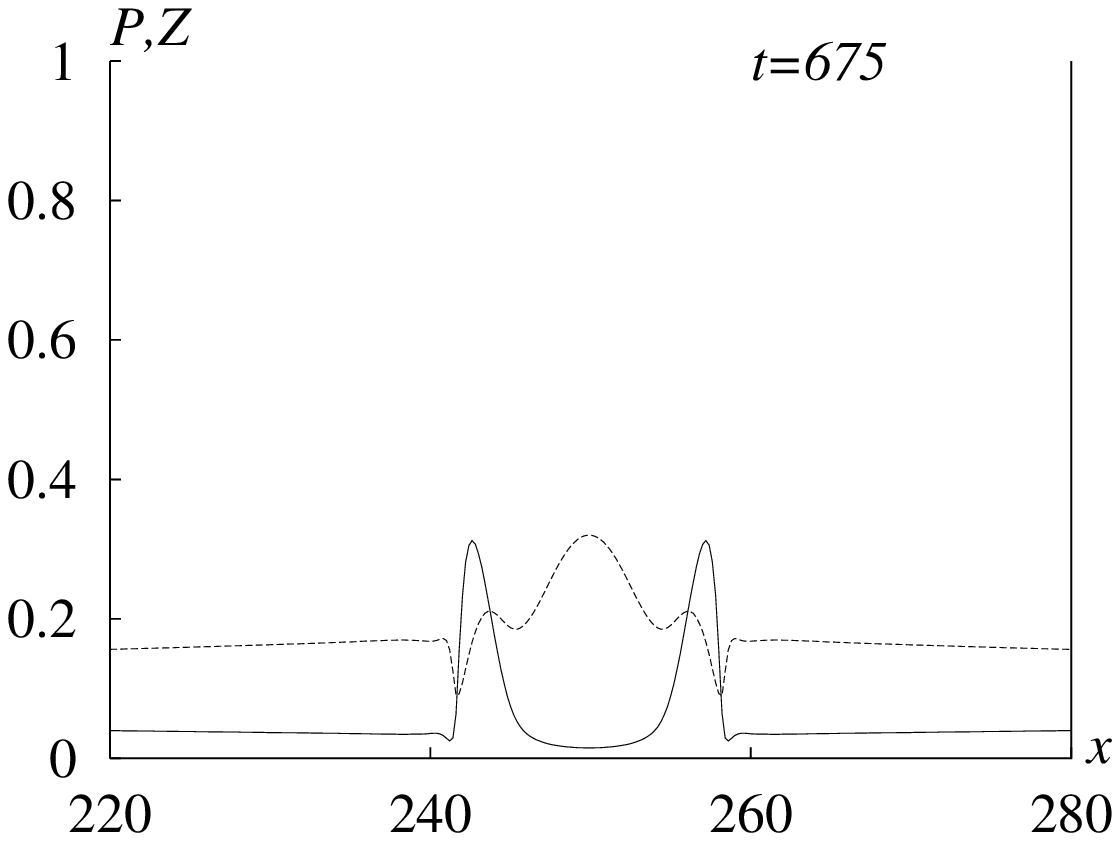} & \panel{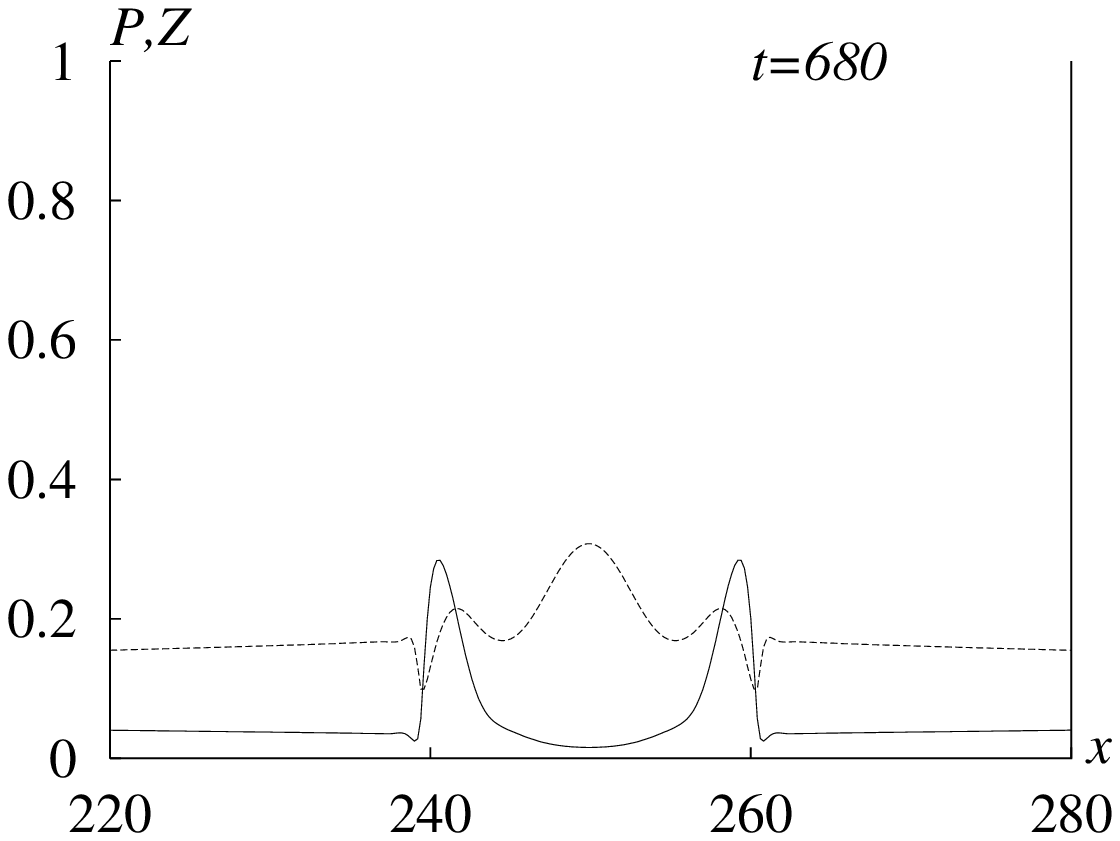} & \panel{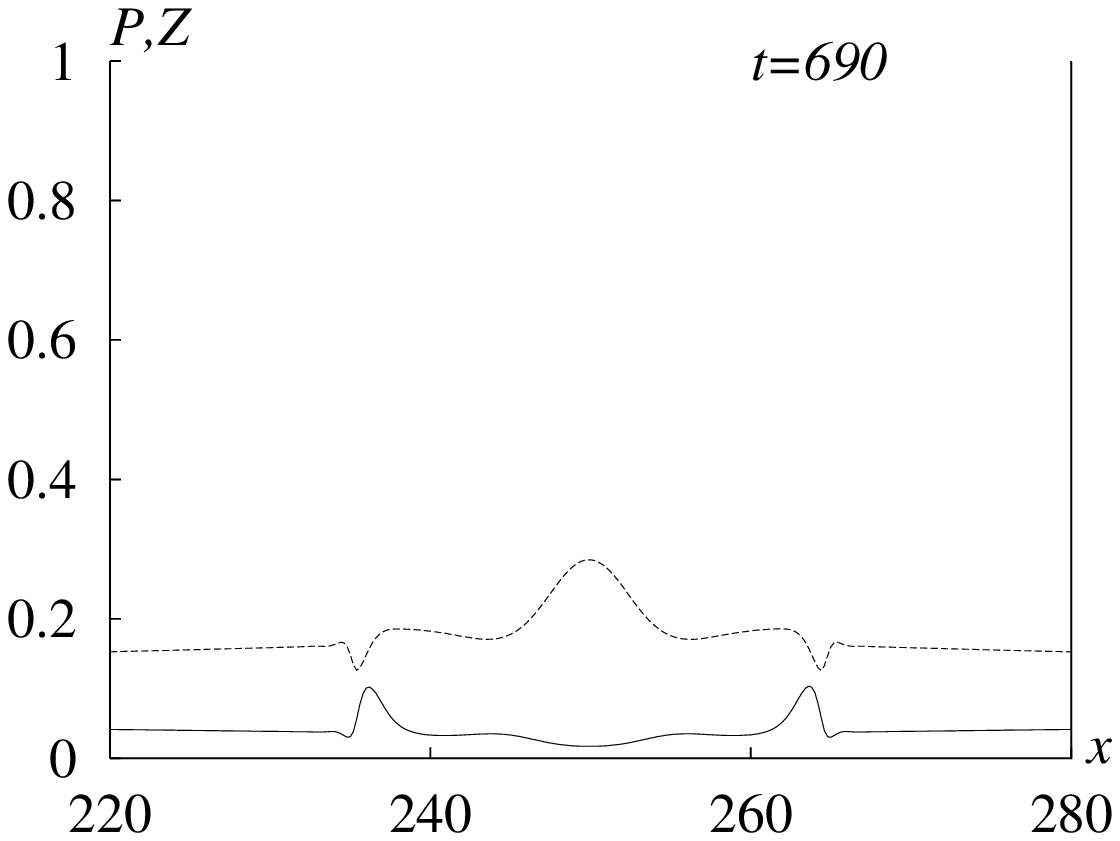} \\
(d)
&
(e)
&
(f)
\end{tabular}}
\caption{ Dynamics of the nonsoliton interaction of taxis waves. $D=0.04, h_-=2, h_+=0.85$
}\label{newAN}
\end{figure*}

Another qualitative difference of interaction of taxis waves from
waves in reaction-diffusion excitable systems is that in the
non-soliton regime (hollow circles on \fig{hbgw}) the waves
demonstrate interference. After the impact (\fig{newAN}a,b,c) they do
not annihilate straight away, but penetrate through each other
(\fig{newAN}d,e) and only decay afterwards (\fig{newAN}f).

% =====================================================NewAN

% -------------------------------------------
\subsection{Wave-splitting}

\begin{figure*}[htbp]
\setlength{\unitlength}{1mm} \newcommand{\panel}[1]{\begin{picture}(42,62)(0,0) % (0,0)
\put(0,0){\vector(1,0){41}} \put(41,1){$x$}
\put(0,0){\vector(0,1){61}} \put(1,61){$t$}
\put(1,1){\setlength{\fboxsep}{0pt}\fbox{\resizebox{38mm}{58mm}{\includegraphics{#1}}}}
\end{picture}}
\newcommand{\panelE}[1]{\begin{picture}(42,33)(0,0) % (0,0)
\put(0,0){\vector(1,0){41}} \put(41,1){$x$}
\put(0,0){\vector(0,1){32}} \put(1,32){$t$}
\put(1,1){\setlength{\fboxsep}{0pt}\fbox{\resizebox{38mm}{29mm}{\includegraphics{#1}}}}
\end{picture}}
\centerline{\begin{tabular}{cccc}
\panel{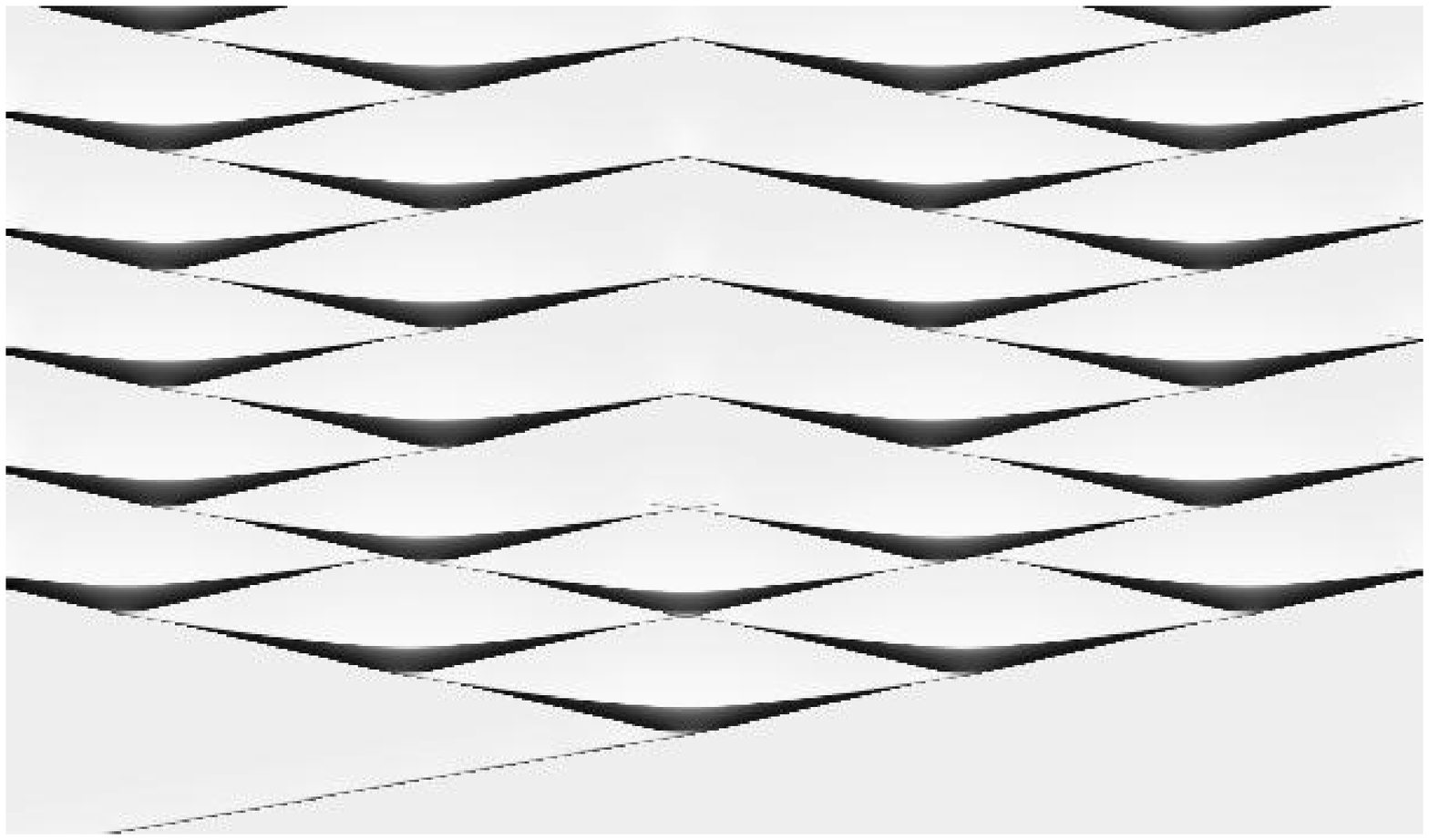} & \panel{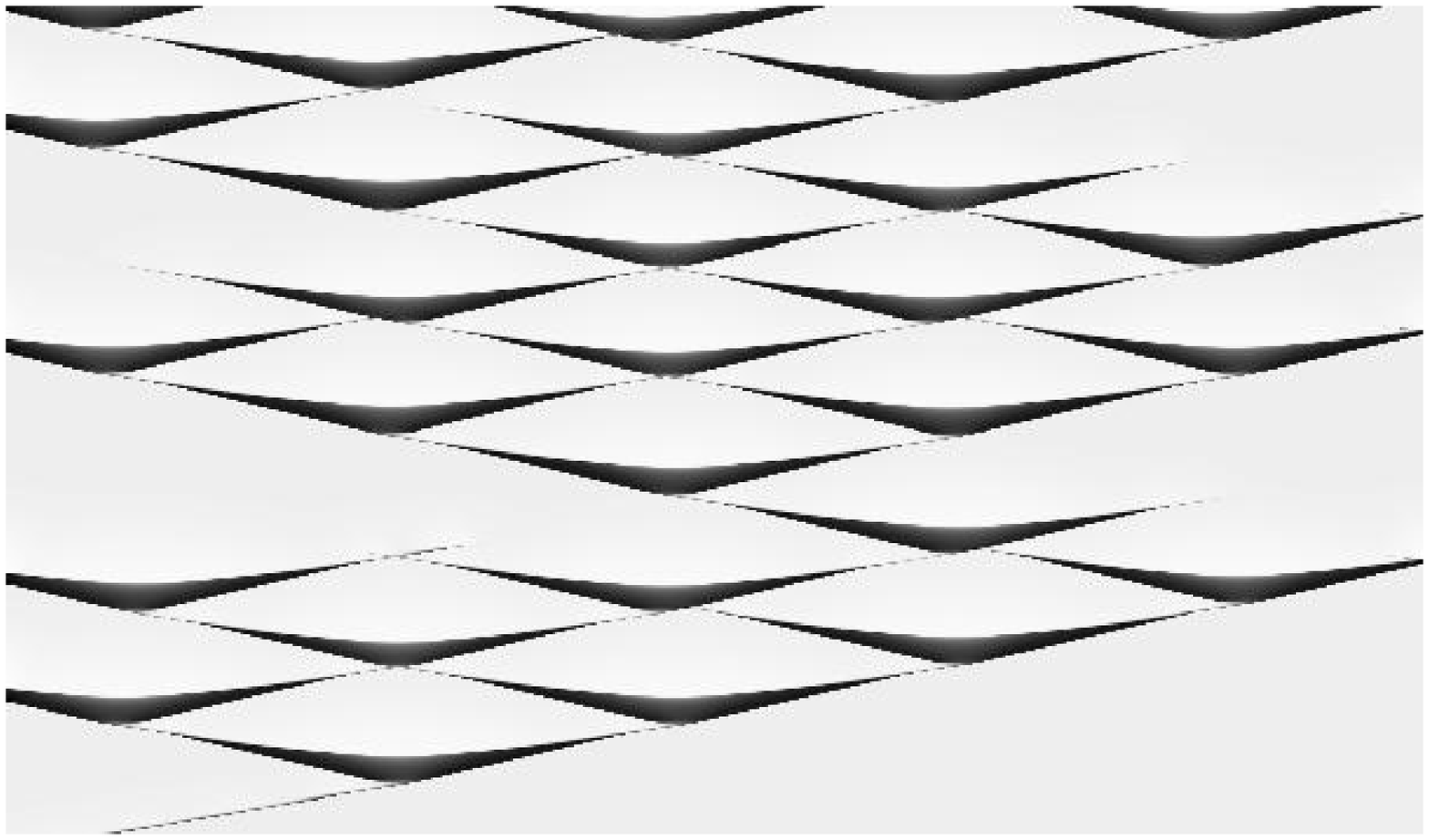} & \panel{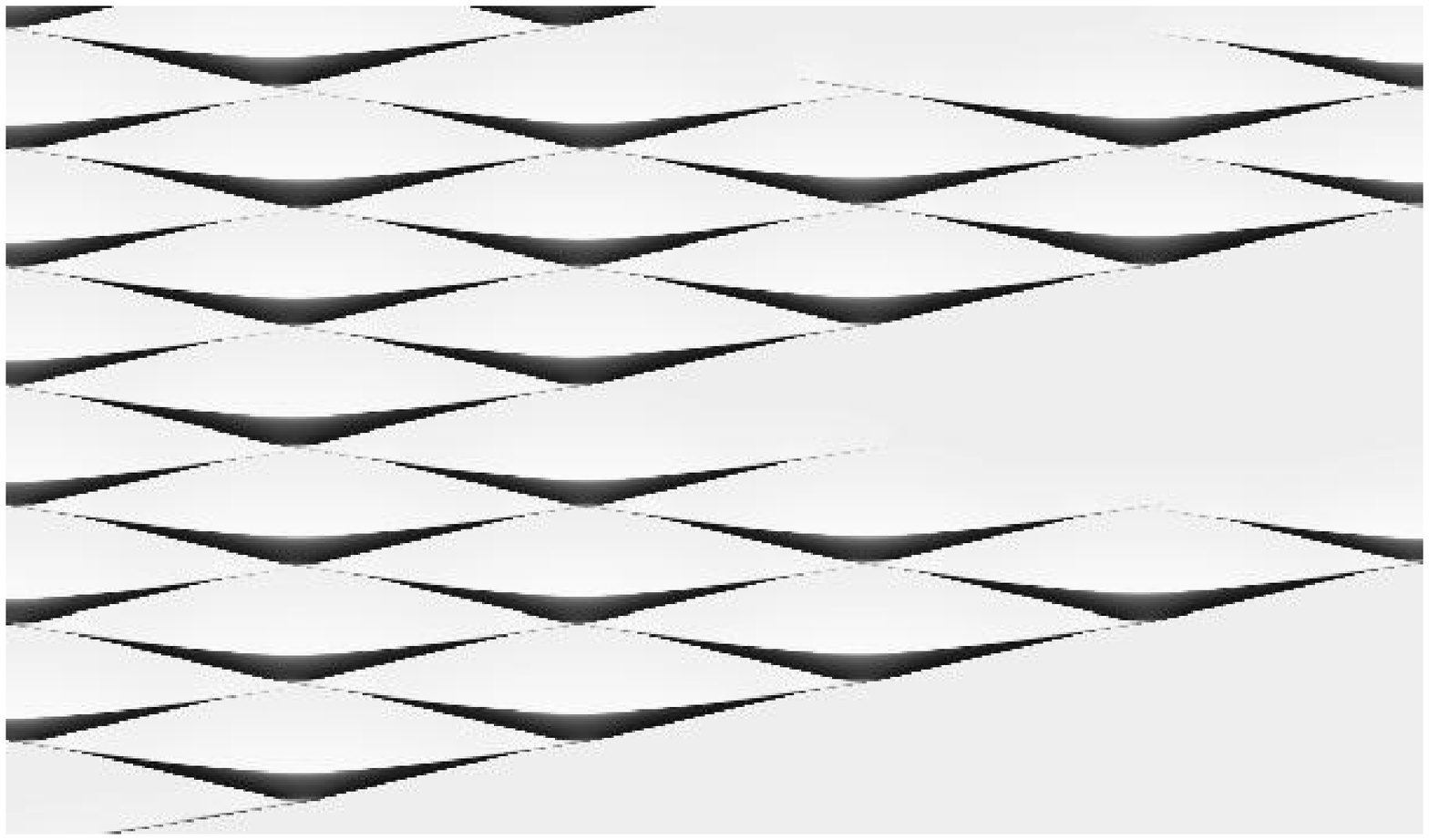} & \panel{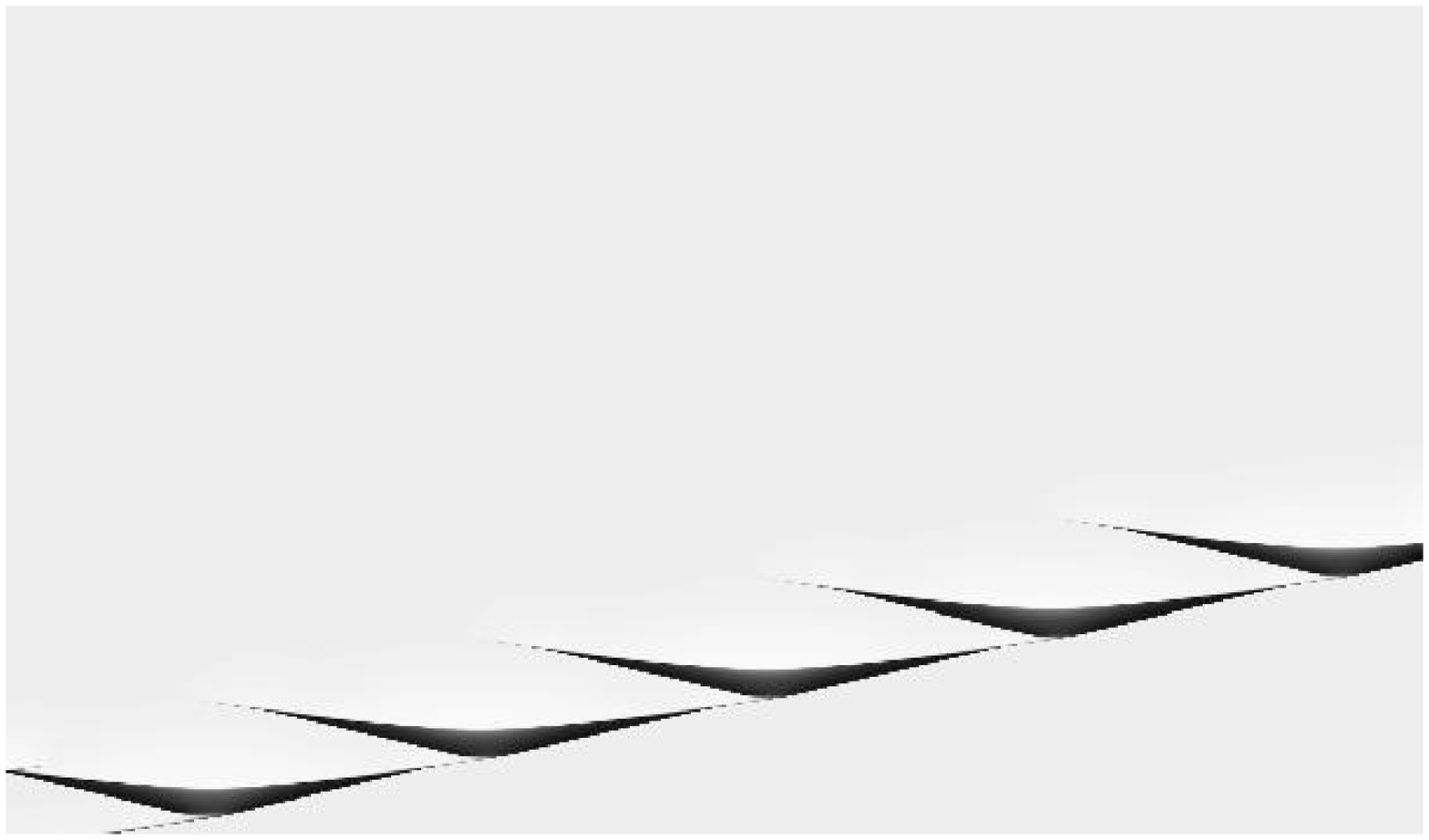} \\
(a) $h_-=2, h_+=2.8$
&
(b) $h_-=2, h_+=2.9$
&
(c) $h_-=2, h_+=3$
&
(d) $h_-=2, h_+=3.1$
\end{tabular}}
\caption{ Space-time density plot propagation and interactions of pulses with
spliting. $L=600$; $t\in[0,3500]$
}\label{birds}
\end{figure*}

The phenomenon of splitting of taxis waves can be observed both in the case
of soliton-like interaction (\fig{hbgw}, solid triangles) and in the
case of non-soliton interaction  (\fig{hbgw}, hollow triangles).
\Fig{birds} shows space-time density plots of various splitting regimes,
depending on $h_+$, observed during propagation of one-dimensional
taxis waves with non-permeable boundaries.

%******************************************************** Birds

\begin{figure}[htbp]
\setlength{\unitlength}{1mm} \newcommand{\panel}[1]{\begin{picture}(42,42)(-6,0) % (0,0)
\put(0,0){\vector(1,0){41}} \put(41,1){$x$}
\put(0,0){\vector(0,1){41}} \put(1,41){$t$}
\put(-6,0.35){$200-$}
\put(-6,17.2){$400-$}
\put(-6,34){$600-$}
\put(1,1){\setlength{\fboxsep}{0pt}\fbox{\resizebox{38mm}{38mm}{\includegraphics{#1}}}}
\end{picture}}
\newcommand{\panelA}[1]{\begin{picture}(42,42)(-6,0) % (0,0)
\put(1,1){\mbox{\resizebox{41mm}{41mm}{\includegraphics{#1}}}}
\end{picture}}
\centerline{\begin{tabular}{cc}
\panel{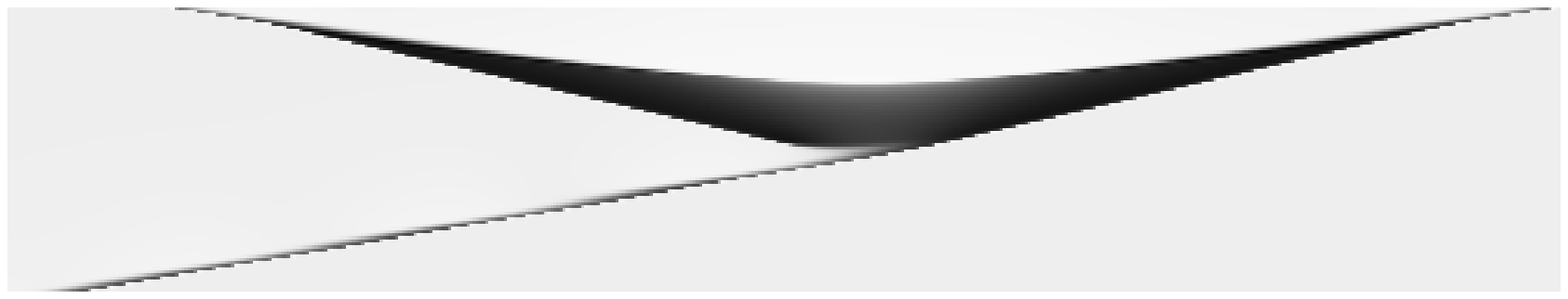} & \panelA{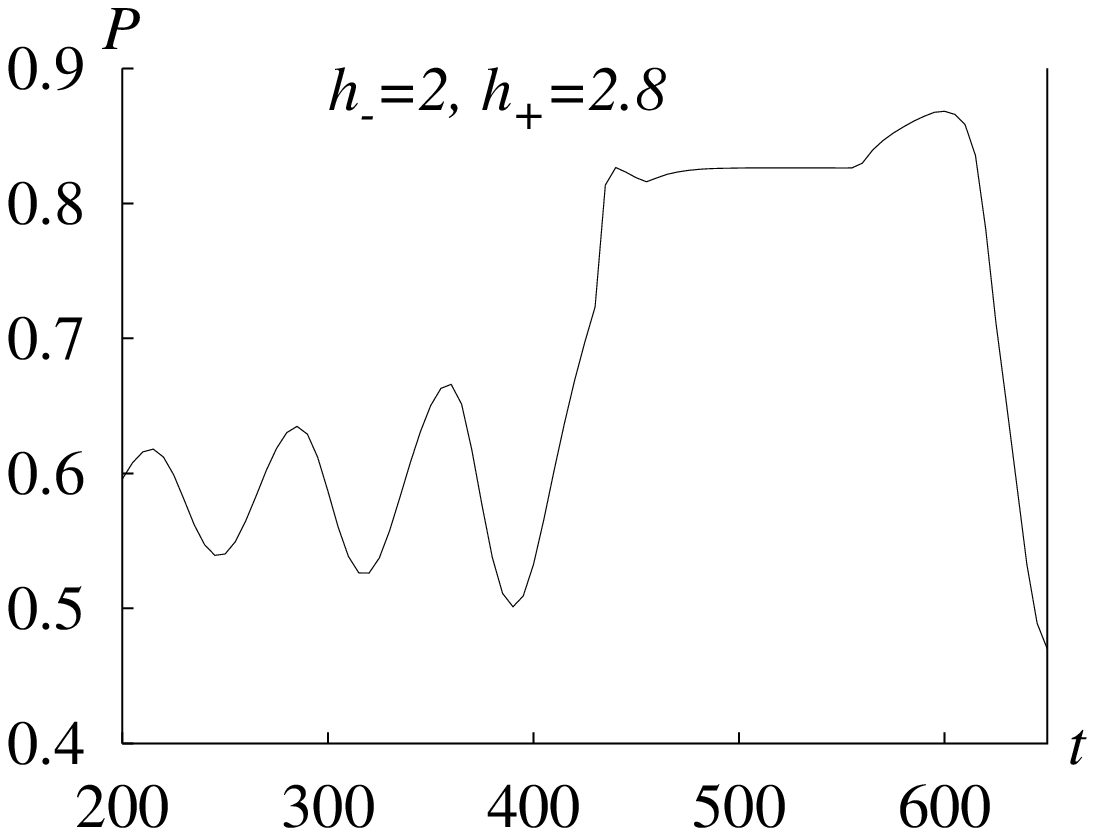} \\
(a) & (b)
\end{tabular}}
\caption{ Dynamics of wave splitting ($h_-=2, h_+=2.8$).
(a) Spacetime density plot ($x\in[150,400]$, $t\in[200,650]$).
(b) $A(t)=\max\limits_{x\in[0,L]}P(x,t)$
}\label{onebird}
\end{figure}

With the increase of $h_+$ (\fig{birds}b-c), the time interval between
subsequent splitting events decreases.
The backward waves splitting from the forward propagating wave may decay,
or split themselves (\fig{birds}). This chain of splitting events can
lead to self-supporting, aperiodic or approximately periodic
activity. A single act of splitting is shown in more detail on
\fig{onebird}. The splitting event is preceded by an oscillatory
instability of the propagating wave, which can be seen as the
growing oscillations of dynamic variables; see e.g. the evolution of the
amplitude of the prey population,
$A(t)=\max\limits_{x\in[0,L]}P(x,t)$, on
\fig{onebird}b.

%***************************************************************** Onebird

\begin{figure}[htbp]
\setlength{\unitlength}{1mm} \newcommand{\panel}[1]{\begin{picture}(42,42)(-6,0) % (0,0)
\put(0,0){\vector(1,0){41}} \put(41,-3){$\xi$}
\put(0,0){\vector(0,1){41}} \put(1,41){$t$}
\put(-6,17){$500-$}
\put(-7,34){$1000-$}
\put(1,1){\setlength{\fboxsep}{0pt}\fbox{\resizebox{38mm}{38mm}{\includegraphics{#1}}}}
\end{picture}}
\newcommand{\panelB}[1]{\begin{picture}(43,43)(-6,0) % (0,-20)
\put(1,1){\mbox{\resizebox{42mm}{42mm}{\includegraphics{#1}}}} \end{picture}}
\centerline{\begin{tabular}{cc}
\panel{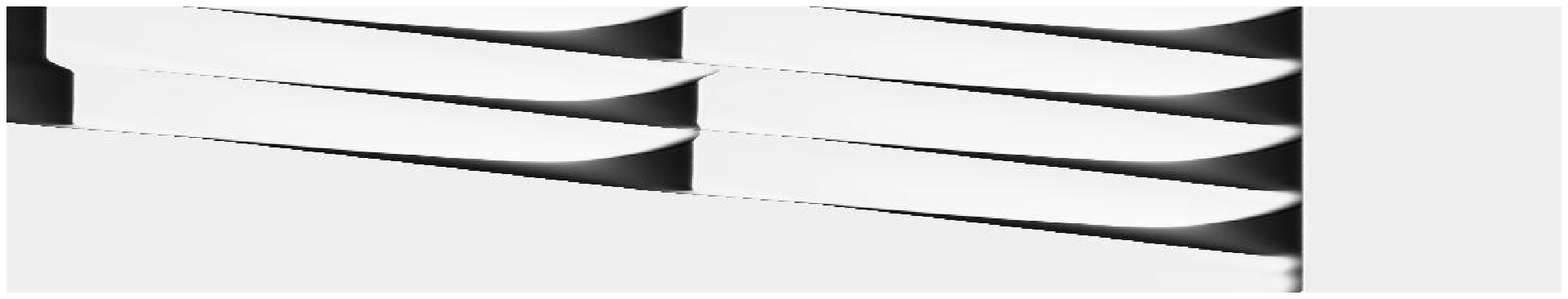} & \panelB{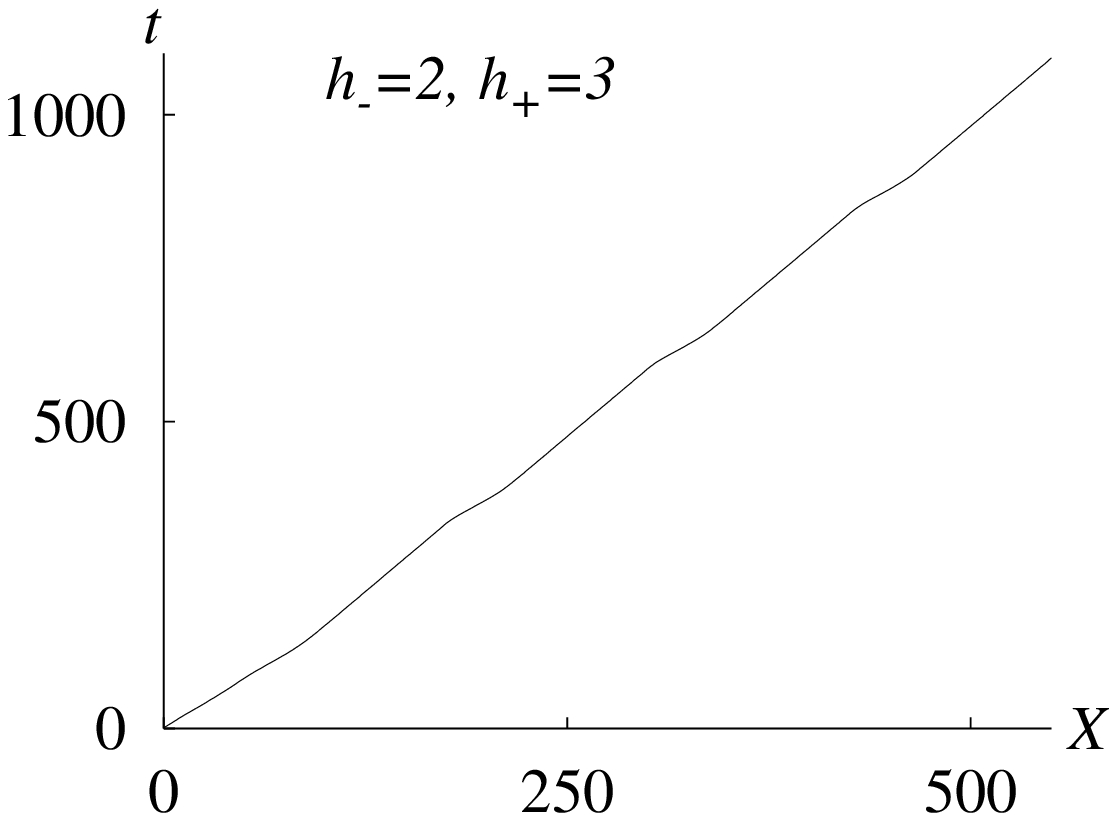} \\
(a)
&
(b)
\end{tabular}}
\caption{
 The results of simulation of \eq{RDTmoving} ($h_-=2$, $h_+=3$).
(a) The splitting waves in a moving frame ($\xi$=$x-X(t)$) of reference attached to their advancing front.
(b) The coordinate of the front as function of time.
}\label{Integral}
\end{figure}

So, on \fig{birds}(c), we observe how a single solitary taxis wave leads to a
self-supporting splitting-wave activity. To see better the properties of the advancing front
of this activity, we have used the explicit decomposition of the system \eq{RDT} with respect
to the translation group, as described in \cite{hypermeander}:
\begin{eqnarray}
\df{P}{t} &=& f(P,Z) + D\ddf{P}{x} + h_-\df{ }{x}P\df{Z}{x} - c(t) \df{P}{x}, \nonumber\\
\df{Z}{t} &=& g(P,Z) + D\ddf{Z}{x} - h_+\df{ }{x}Z\df{P}{x} - c(t) \df{Z}{x} , \nonumber\\
P(x_*,t) &=& P_*, \nonumber\\
\Df{X}{t} &=& c(t), \label{RDTmoving}
\end{eqnarray}
where $P_*$ has been chosen equal to $0.5$, and $x_*$ well within the simulation interval,
to minimize the effect of boundaries.
In other words, we consider \eq{RDT} in a moving frame of reference, whose origin has
coordinate $X(t)$ in the laboratory frame of reference.
This frame of reference
is chosen so that the position of the wave front defined
as $P(x,t)=P_*$ is always at a fixed selected point $x=x_*$ in it.
The advantage of this approach is that we can observe the evolution
of the front for a long time interval, unrestricted by the medium size.
The results of simulation of \eq{RDTmoving} are shown on \fig{Integral} (this is a small
part, actual simulation was much longer).
We can see that each backfiring wave only slightly affects the propagation velocity of
the first front.
Also, in this way we have ensured that this advancing front continued for a very
long time (up to $t=7000$), thus presumably is stable, and the process of back-splitting
remained approximately periodic for all that time.

Similarly looking splitting waves have been observed in excitable
Belousov-Zhabotinskii chemical system \cite{Guriy93}, where they were
due to the medium anisotropy induced by atmospheric oxygen, in
numerical simulations of FHN-type equations \cite{Rubin95} and
in a three-component reaction-diffusion
model describing blood clotting \cite{Fazli03}.

%===========================================================Velocity
\subsection{Propagation velocity of taxis waves}

\begin{figure*}[htbp]
\setlength{\unitlength}{1mm} \newcommand{\panel}[1]{\begin{picture}(44,44)(0,0) % (0,-20)
\put(1,1){\mbox{\resizebox{43mm}{!}{\includegraphics{#1}}}}
\end{picture}}
\centerline{\begin{tabular}{cccc}
\panel{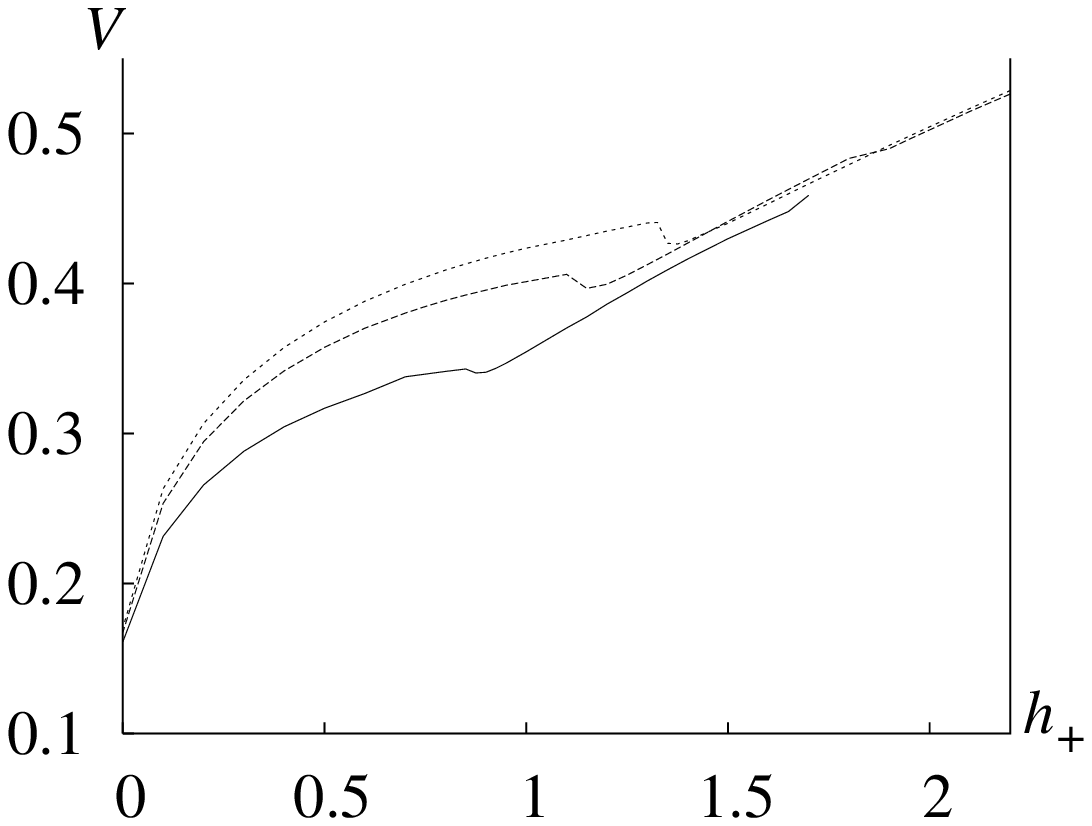} & \panel{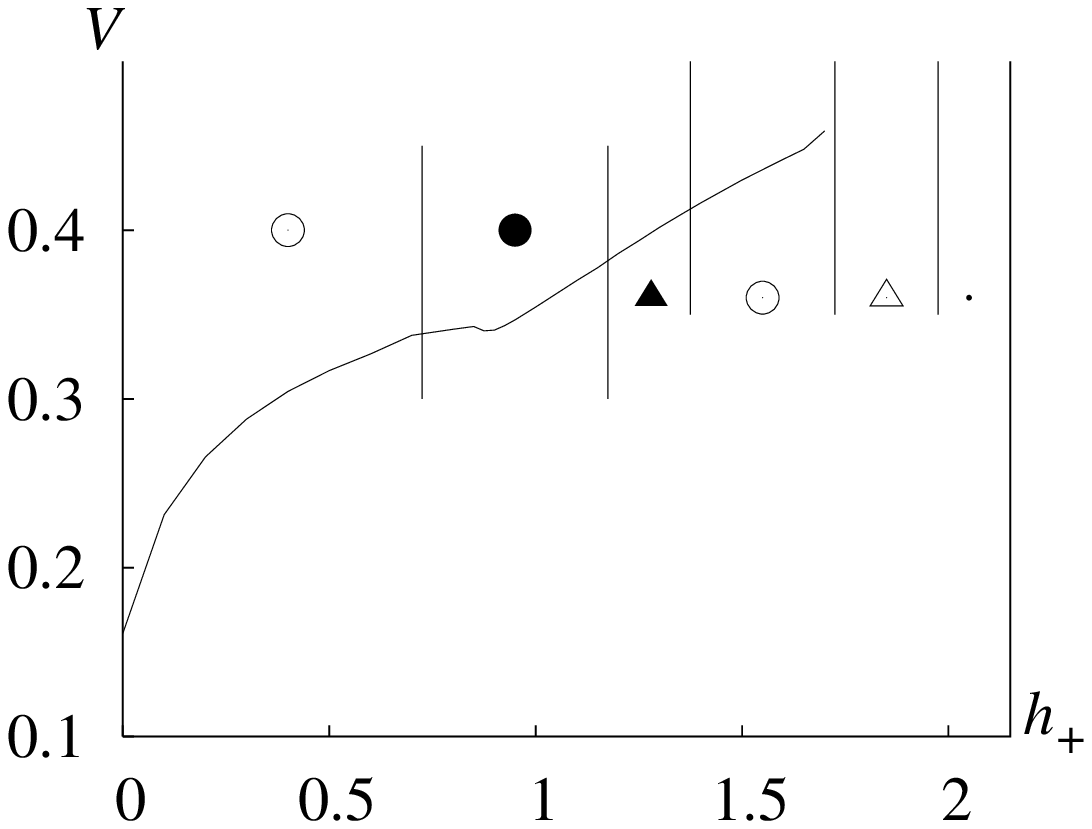} & \panel{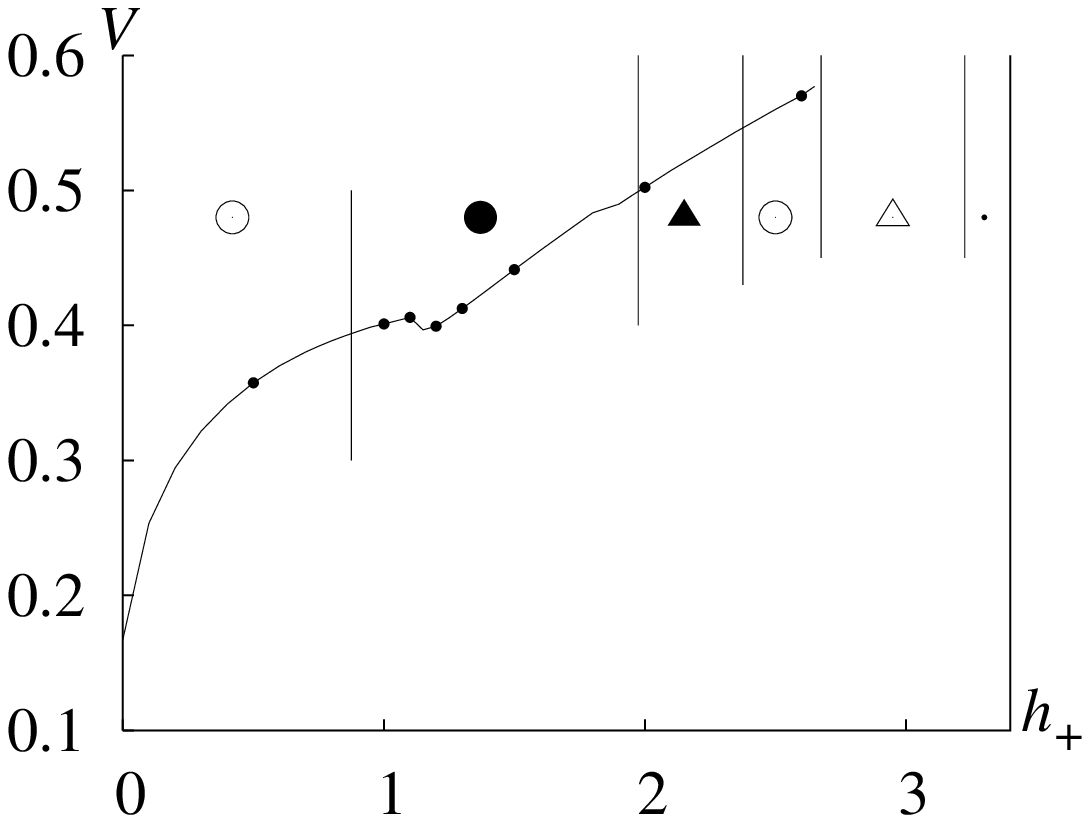} & \panel{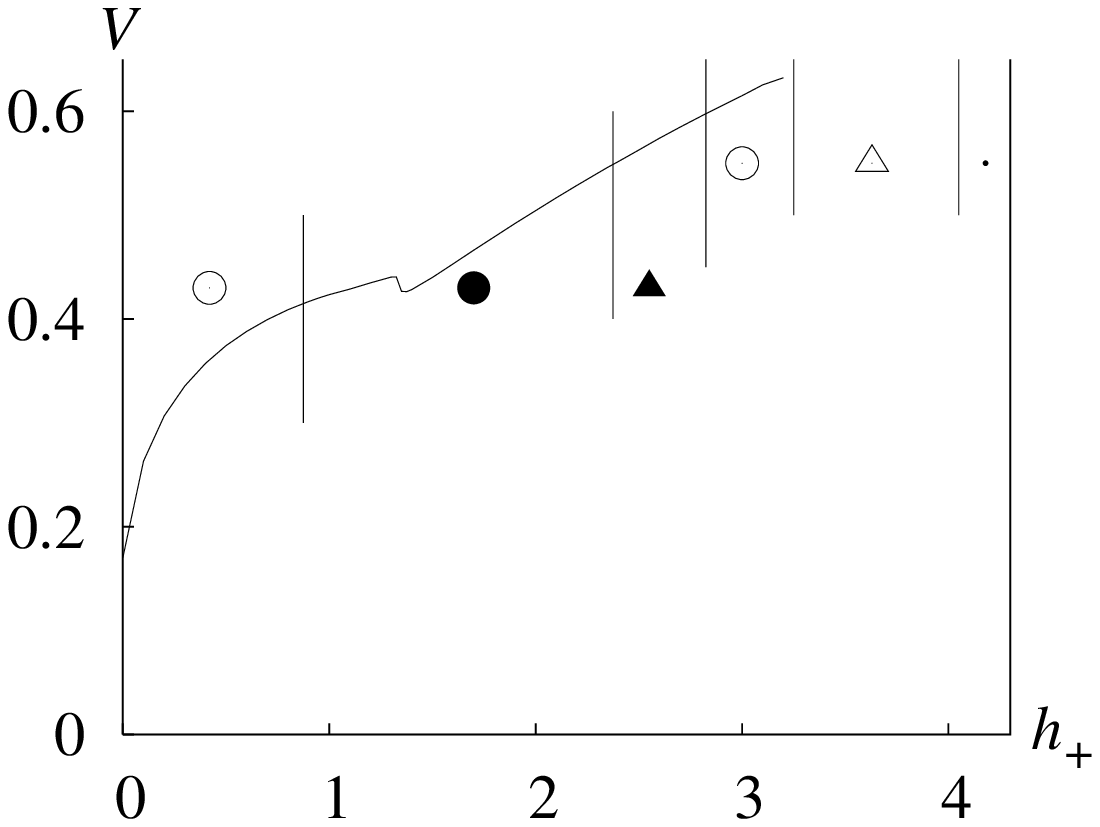} \\
(a) $h_-=1$, $h_-=2$, $h_-=2.5$
&
(b) $h_-=1$
&
(c) $h_-=2$
&
(d) $h_-=2.5$
\end{tabular}}
\caption{ Dependence of the propagation velocity of taxis waves on
the pursuit coefficient $h_+$ for various values of the
evasion coefficient $h_-$. Vertical bars and symbols on panels (b), (c) and (d) designate
regions corresponding to the different propagation and interaction
regimes.
Solid circle: quasisolitons pulses with soliton interaction.
Solid triangle: quasisolitons with the wave splitting.
Hollow circles: stable propagation of pulses with nonsoliton interaction on collision.
Hollow triangle: propagation of pulses with/by the wave splitting.
Dot: there is no stable propagation of pulses.
Dots on the graph of panel (c) correspond to the profiles shown on \fig{profV}.
}
\label{Velocity}
\end{figure*}

\begin{figure*}[htbp]
\setlength{\unitlength}{1mm} \newcommand{\panel}[1]{\begin{picture}(44,44)(0,0) % (0,-20)
\put(1,1){\mbox{\resizebox{43mm}{!}{\includegraphics{#1}}}}
\end{picture}}
\centerline{\begin{tabular}{cccccccc}
\panel{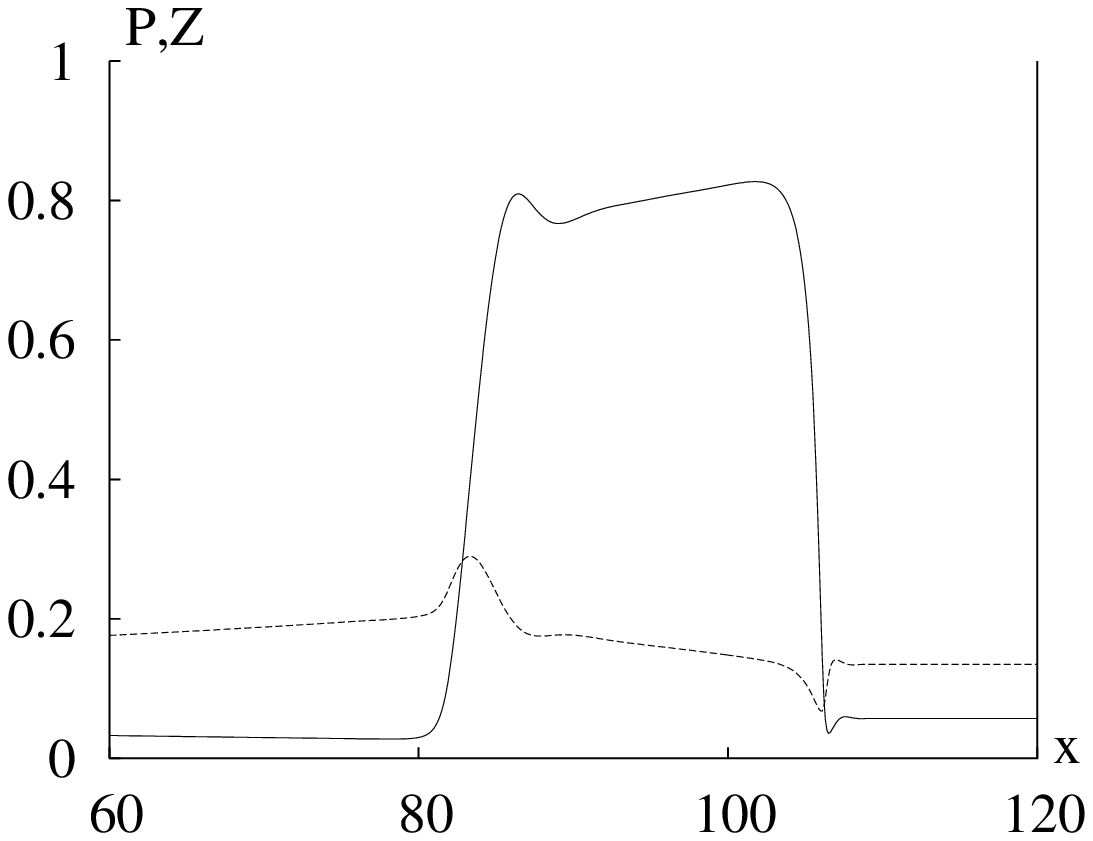} & \panel{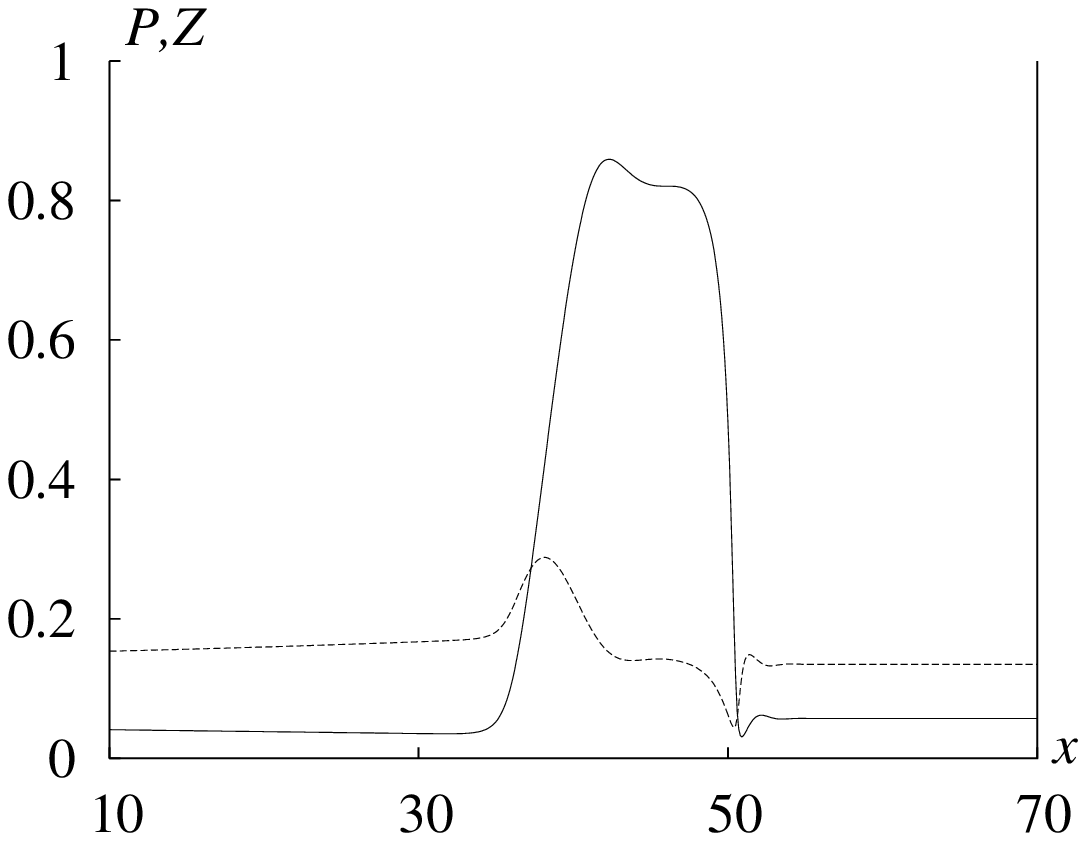} & \panel{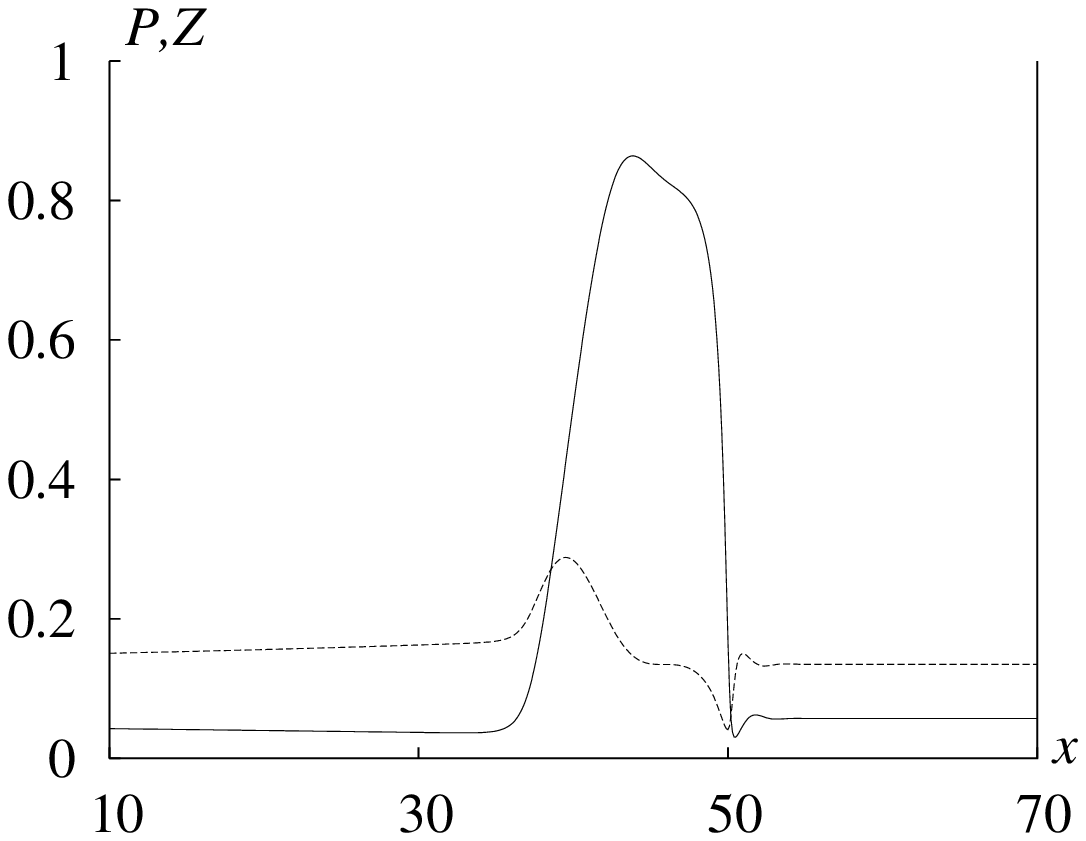} & \panel{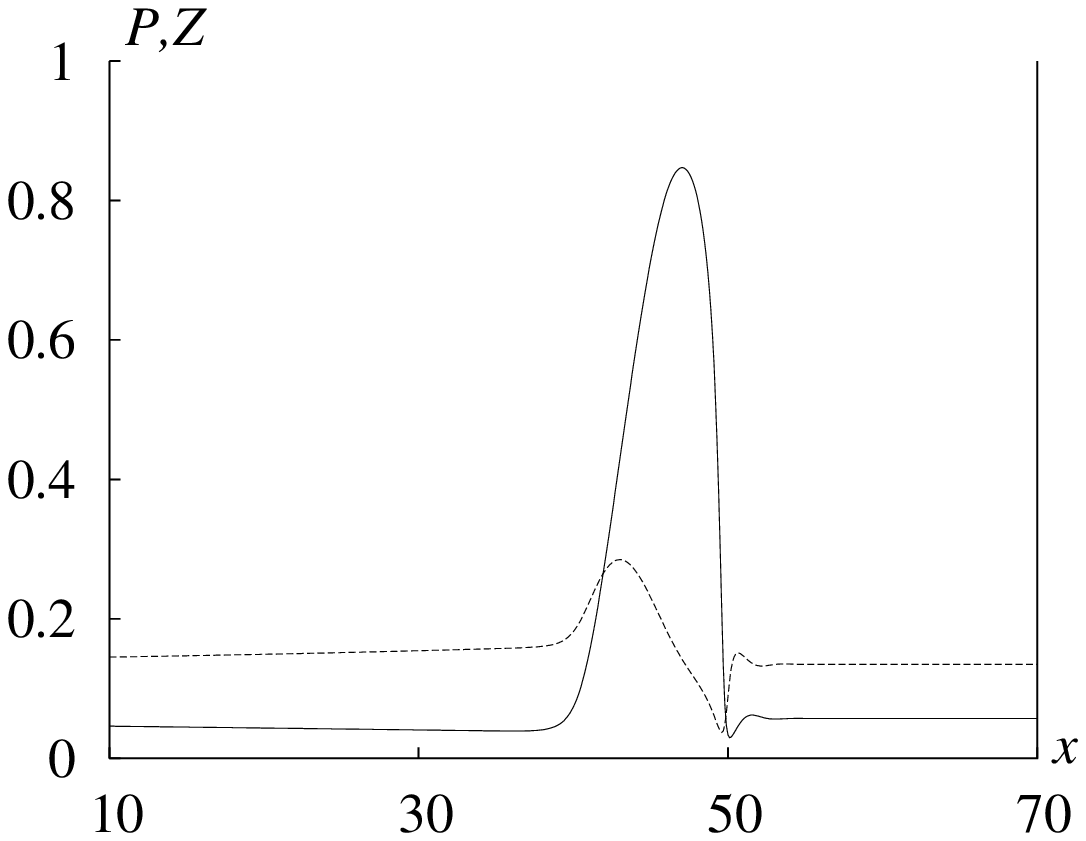} \\
(a)$h_+=0.5$ & (b)$h_+=1$ & (c)$h_+=1.1$ & (d)$h_+=1.2$ \\
\panel{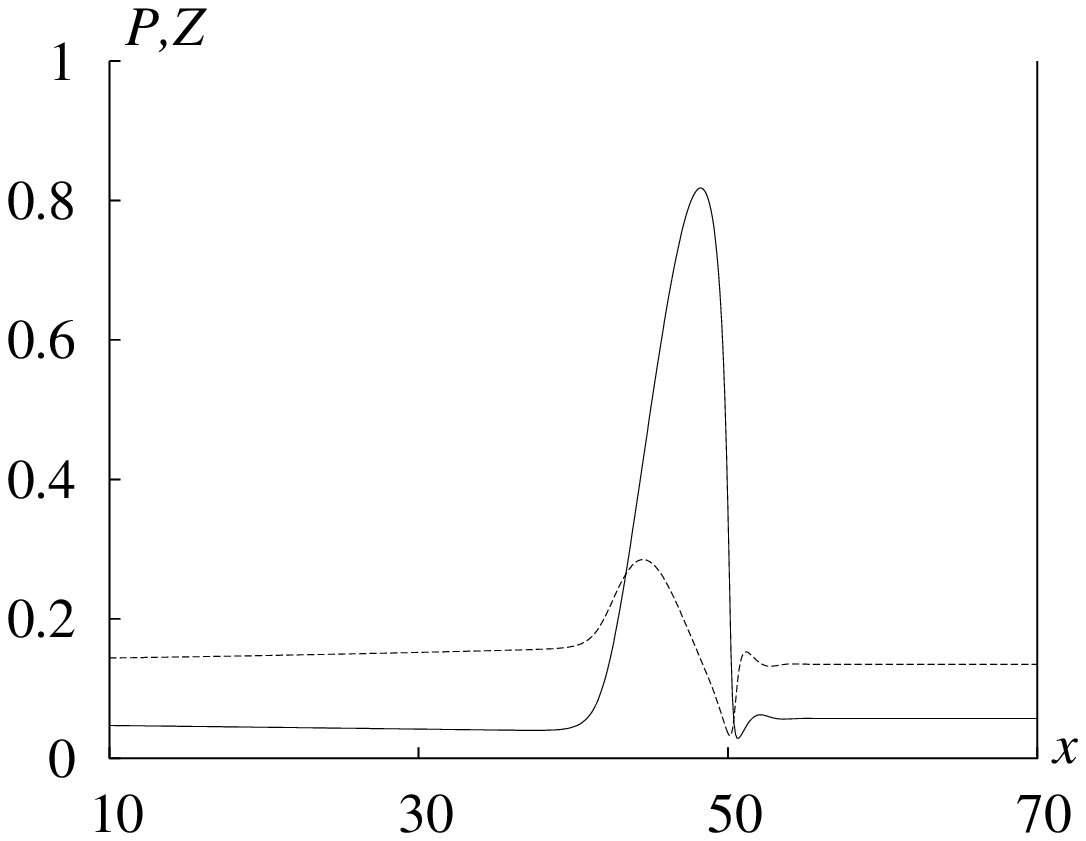} & \panel{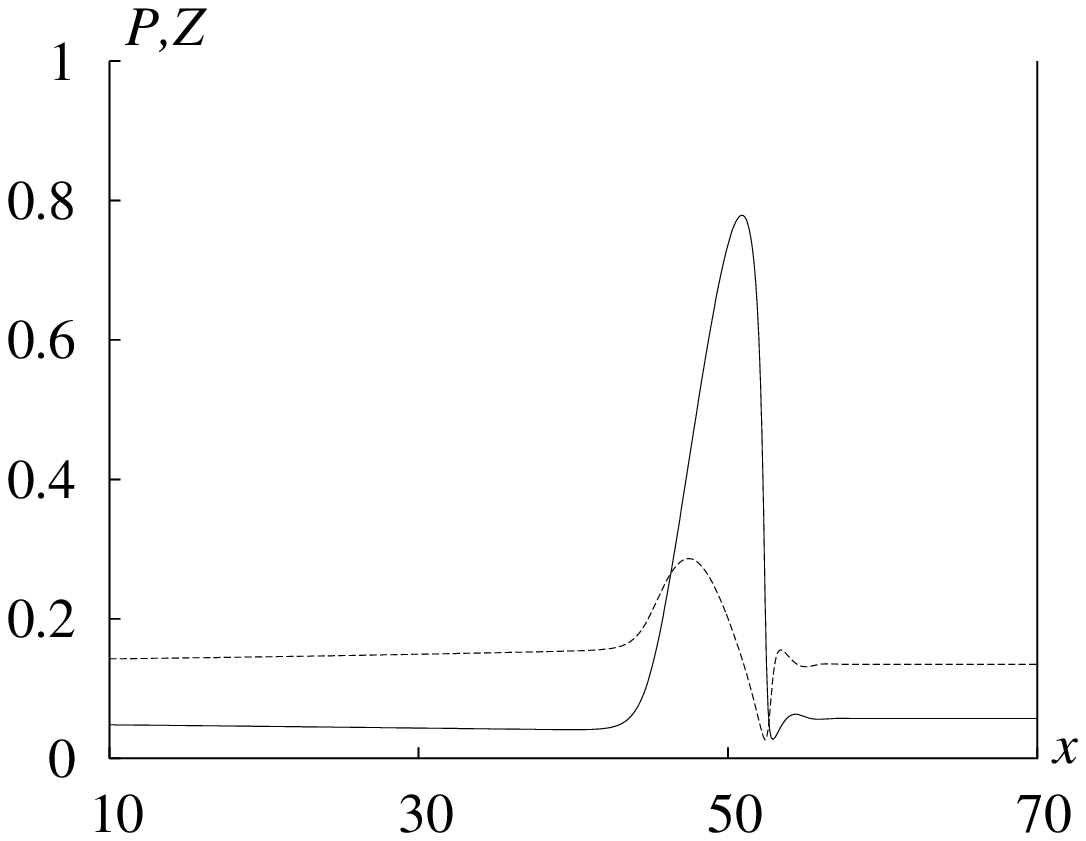} & \panel{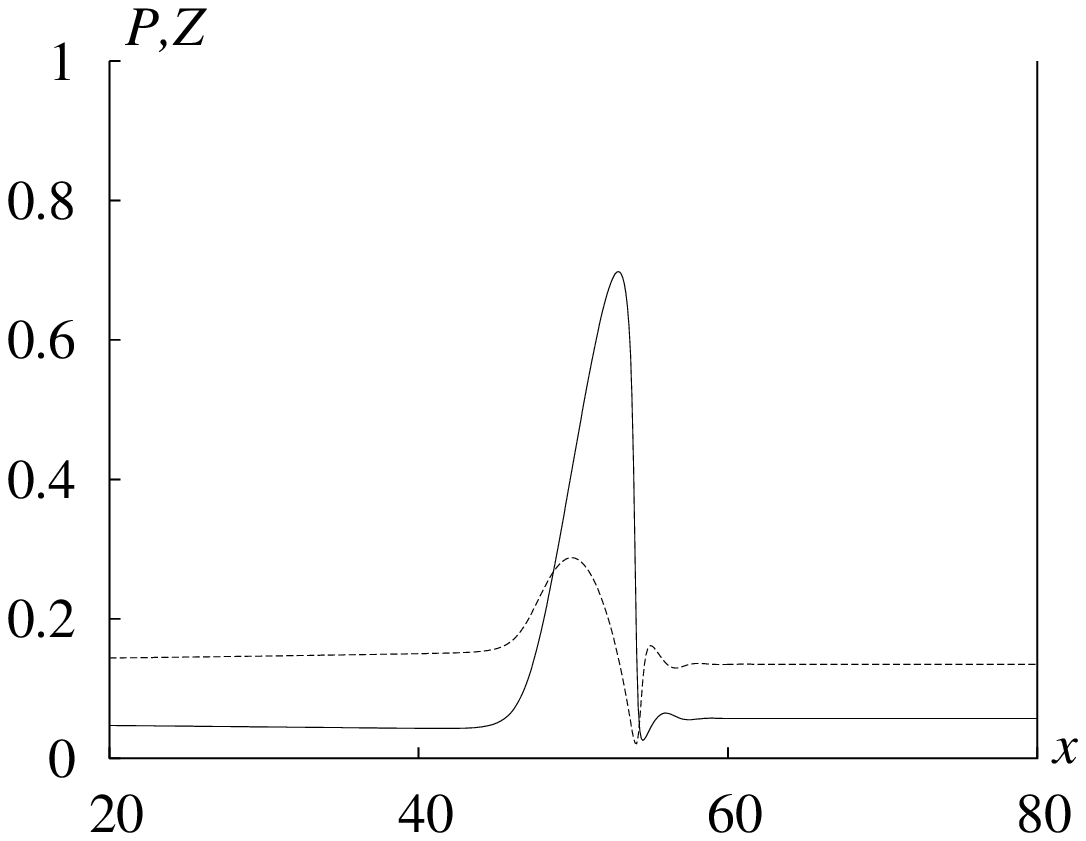} & \panel{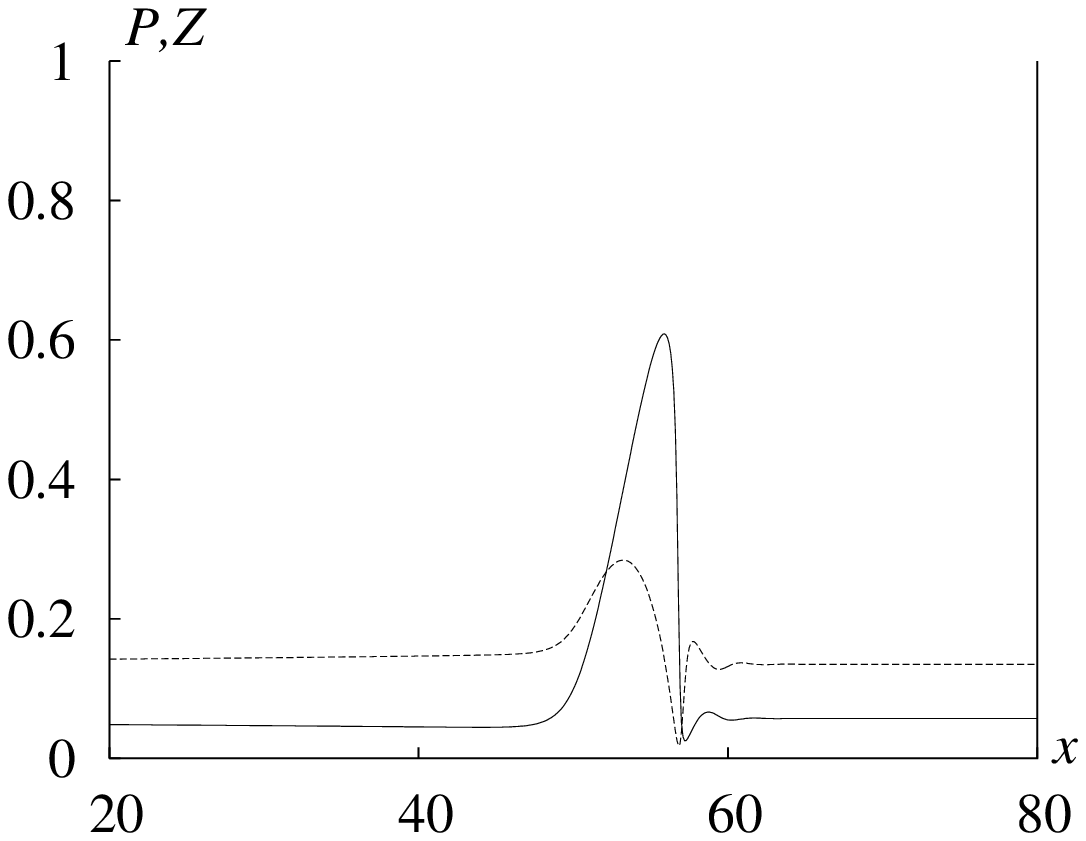} \\
(e)$h_+=1.3$
&
(f)$h_+=1.5$
&
(g)$h_+=2$
&
(h)$h_+=2.6$
\end{tabular}}
\caption{
The profiles of taxis waves for $h_-=2$  and  different  $h_+$,
corresponding to various parts of the velocity graph, marked
by dots on \fig{Velocity}c).
}
\label{profV}
\end{figure*}

\Fig{Velocity} shows dependence of the propagation velocity of taxis waves
on $h_+$ at various fixed values of $h_-$. It also shows symbols of the
corresponding interaction and propagation regimes.
The graphs of the velocity have two branches, ``parabolic'' and ``linear''.
Above a certain value of $h_+$, the linear branches are almost independent of $h_-$ (\fig{Velocity}a).
The transition, with $h_+$ increasing, from the parabolic to the linear branch correlates
with a qualitative change of the shape of the wave.
The dots on \fig{Velocity}c graph mark selected values of $h_+$, the wave profiles
for which are shown on \fig{profV}.
The parabolic branch of this graph correspond to ``double-hump''
shape of the $P(x)$ profile, and the linear branch corresponds to a ``single-hump'' shape.

%================================================================Profil-V
\begin{figure*}[htbp]
\setlength{\unitlength}{1mm} \newcommand{\panel}[1]{\begin{picture}(52,52)(0,0) % (0,-20)
\put(1,1){\mbox{\resizebox{49mm}{49mm}{\includegraphics{#1}}}}
\end{picture}}
\centerline{\begin{tabular}{ccc}
\panel{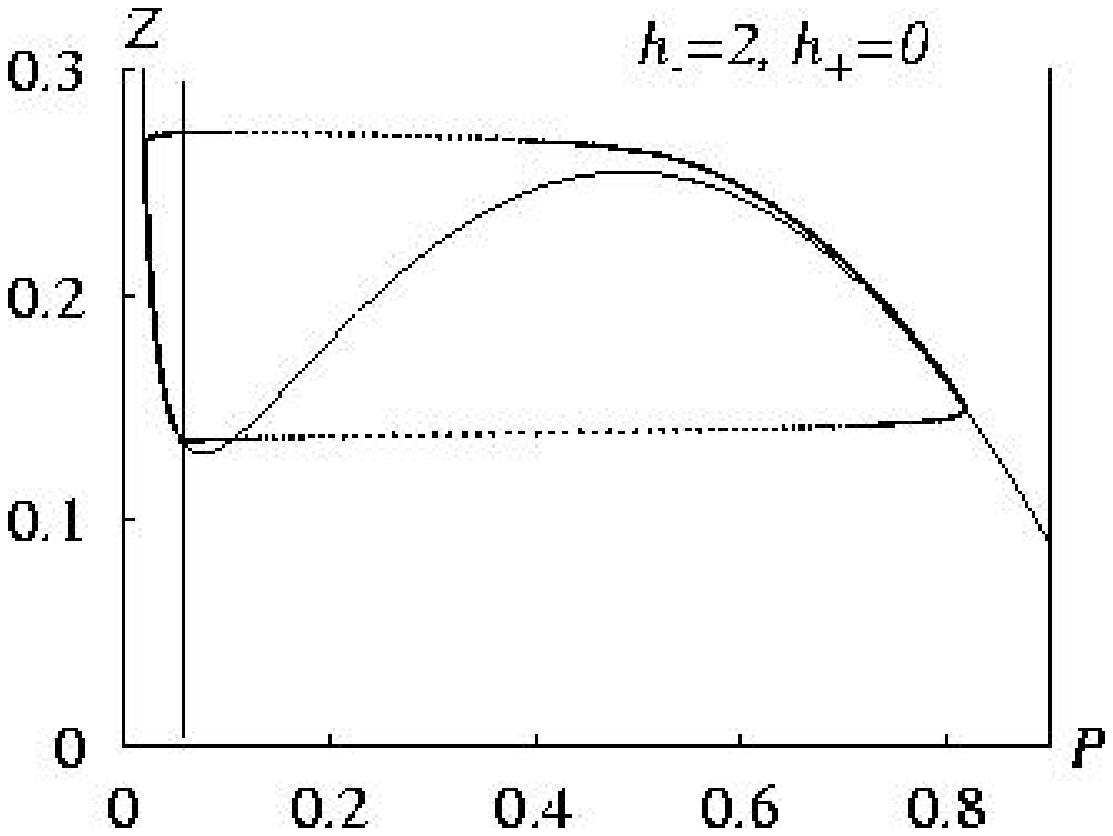} & \panel{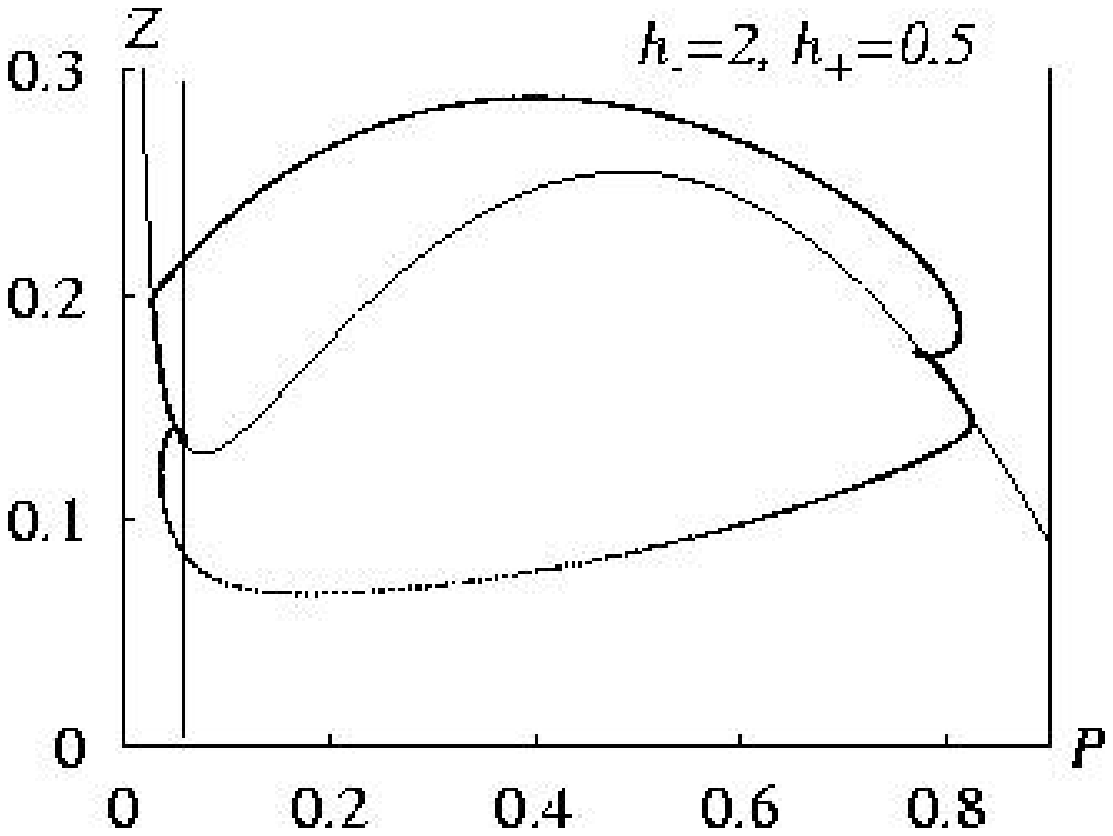} & \panel{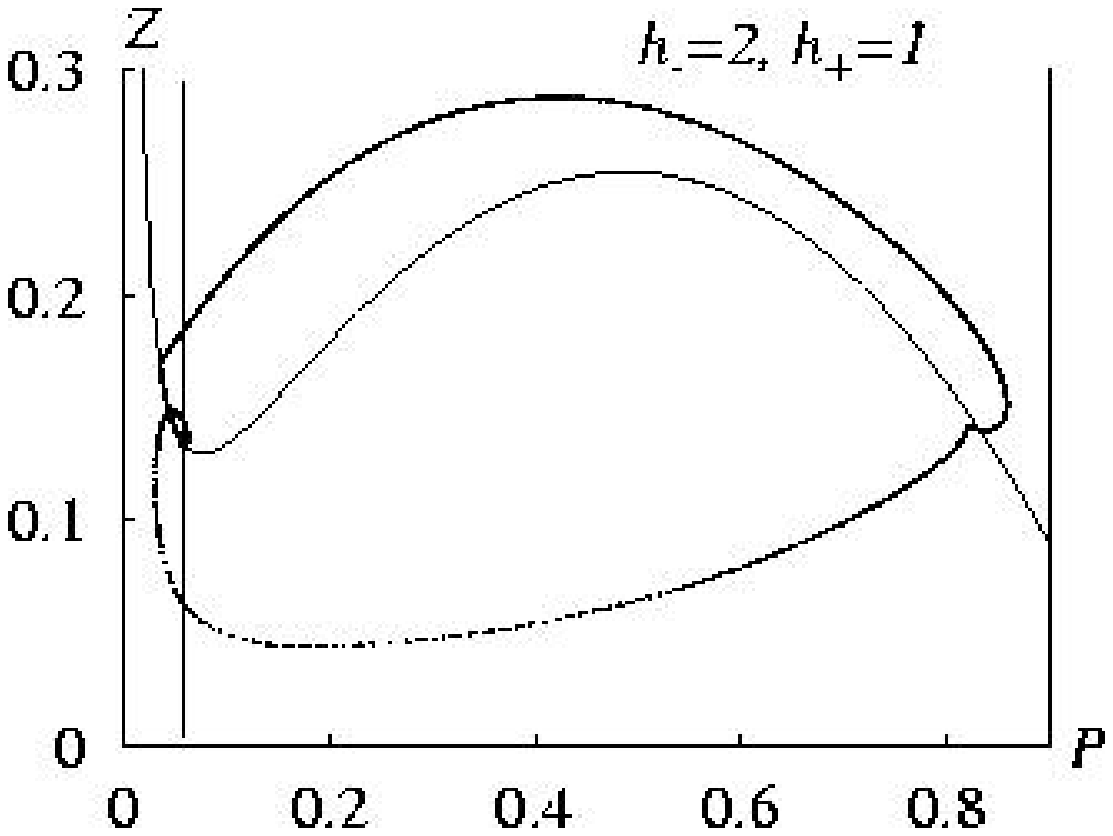} \\
(a) %$h_-=2, h_+=0$
&
(b) %$h_-=2, h_+=0.5$
&
(c) %$h_-=2, h_+=1$
\end{tabular}}
\centerline{\begin{tabular}{ccc}
\panel{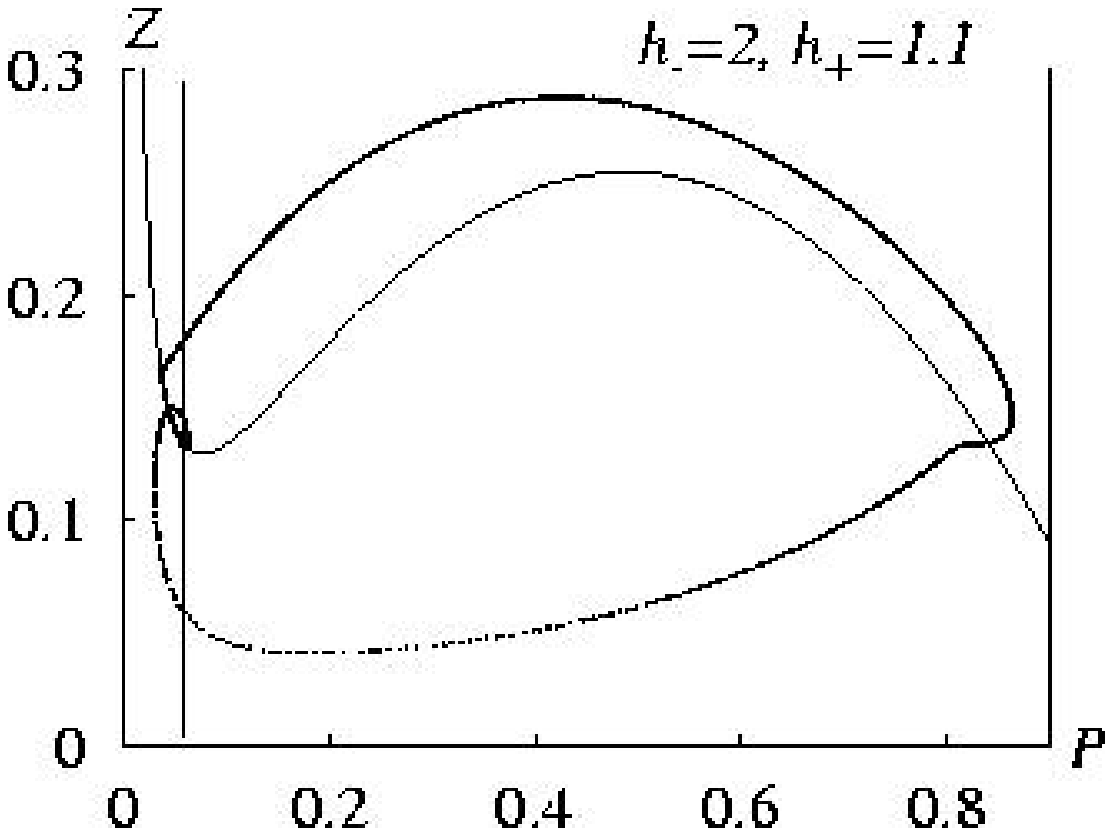} & \panel{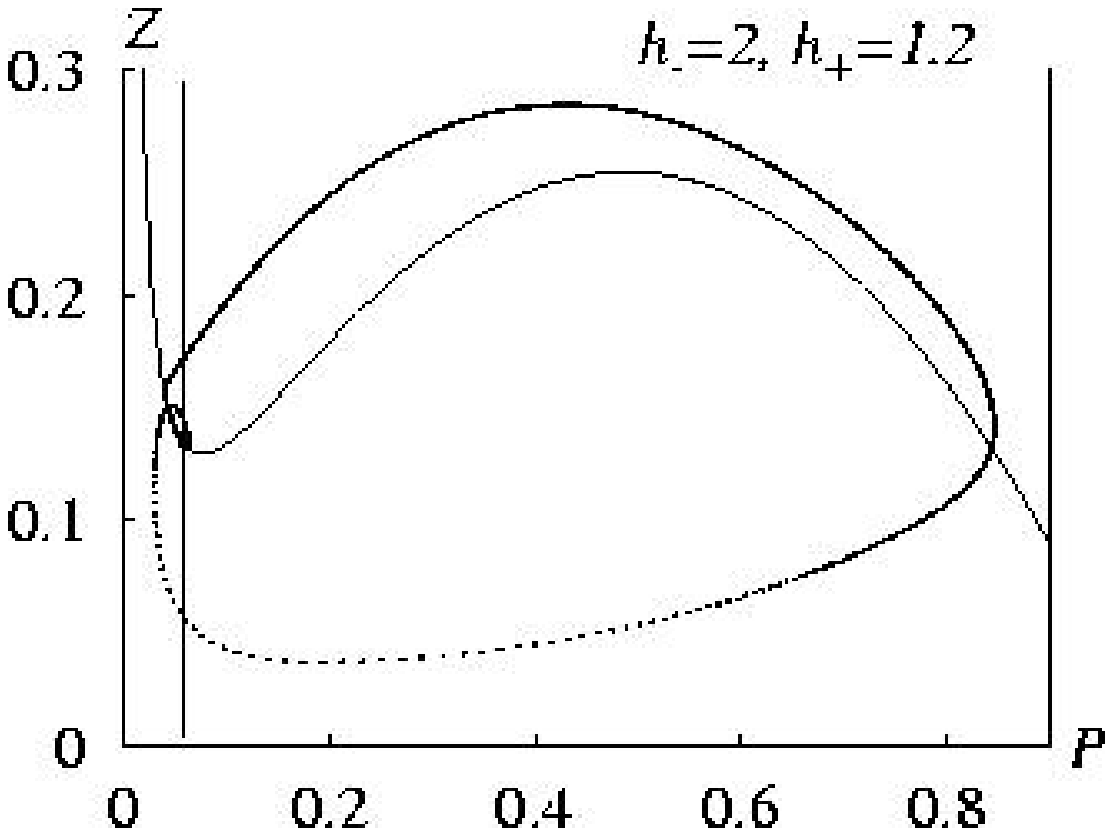} & \panel{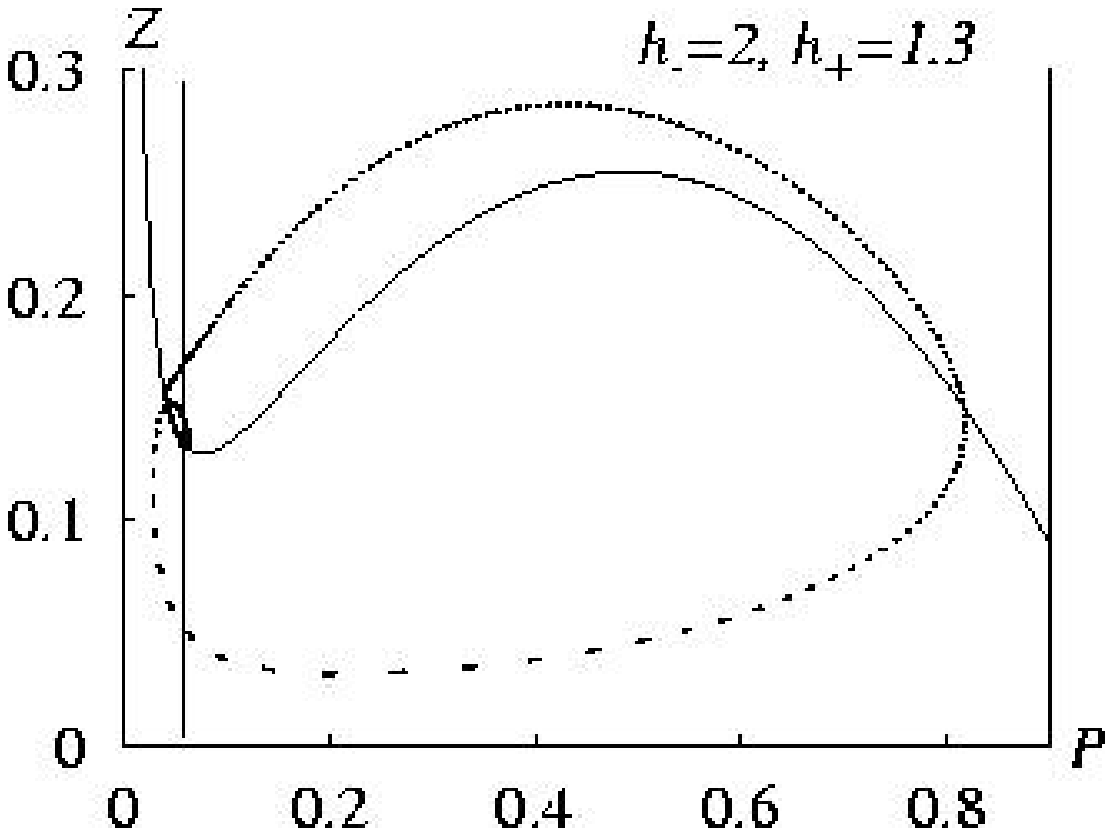} \\
(d) %$h_-=2, h_+=1.1$
&
(e) %$h_-=2, h_+=1.2$
&
(f) %$h_-=2, h_+=1.3$
\end{tabular}}
\centerline{\begin{tabular}{ccc}
\panel{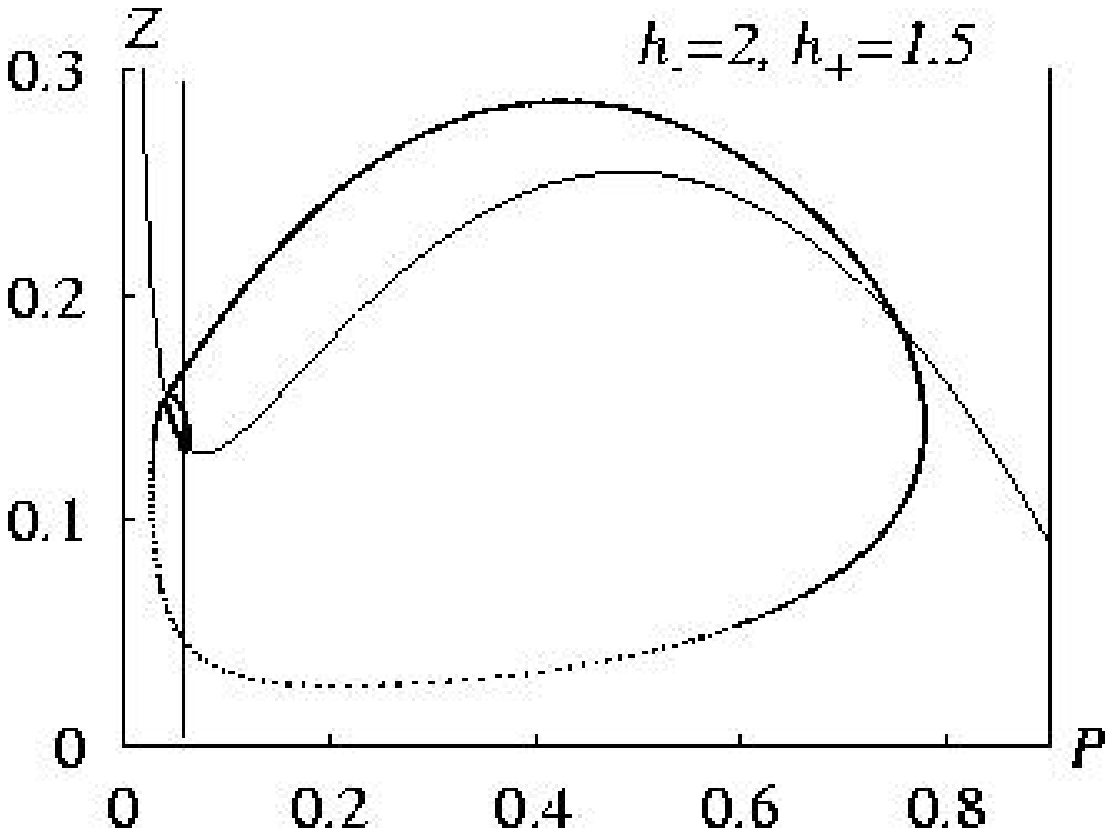} & \panel{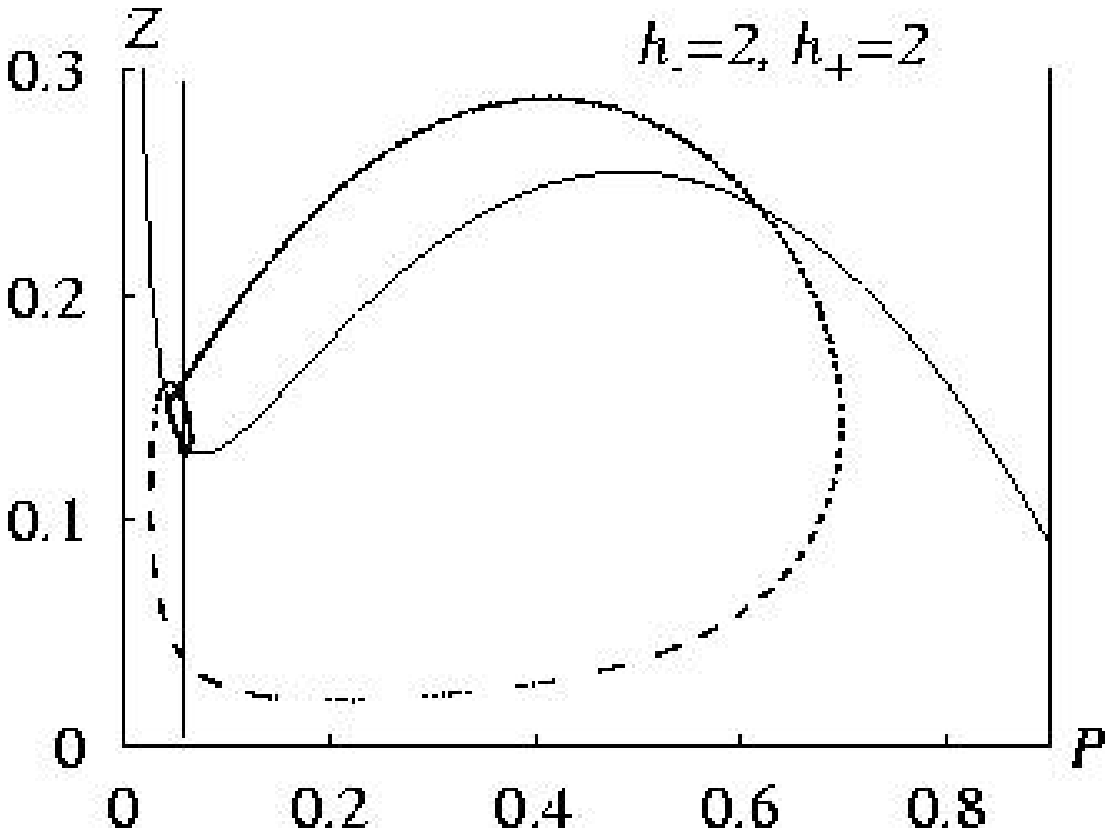} & \panel{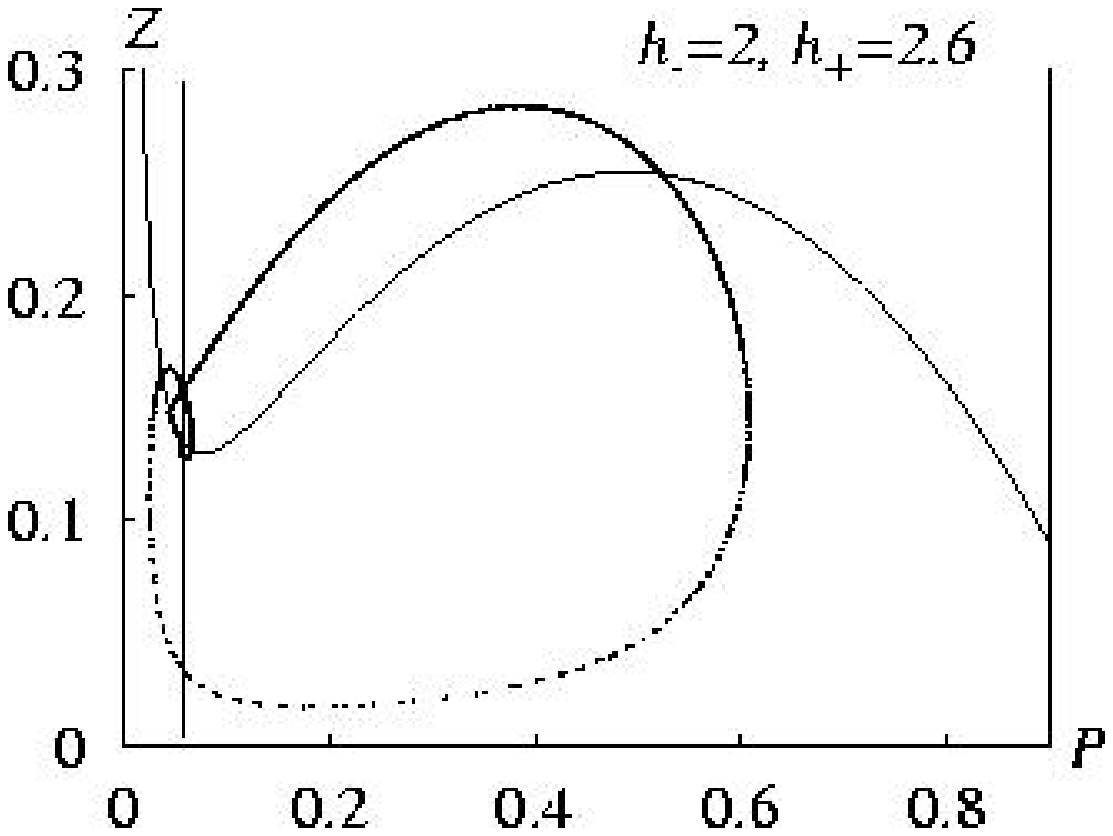} \\
(g) %$h_-=2, h_+=1.5$
&
(h) %$h_-=2, h_+=2$
&
(i) %$h_-=2, h_+=2.6$
\end{tabular}}
\caption{ Phase planes corresponding to various taxis waves, corresponding
to the profiles shown on \fig{profV}.
}
\label{PZ-V}
\end{figure*}

%===============================================================PZ plane-V
This result is confirmed by phase plane figures (\fig{PZ-V}). One can see that for
$h_-=2$, the transition from the parabolic to the linear branch
takes places at $h_+=1.2$, which agrees with \fig{profV}(d).

% ---------------------------------------------------------------------
% *****************************************************************PartB

\section{Case II: the reaction-taxis waves for $\gamma=0.016$ }

In the previous section, we studied properties of taxis waves when
the local kinetics admitted purely diffusive waves in \eq{RDT},
i.e. for $h_-=h_+=0$ and $D>0$.
When the parameters of the local kinetics are changed, such propagation
may become impossible.
We have found that the taxis waves may still be possible in such cases.
Specifically, we increased parameter $\gamma$, and found regimes for which waves can propagate only
with the taxis terms, but not without them.

Note that the central difference scheme (A) only allows computations for $D>0$,
whereas our ``upwind'' schemes (B) and (C) can work for $D=0$ as well.

\begin{figure*}[htbp]
\setlength{\unitlength}{1mm} \newcommand{\panel}[1]{\begin{picture}(82,82)(0,0)
\put(0,0){\mbox{\resizebox{81mm}{!}{\includegraphics{#1}}}}
\end{picture}}
\centerline{\begin{tabular}{cc}
\panel{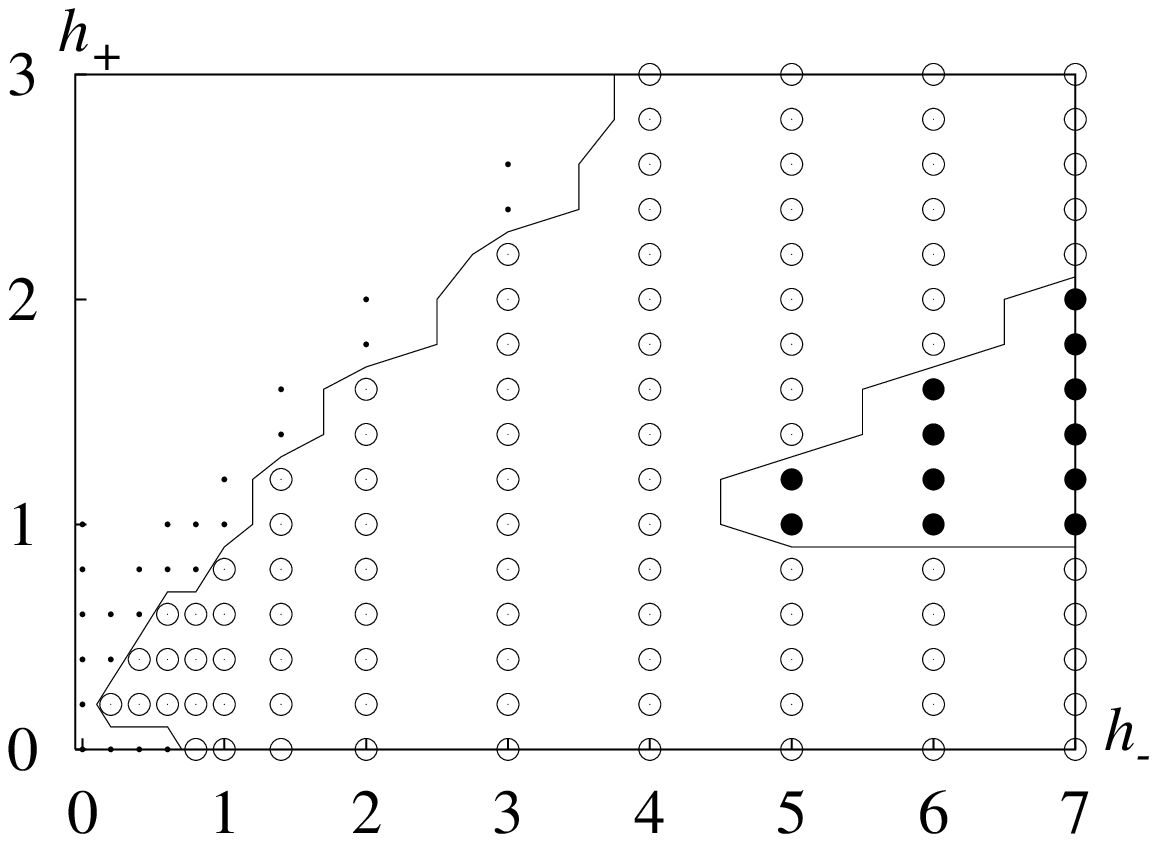} & \panel{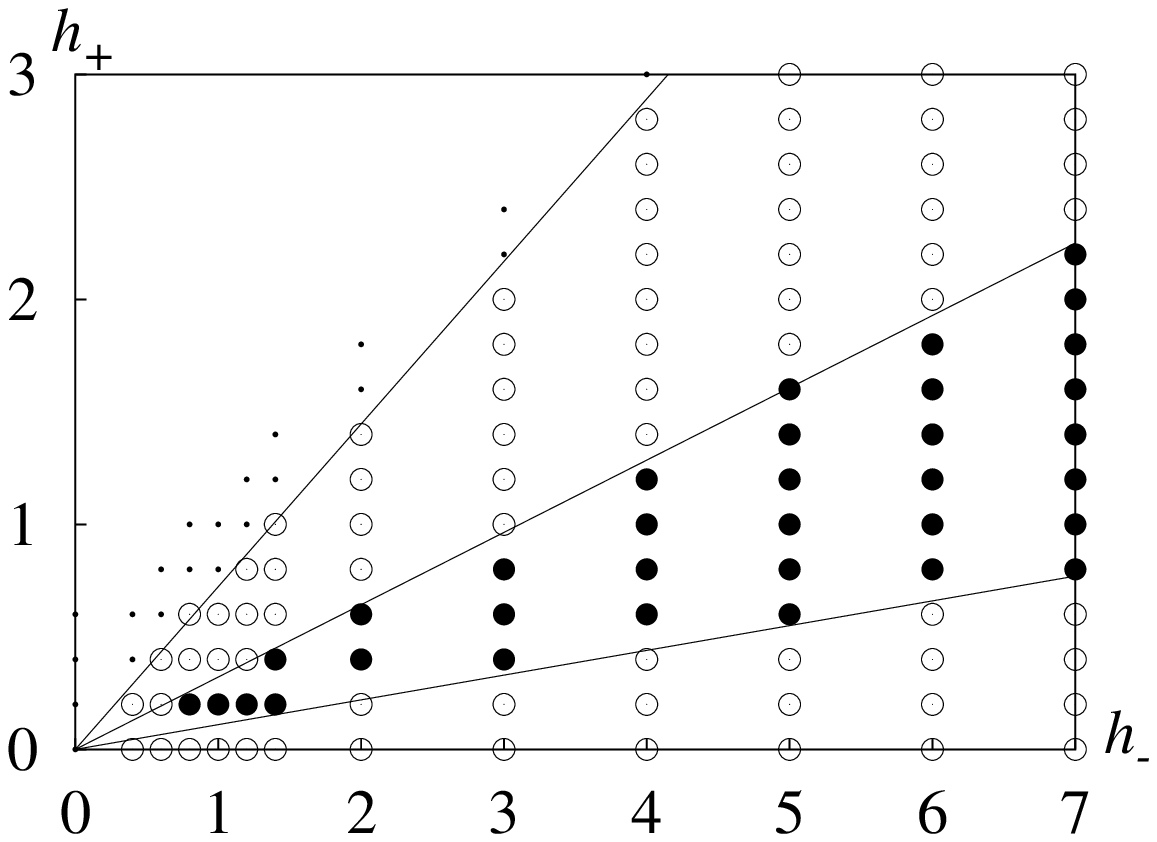}  \\
(a) $D=0.04$
&
(b) $D=0$
\end{tabular}}
\caption{Parameter regions corresponding to different regimes of taxis waves
for $\gamma=0.016$}
\label{hhU}
\end{figure*}

\Fig{hhU}a shows the dependence of the behaviour of taxis waves on $h_-$ and $h_+$,
for fixed $\gamma=0.016$ and $D=0.04$. The point $h_+=h_-=0$ is in the
region where stationary propagation of waves is not possible. Yet, for
this local kinetics, in the $(h_-,h_+)$ plane there are regions of
both quasisoliton and non-soliton waves. Unlike the local kinetics of
Part I, for this $\gamma=0.016$ we did not see any splitting waves in
the explored parameter region.

Now, we consider the case of no diffusion, $D=0$, at
$\gamma=0.016$ (\fig{hhU}b).

% ==============================================================Param.Tab.U

%==============================================================Velocity-U

\begin{figure}[htbp]
\setlength{\unitlength}{1mm} \newcommand{\panel}[1]{\begin{picture}(54,54)(0,0) % (0,-20)
\put(1,1){\mbox{\resizebox{53mm}{!}{\includegraphics{#1}}}}
\end{picture}}
\centerline{\begin{tabular}{cc}
\panel{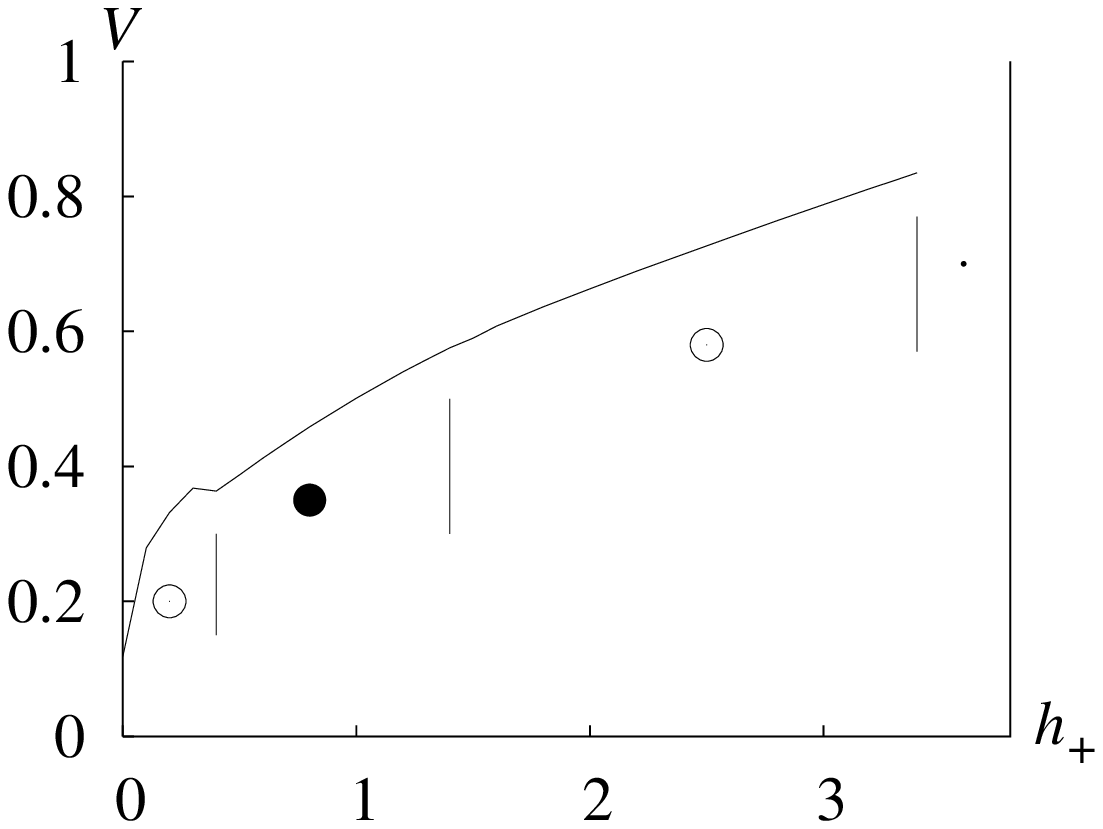} & \panel{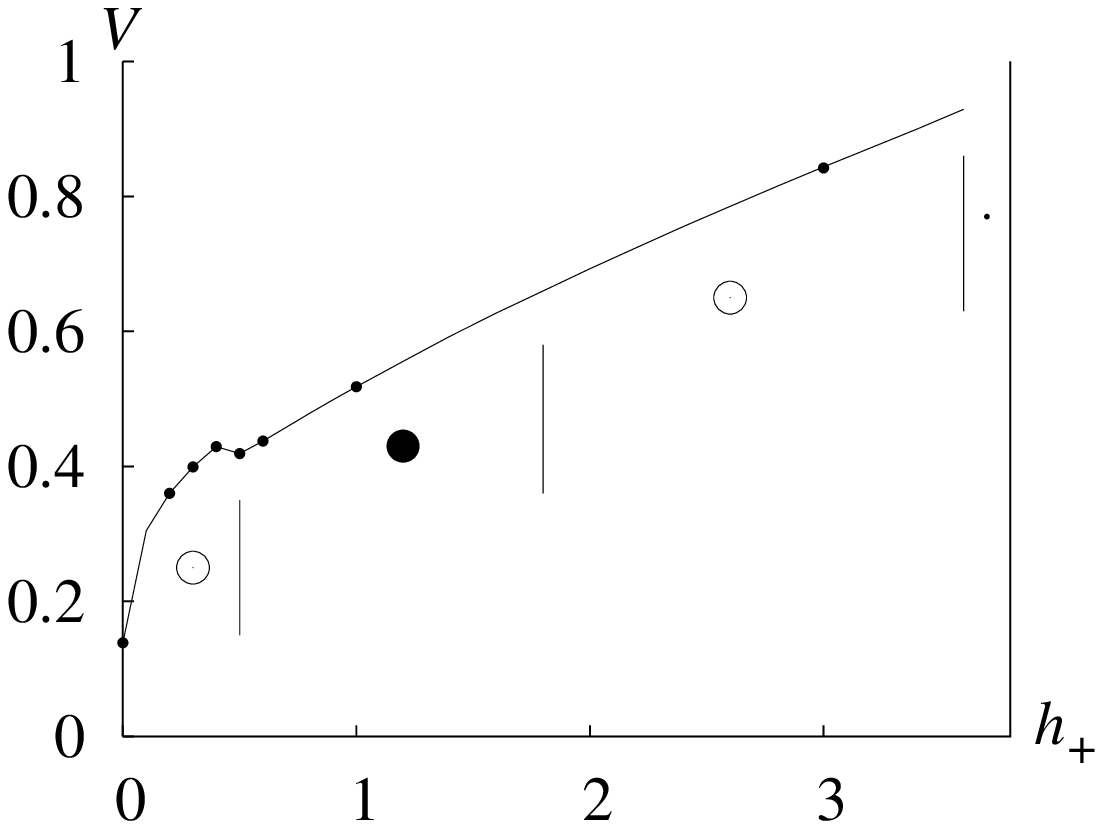} \\
(b) $h_-=5$
&
(c) $h_-=7$
\end{tabular}}
\caption[]{ Dependence of propagation velocity of taxis waves
on $h_+$ for selected values of $h_-$, for $\gamma=0.016$ and $D=0$.
The vertical bars and symbols on the graphs designate
different regimes of propagation and interaction.
Solid circle: quasisolitons pulses with soliton interaction.
Hollow circles: stable propagation of pulses with nonsoliton interaction on collision.
Dot: there is no stable propagation of pulses.
The dots on the graph of panel (b) correspond to wave profiles shown on \fig{profVu}
for various values of $h_+$.
}
\label{VelU}
\end{figure}

Dependence of the propagation velocity on $h_+$ for this case is
shown on \fig{VelU}.  Here we also see two distinct branches of the
graph, and the transition between them correlates also with the
transition from quasisoliton to non-soliton regimes of interaction.

%================================================================Profil-U

\begin{figure*}[htbp]
\setlength{\unitlength}{1mm} \newcommand{\panel}[1]{\begin{picture}(42,42)(0,0) % (0,-20)
\put(1,1){\mbox{\resizebox{41mm}{!}{\includegraphics{#1}}}}
\end{picture}}
\centerline{\begin{tabular}{cccccccc}
\panel{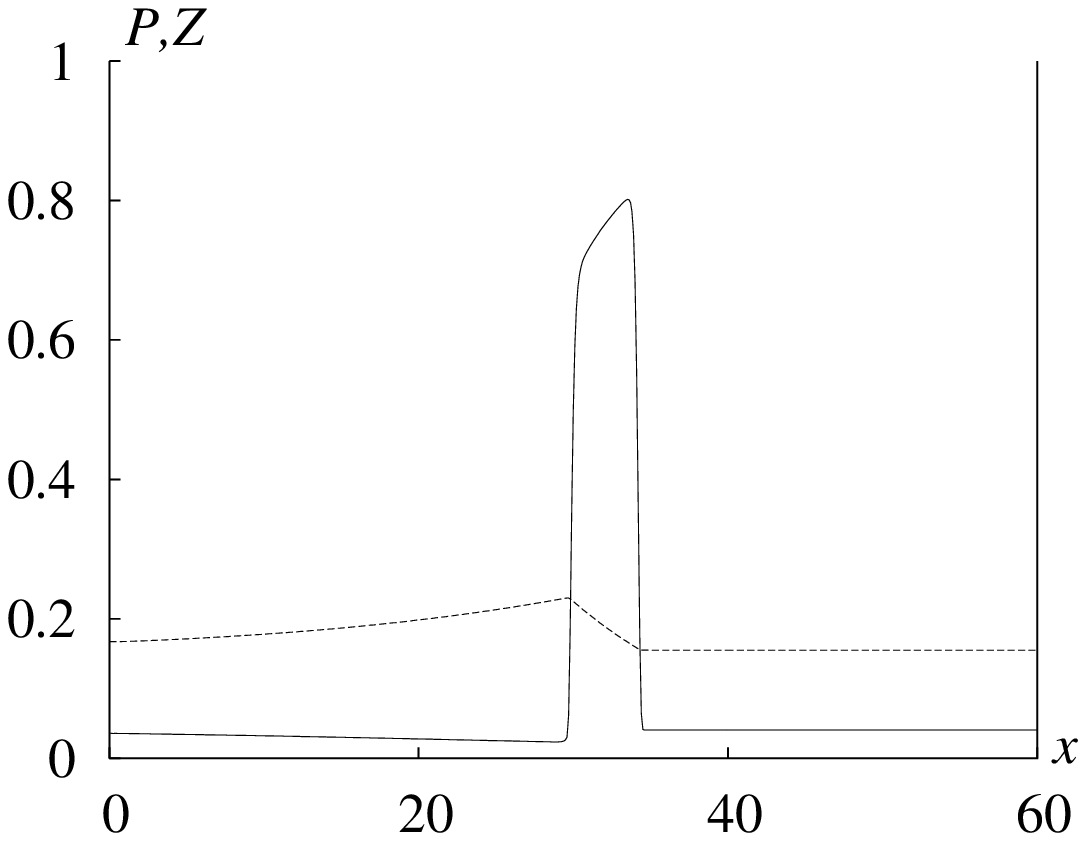} & \panel{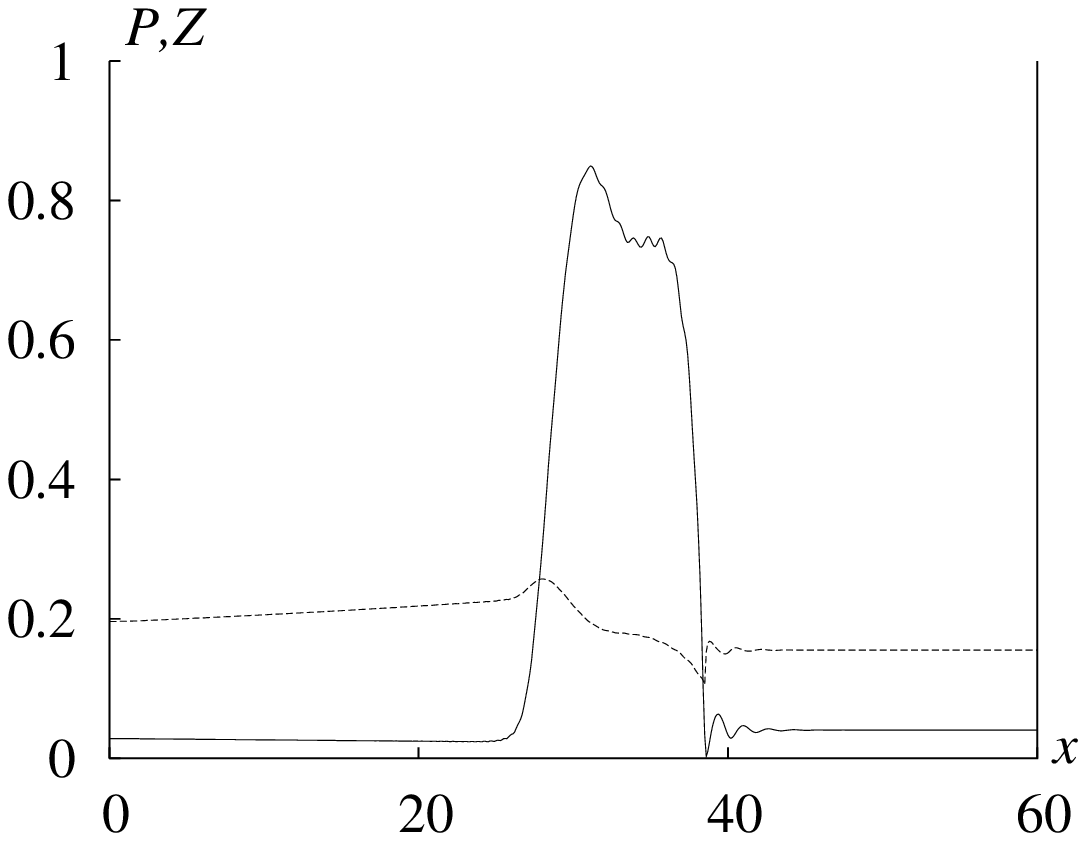} & \panel{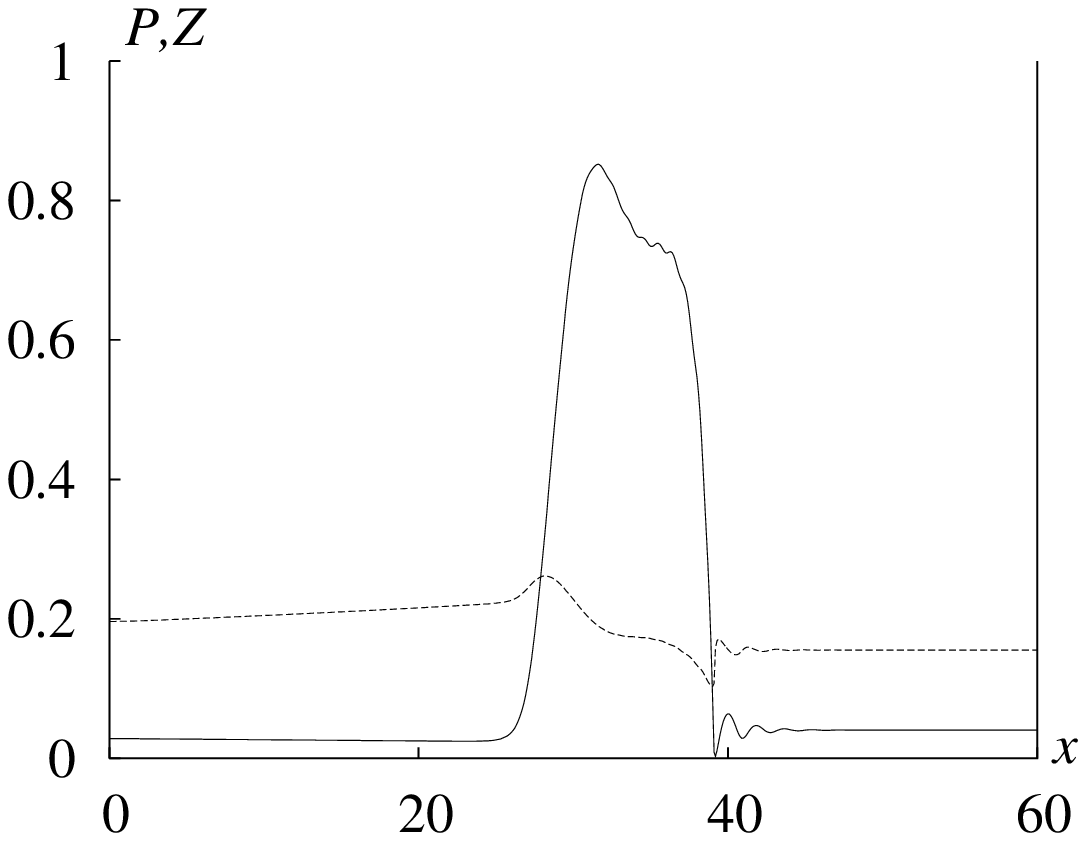} & \panel{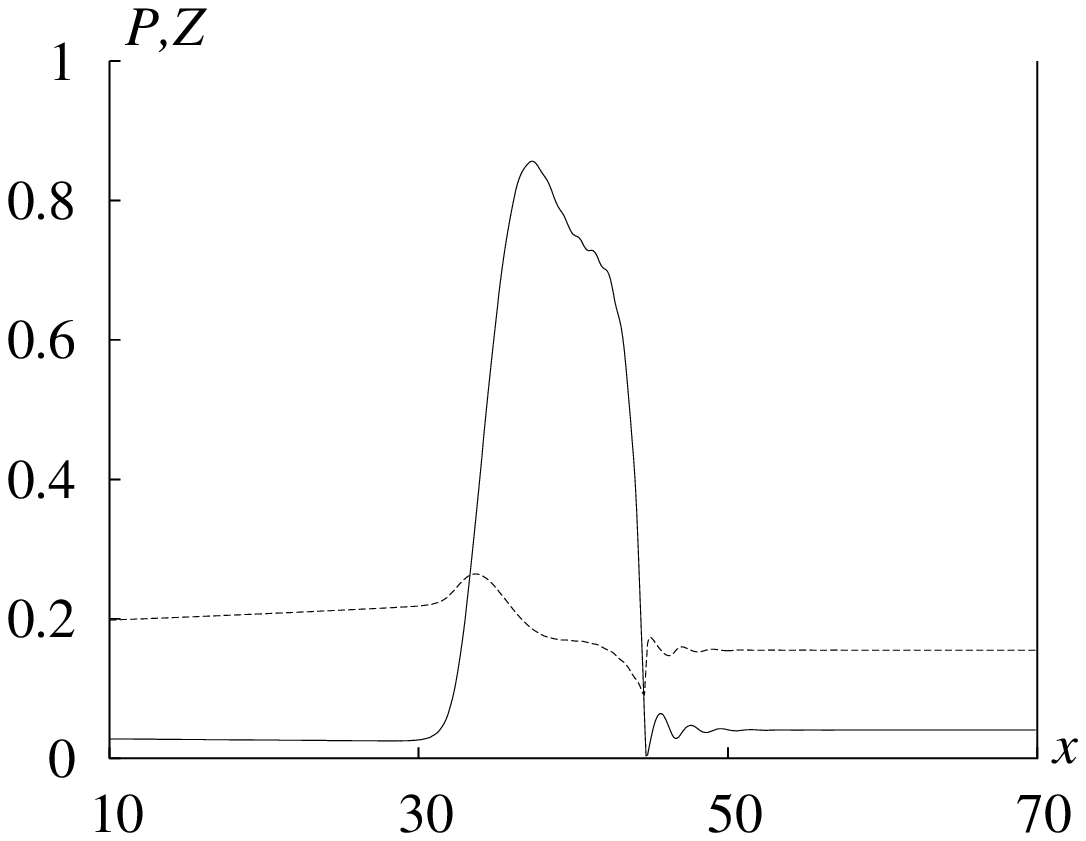} \\
(a)$h_+=0$
&
(b)$h_+=0.2$
&
(c)$h_+=0.3$
&
(d)$h_+=0.4$ \\
\panel{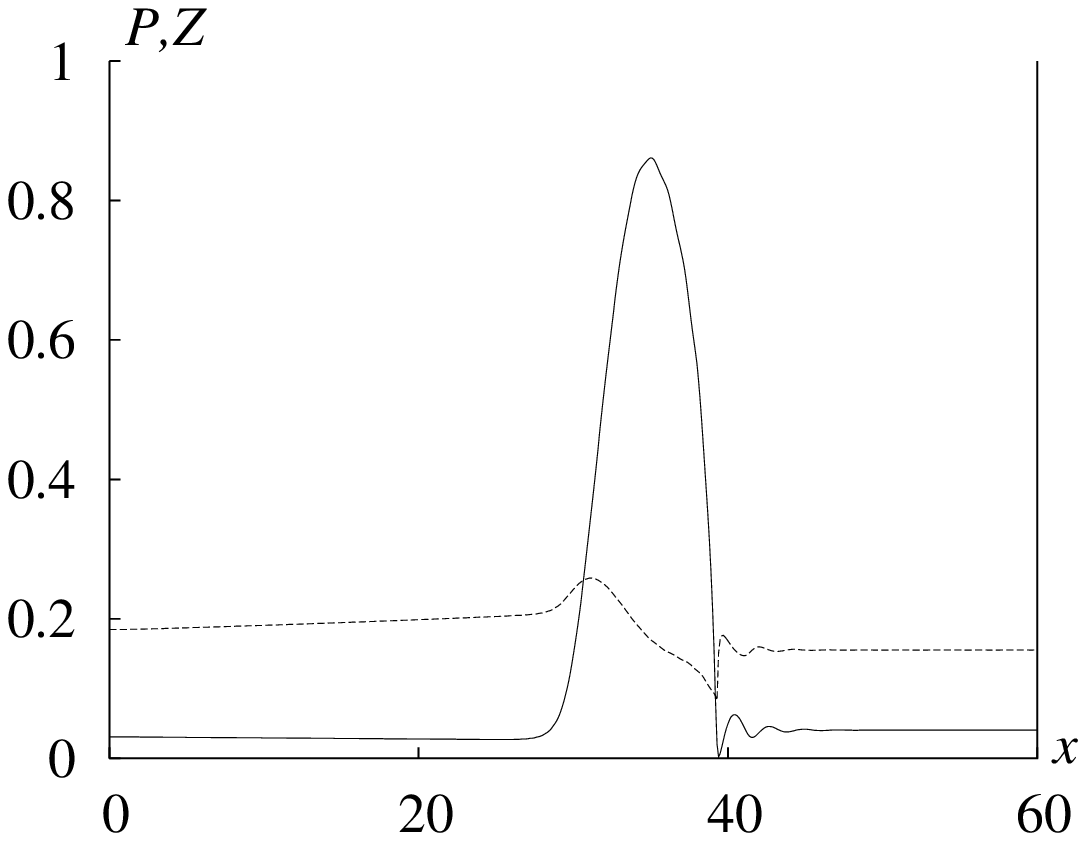} & \panel{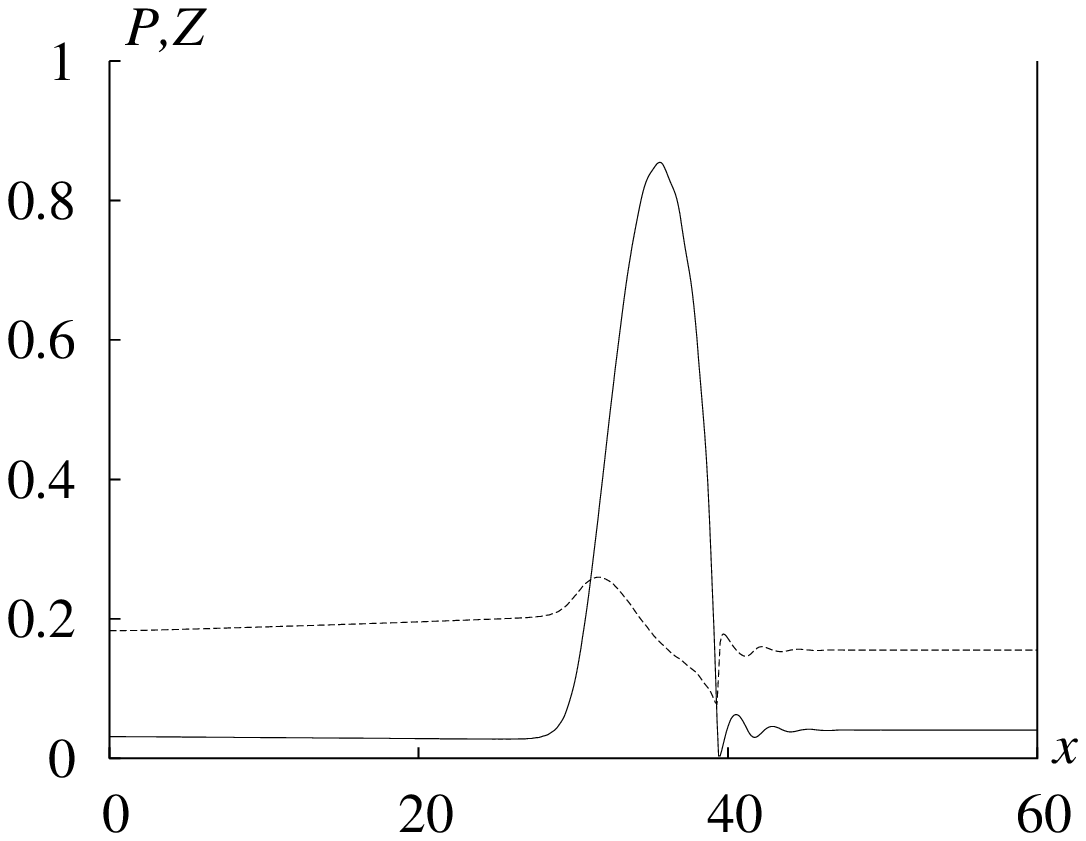} & \panel{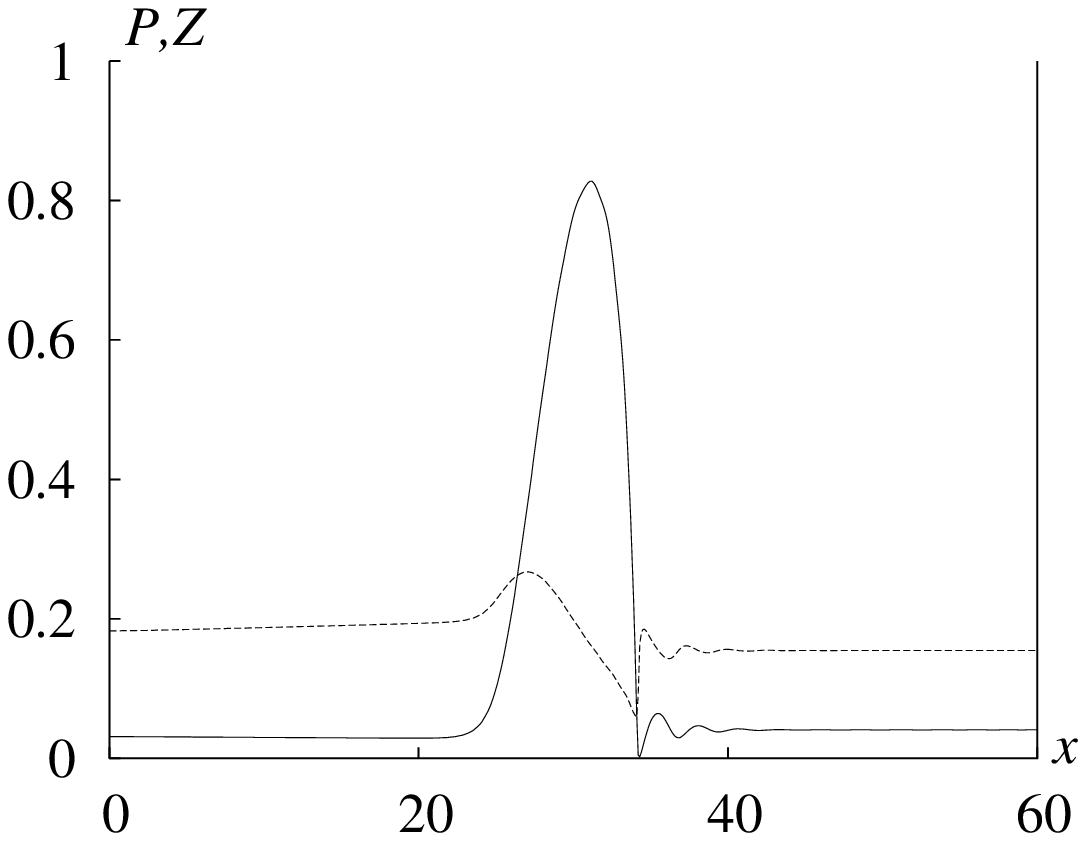} & \panel{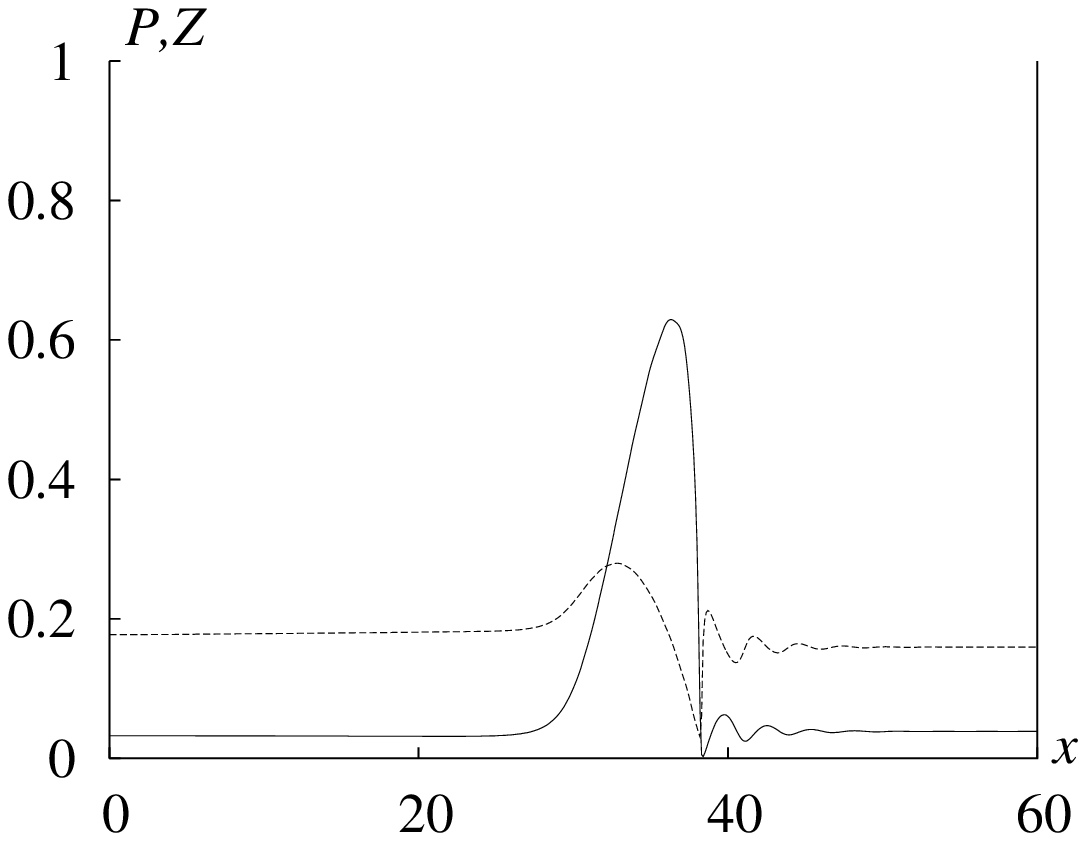} \\
(e)$h_+=0.5$
&
(f)$h_+=0.6$
&
(g)$h_+=1$
&
(h)$h_+=3$
\end{tabular}}
\caption{ .
Profiles of taxis waves for $h_-=7$ and different $h_+$, corresponding
to the selected points shown by dots on the graph \fig{VelU}b).
}
\label{profVu}
\end{figure*}

\begin{figure*}[htbp]
\setlength{\unitlength}{1mm}
\newcommand{\panel}[1]{\begin{picture}(54,54)(0,0) % (0,-20)
\put(1,1){\mbox{\resizebox{53mm}{53mm}{\includegraphics{#1}}}}
\end{picture}}
\centerline{\begin{tabular}{ccc}
\panel{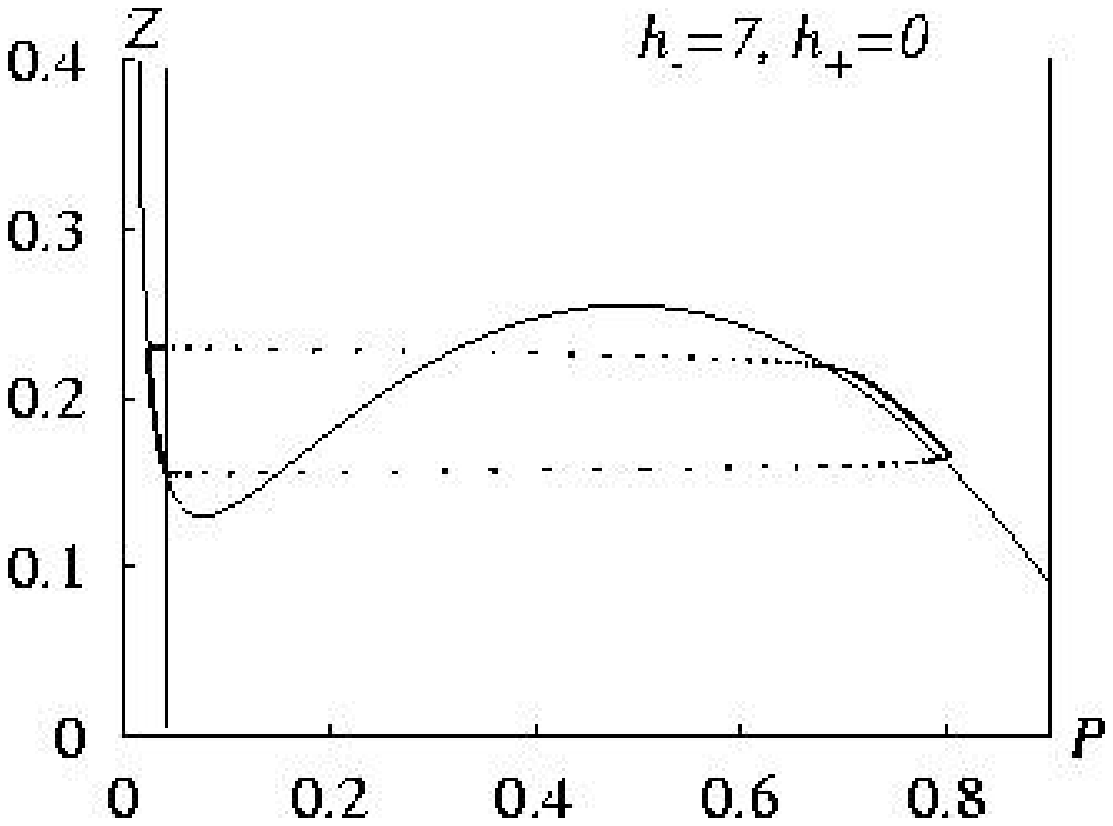} & \panel{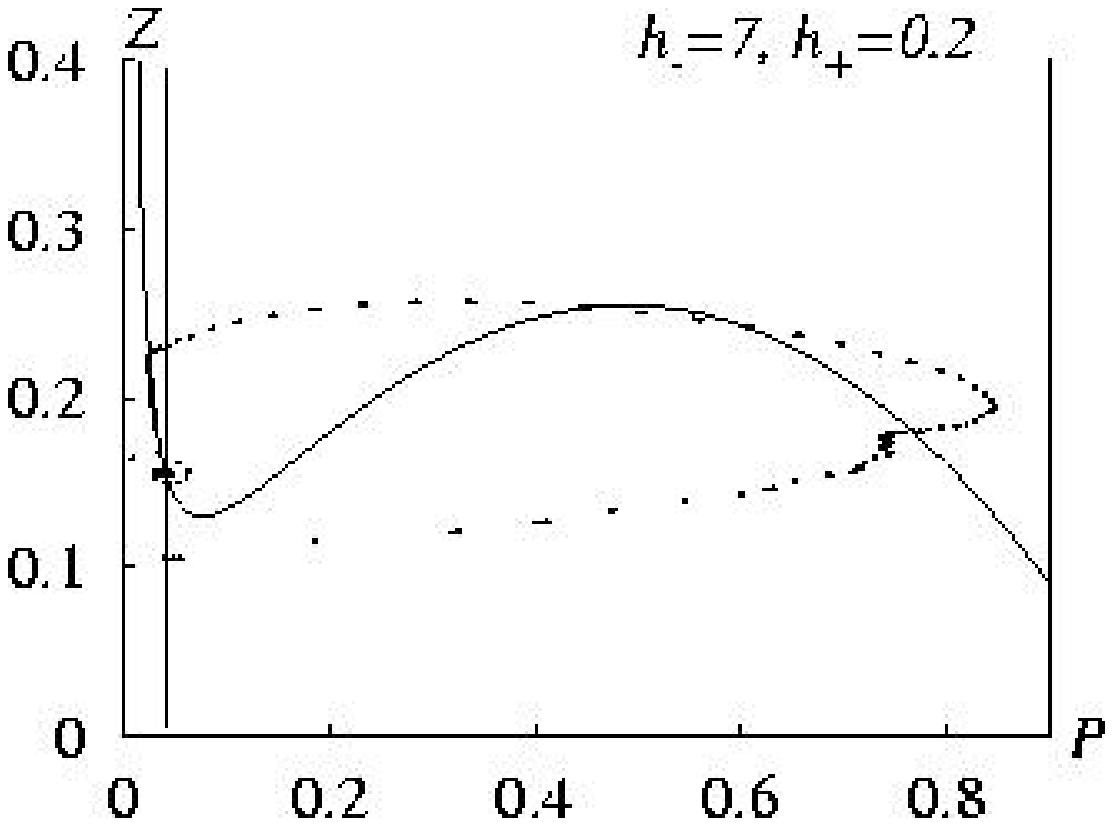} & \panel{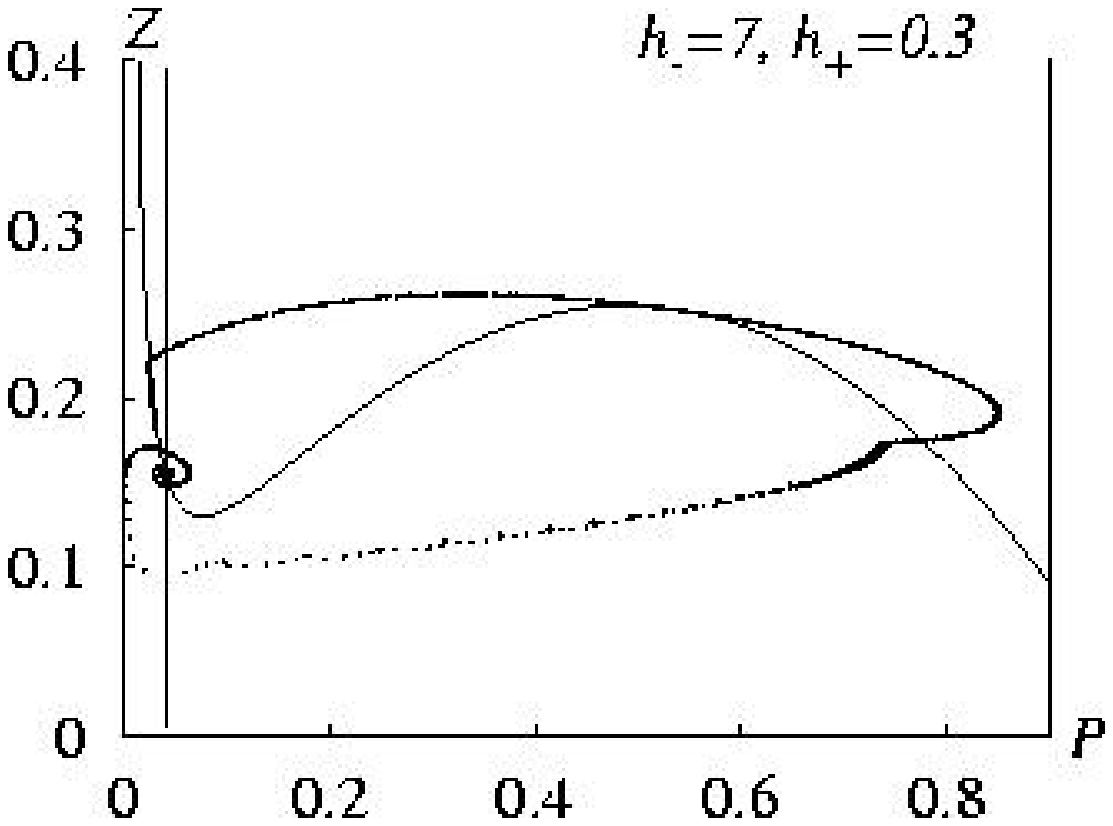} \\
(a) %$h_-=7, h_+=0$
&
(b) %$h_-=7, h_+=0.2$
&
(c) %$h_-=7, h_+=0.3$
\end{tabular}}
\centerline{\begin{tabular}{ccc}
\panel{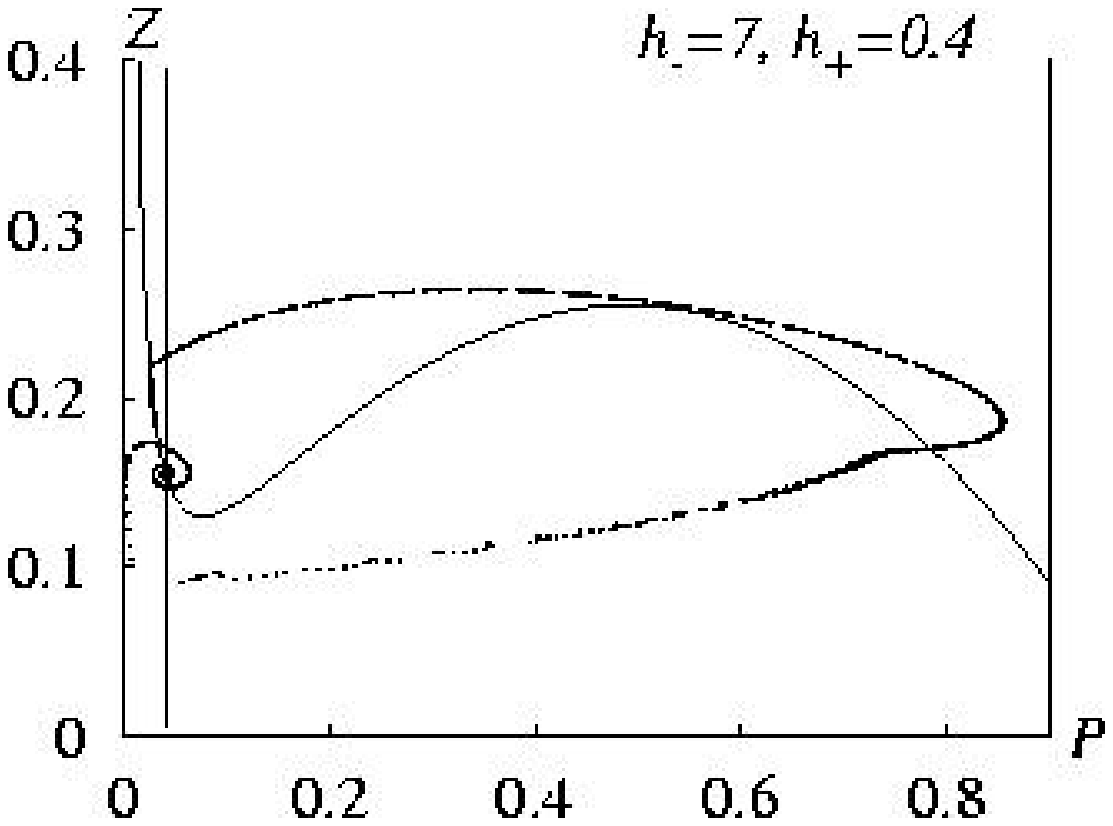} & \panel{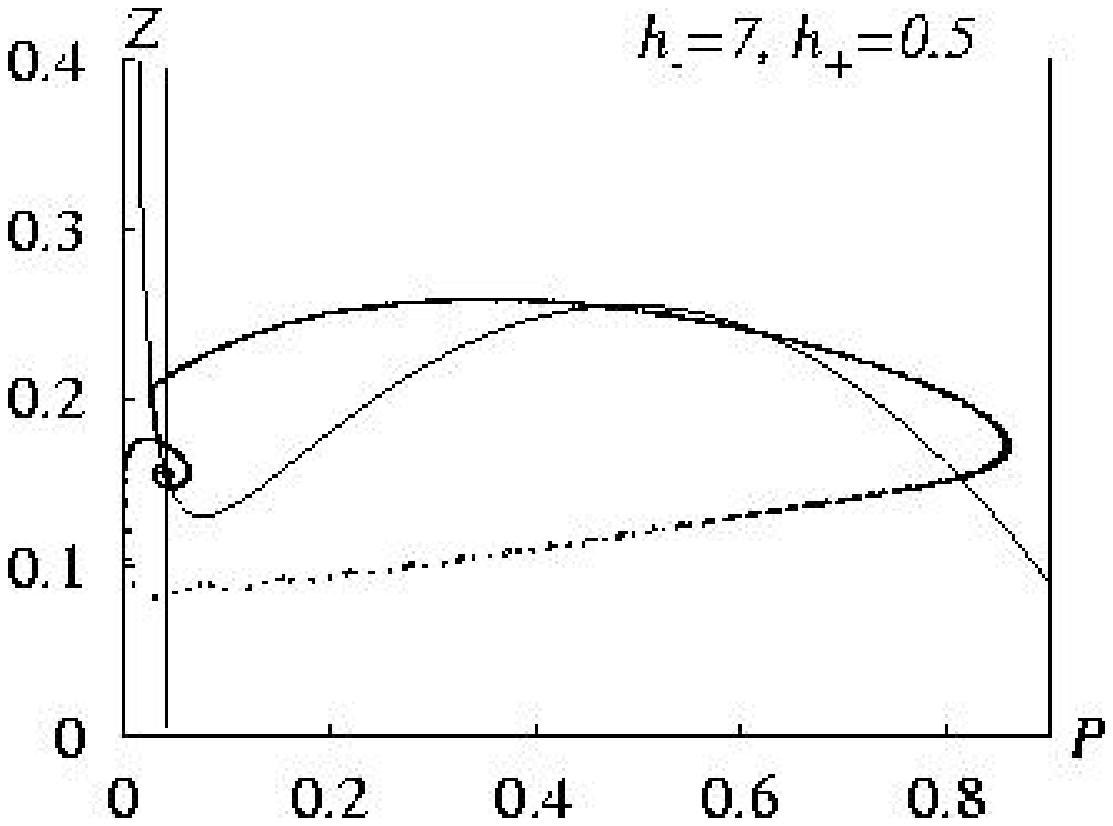} & \panel{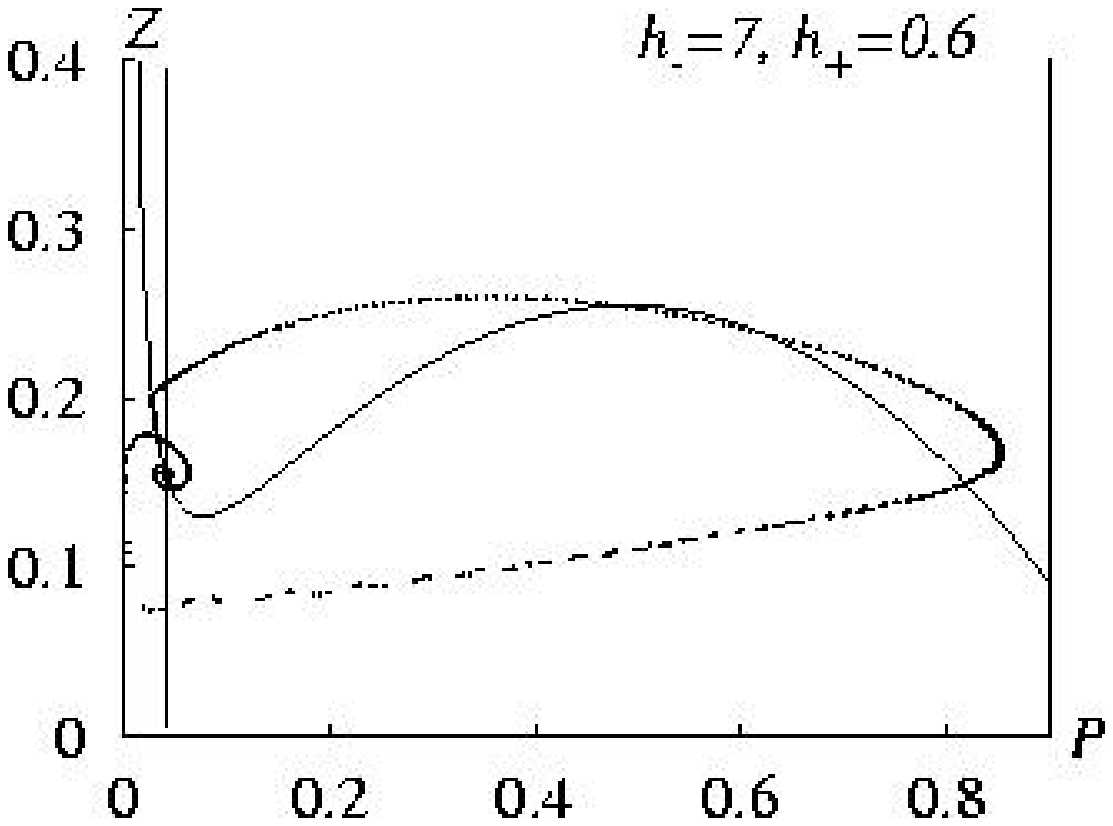} \\
(d) %$h_-=7, h_+=0.4$
&
(e) %$h_-=7, h_+=0.5$
&
(f) %$h_-=7, h_+=0.6$
\end{tabular}}
\centerline{\begin{tabular}{cc}
\panel{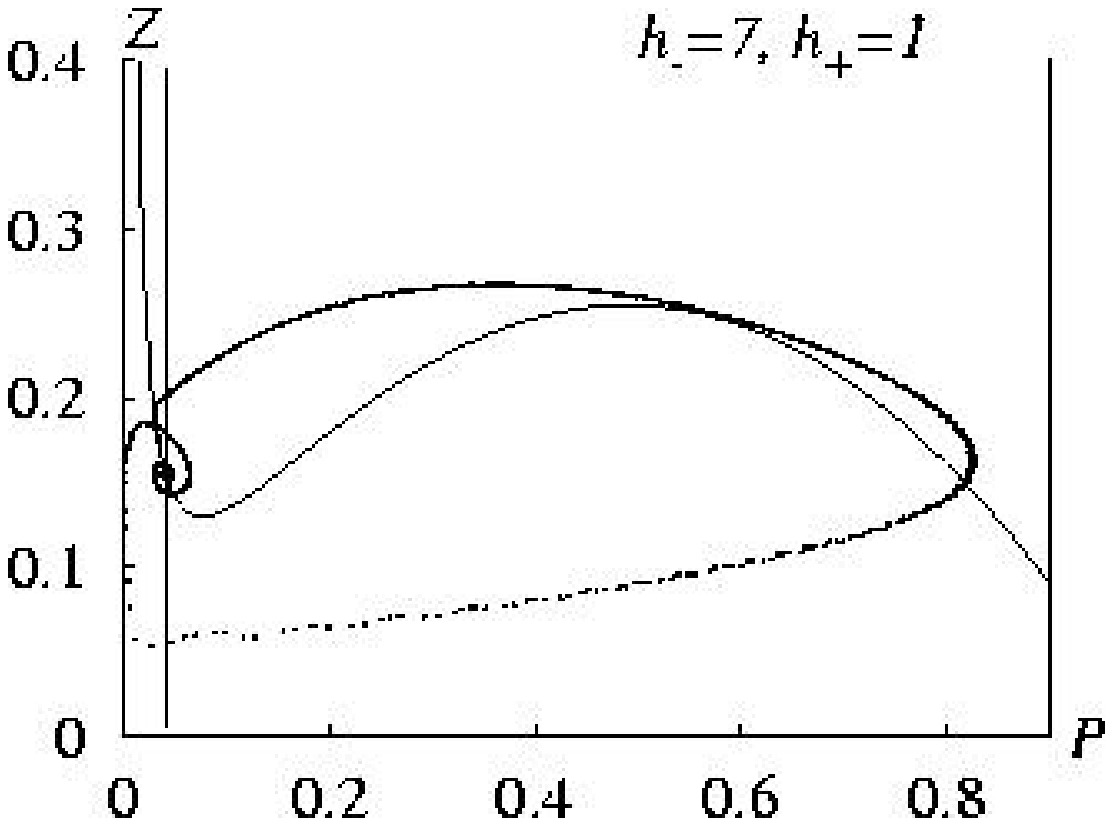} & \panel{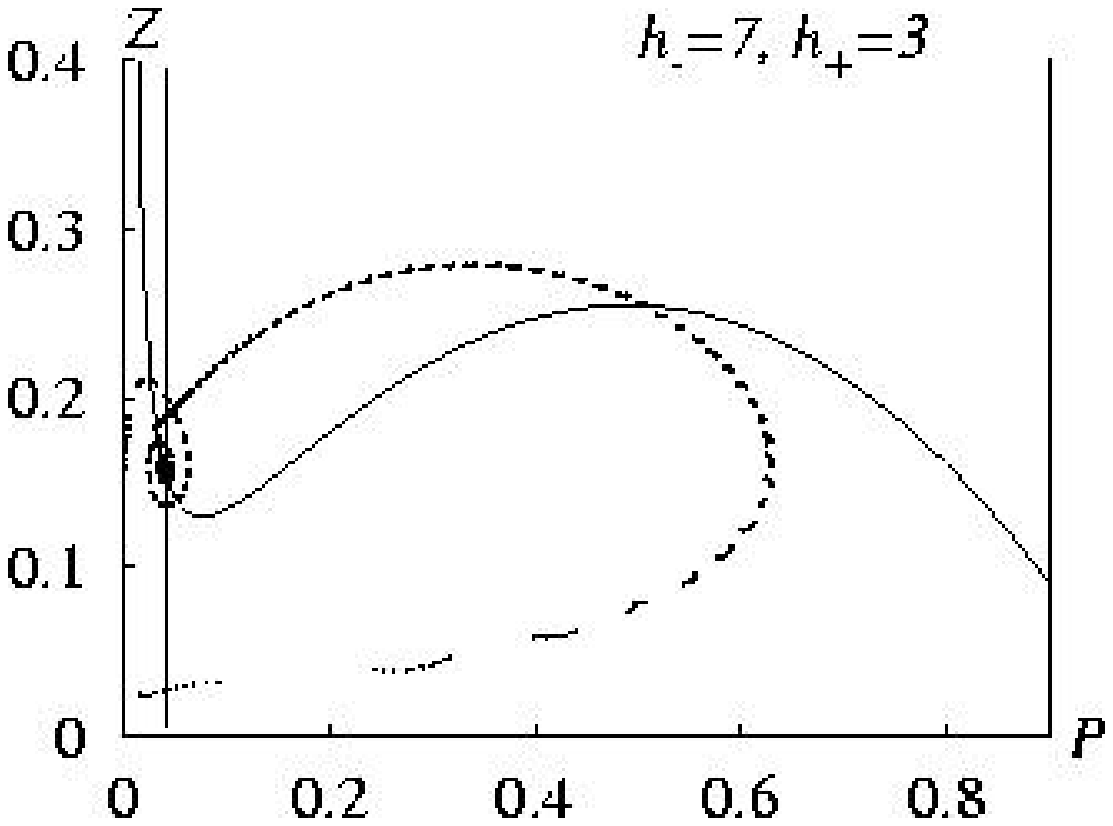} \\
(g) %$h_-=7, h_+=1$
&
(h) %$h_-=7, h_+=1$
\end{tabular}}
\caption[]{
Phase planes corresponding to different taxis waves, shown on \fig{profVu},
for $\gamma=0.016$, $D=0$.
}
\label{PZ-Vu}
\end{figure*}

The correlation between parabolic-linear transition of the speed vs $h_+$ dependence
with double-single hump transition is illustrated by \fig{profVu} and corresponding
phase planes on \fig{PZ-Vu}.

%===============================================================PZplane-U

%===============================================================NewAN-5-2

\begin{figure*}[htbp]
\setlength{\unitlength}{1mm} \newcommand{\panel}[1]{\begin{picture}(44,44)(0,0) % (0,-20)
\put(1,1){\mbox{\resizebox{43mm}{!}{\includegraphics{#1}}}}
\end{picture}}
\centerline{\begin{tabular}{cccc}
\panel{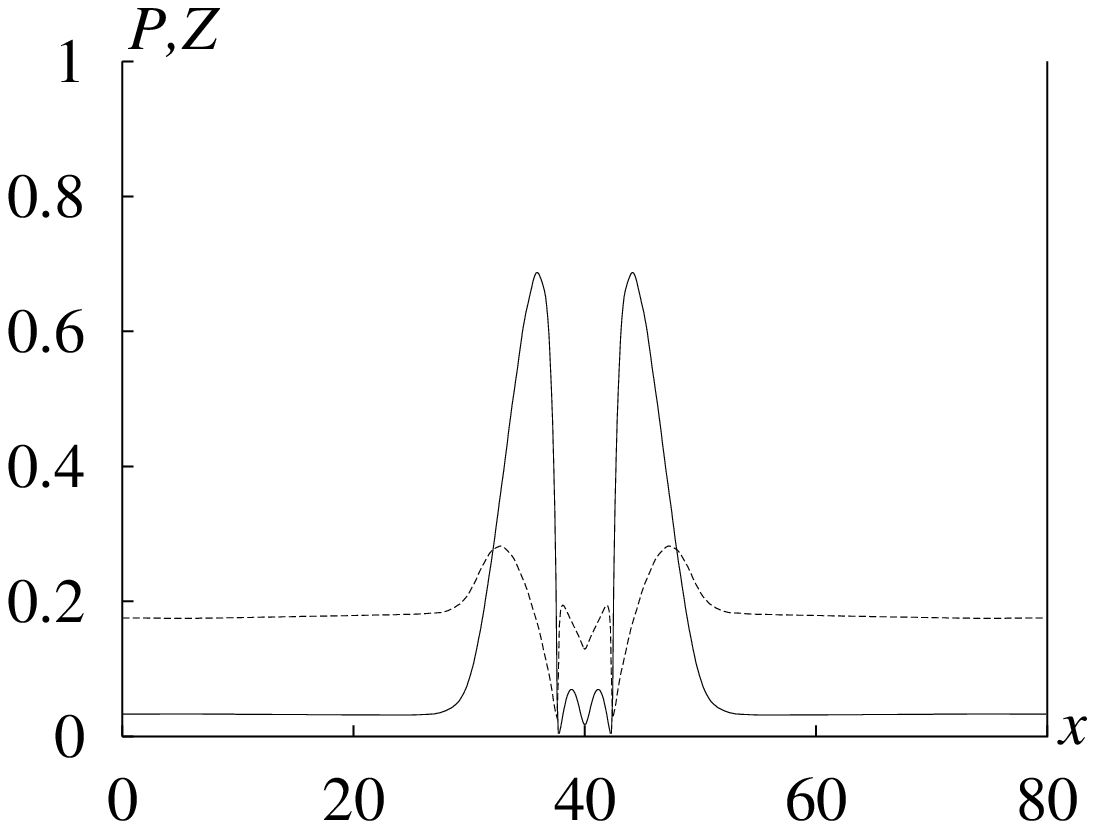} & \panel{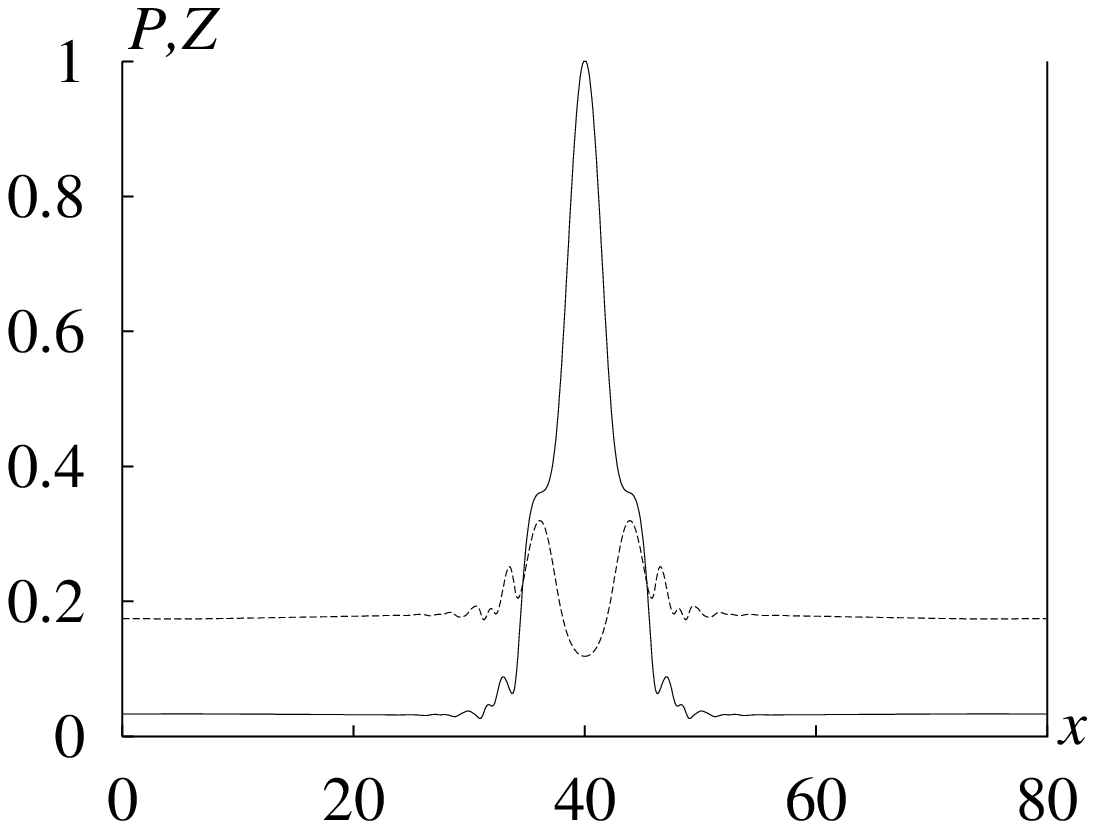} & \panel{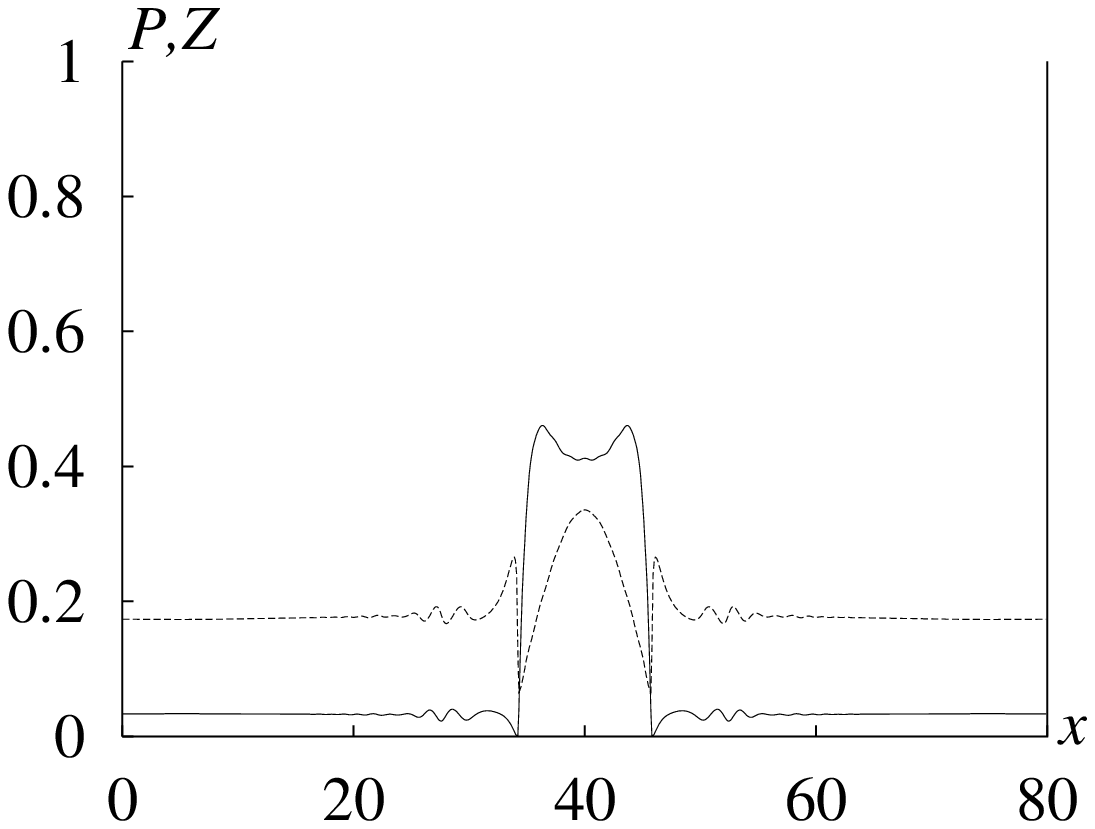} & \panel{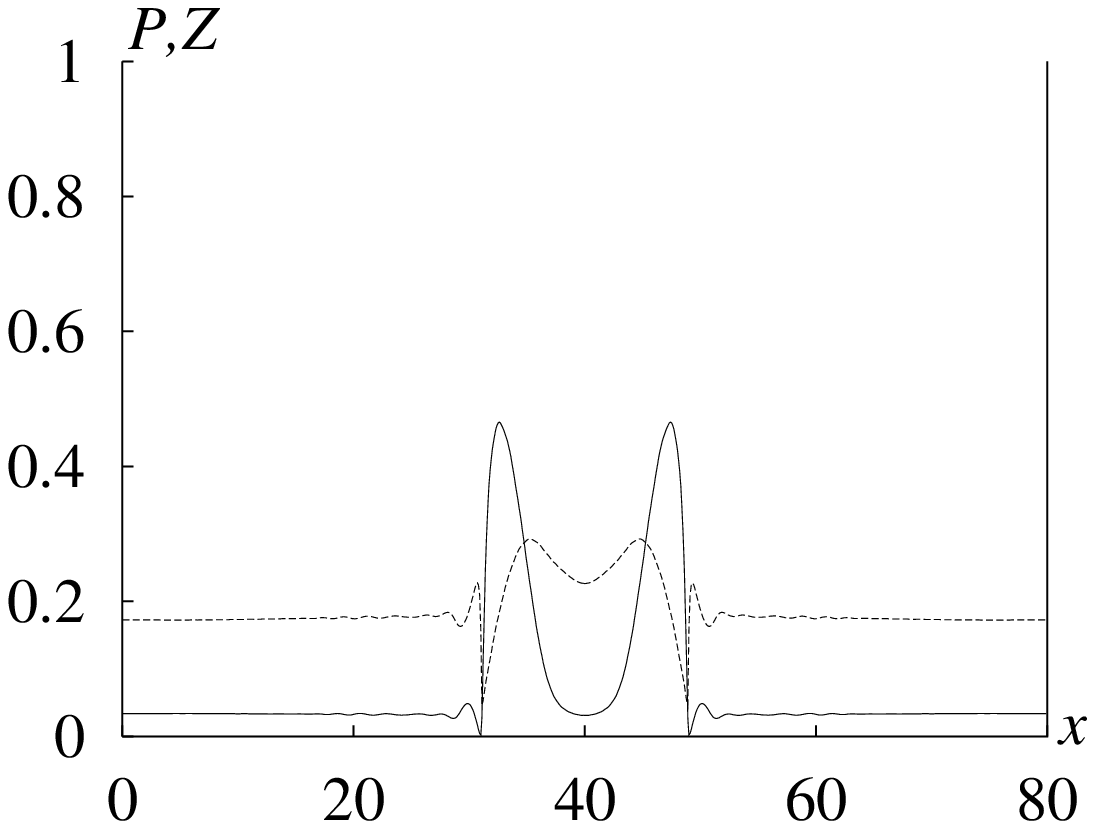} \\
(a)
&
(b)
&
(c)
&
(d)
\end{tabular}}
\centerline{\begin{tabular}{cccc}
\panel{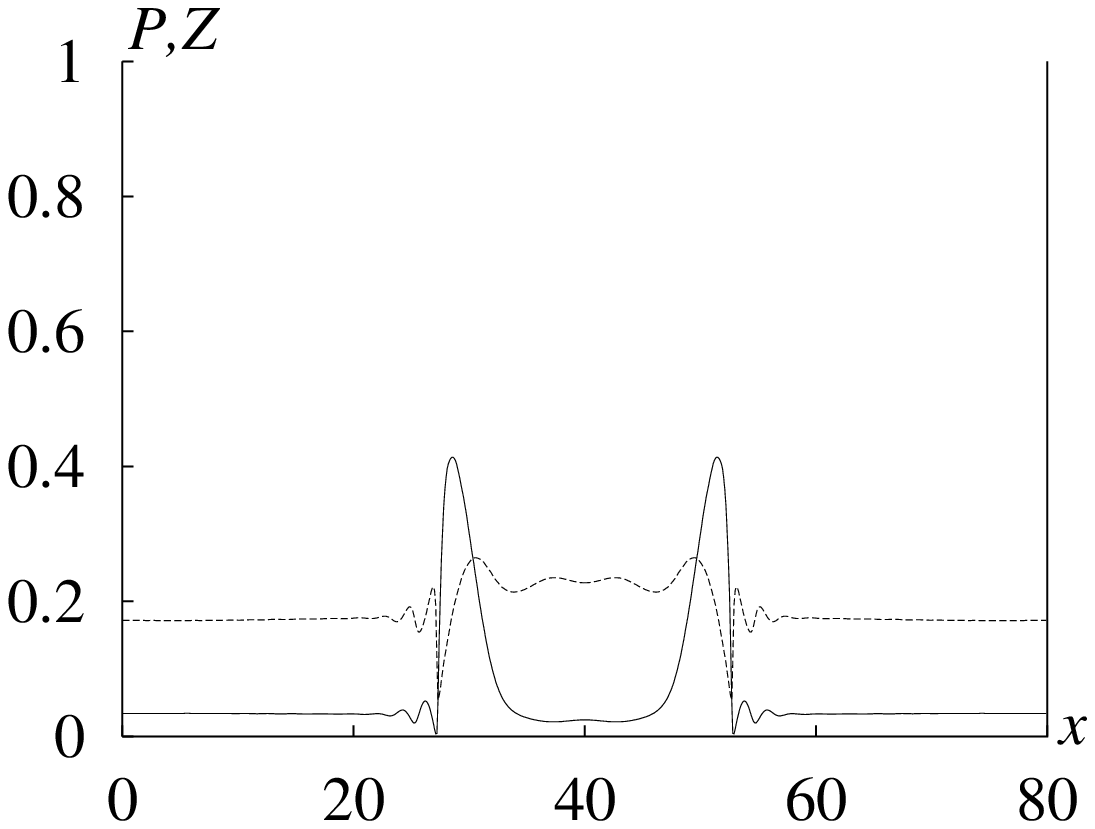} & \panel{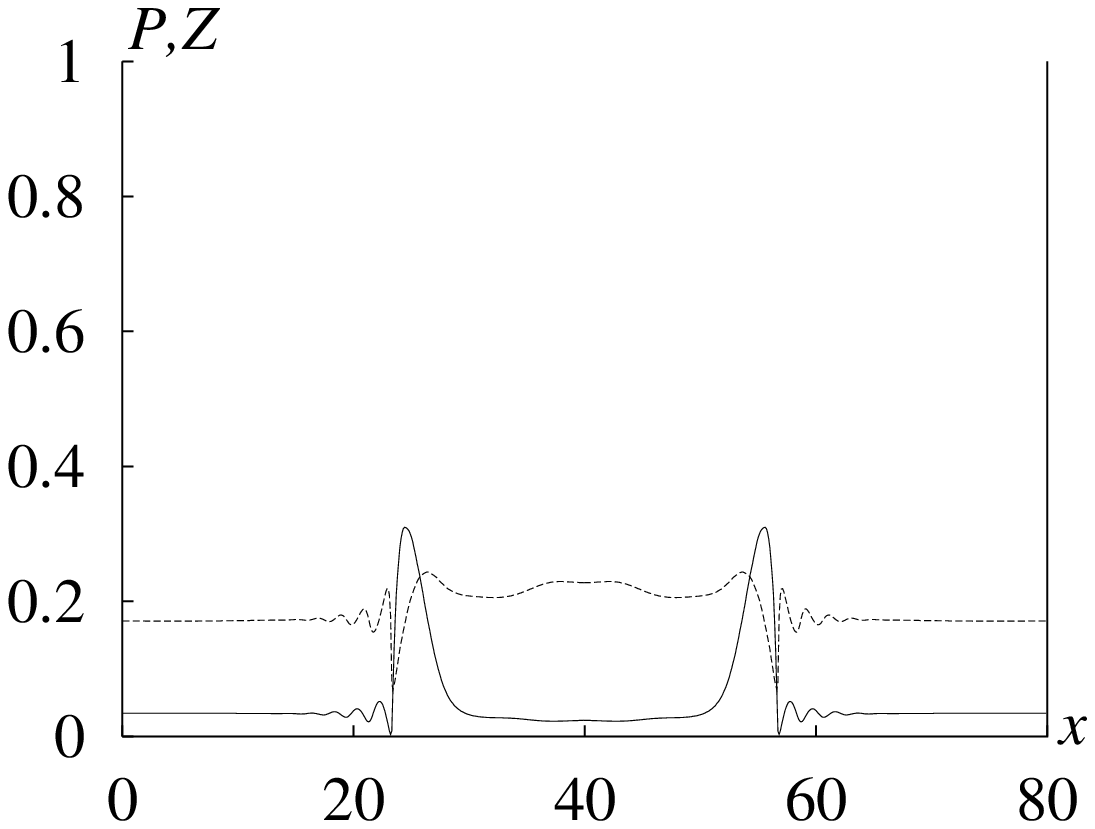} & \panel{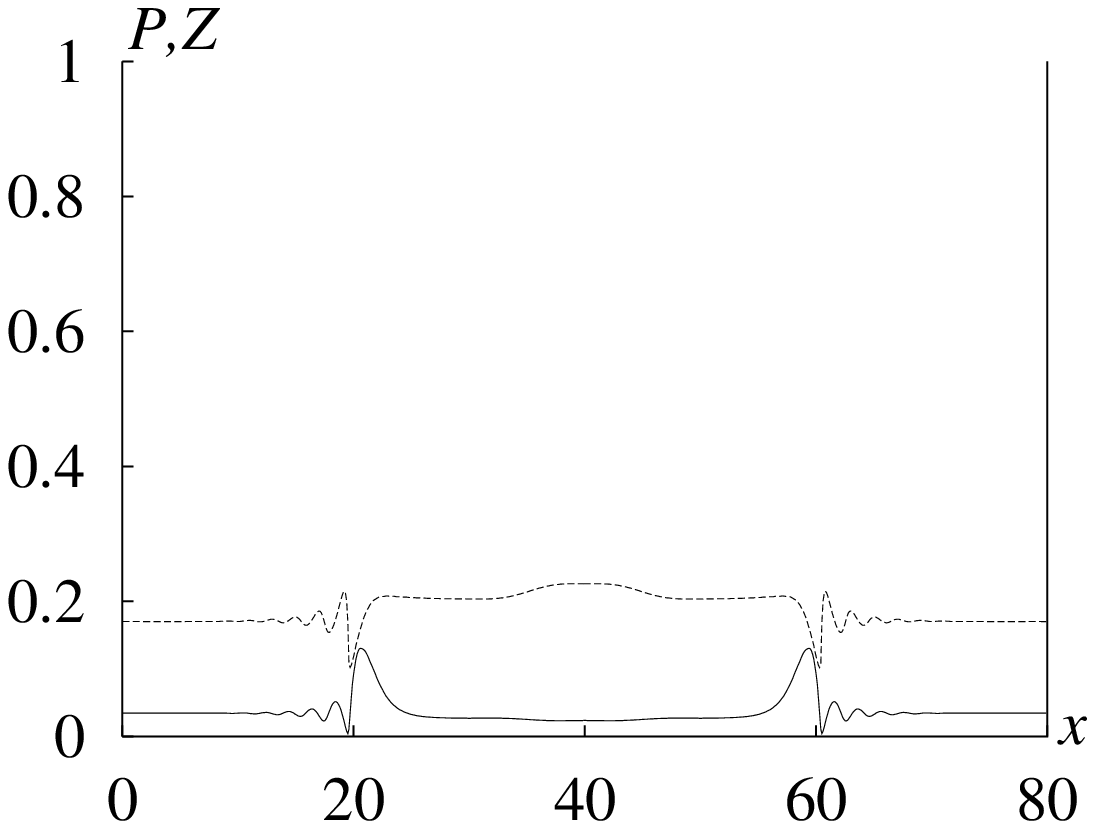} & \panel{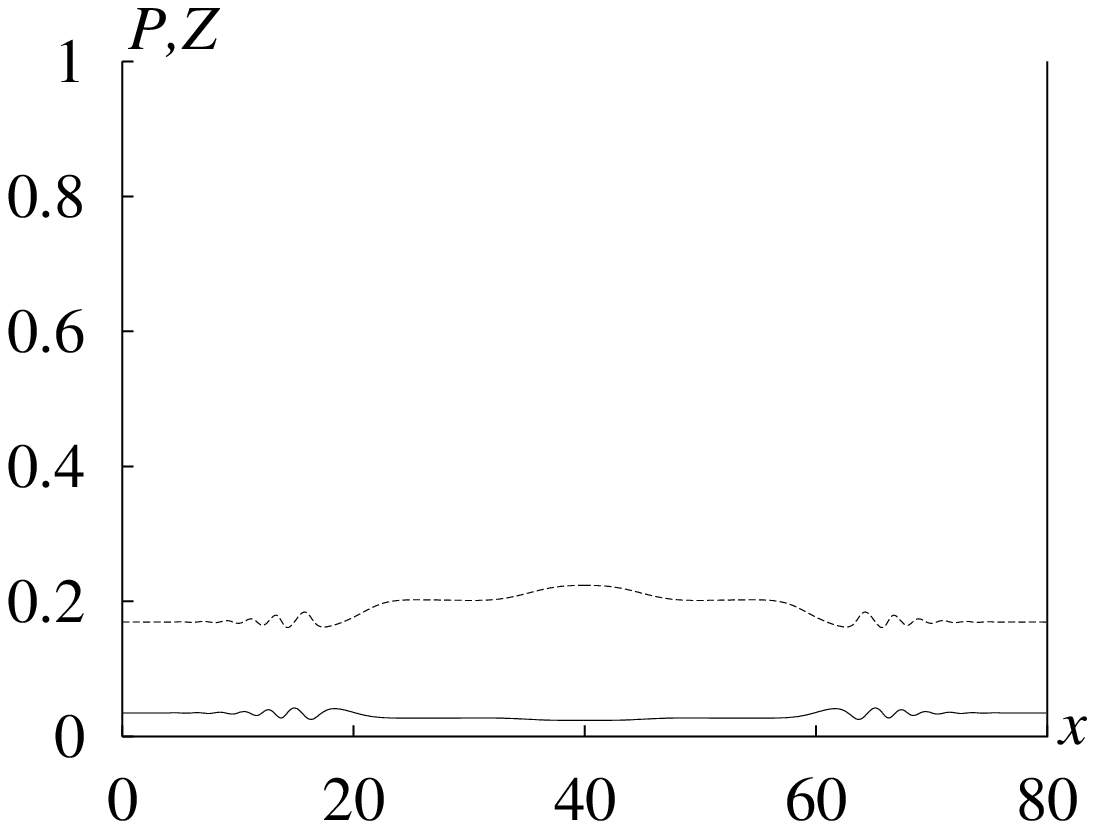}  \\
(e)
&
(f)
&
(g)
&
(h)
\end{tabular}}
\caption{ Dynamics of the nonsoliton interaction for $\gamma=0.016$ and $D=0$,
 $h_-=5, h_+=2$, time interval =5.
}
\label{newANu}
\end{figure*}

As in the previously considered case of $\gamma=0.01$ and
$D=0.04$, the non-soliton waves at $\gamma=0.016$ and $D=0$
demonstrate partial interference
(\fig{newANu}).
After the collision (\fig{newANu}a,b), the waves penetrate through each other
(\fig{newAN}c-e), and then decay (\fig{newAN}f-h, cf. \fig{newAN}).

% ---------------------------------------------------------------------
%===============================================================Solit.Dynam.-5-1

\section{Mechanism of quasisoliton interaction}

\begin{figure*}[htbp]
\setlength{\unitlength}{1mm} \newcommand{\panel}[1]{\begin{picture}(44,44)(0,0) % (0,-20)
\put(1,1){\mbox{\resizebox{43mm}{!}{\includegraphics{#1}}}}
\end{picture}}
\centerline{\begin{tabular}{cccc}
\panel{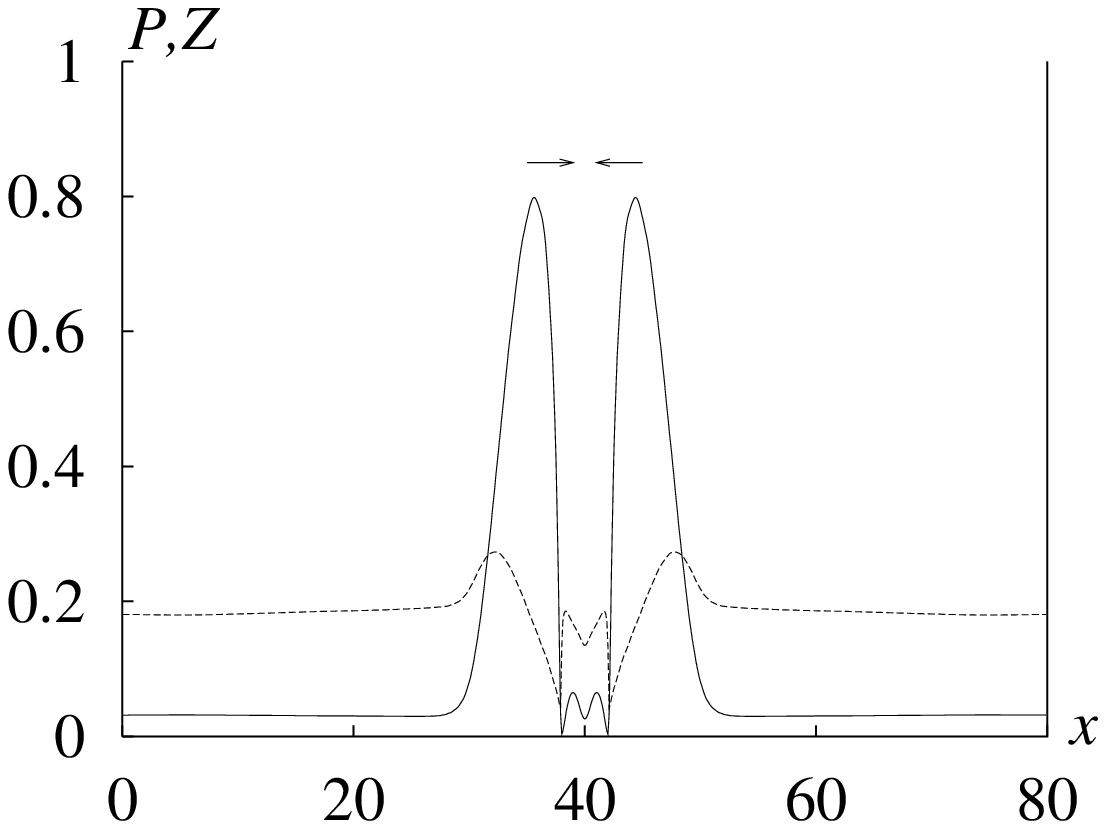} & \panel{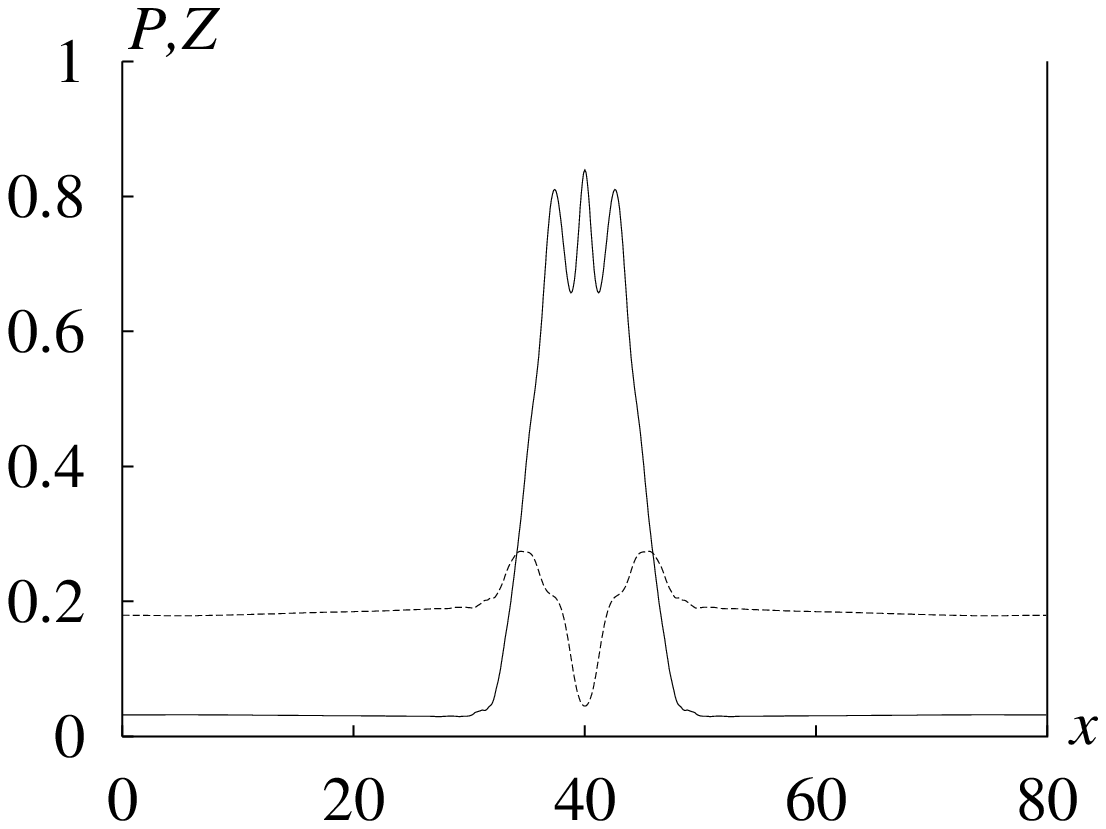} & \panel{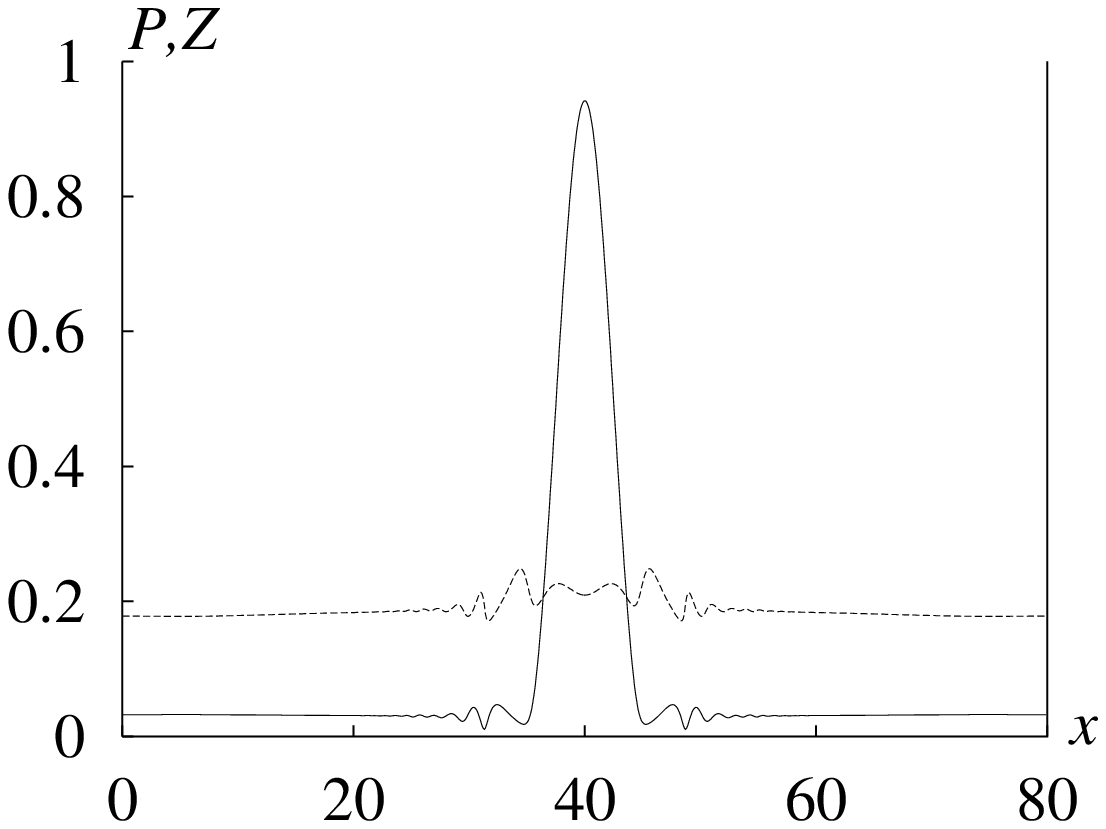} & \panel{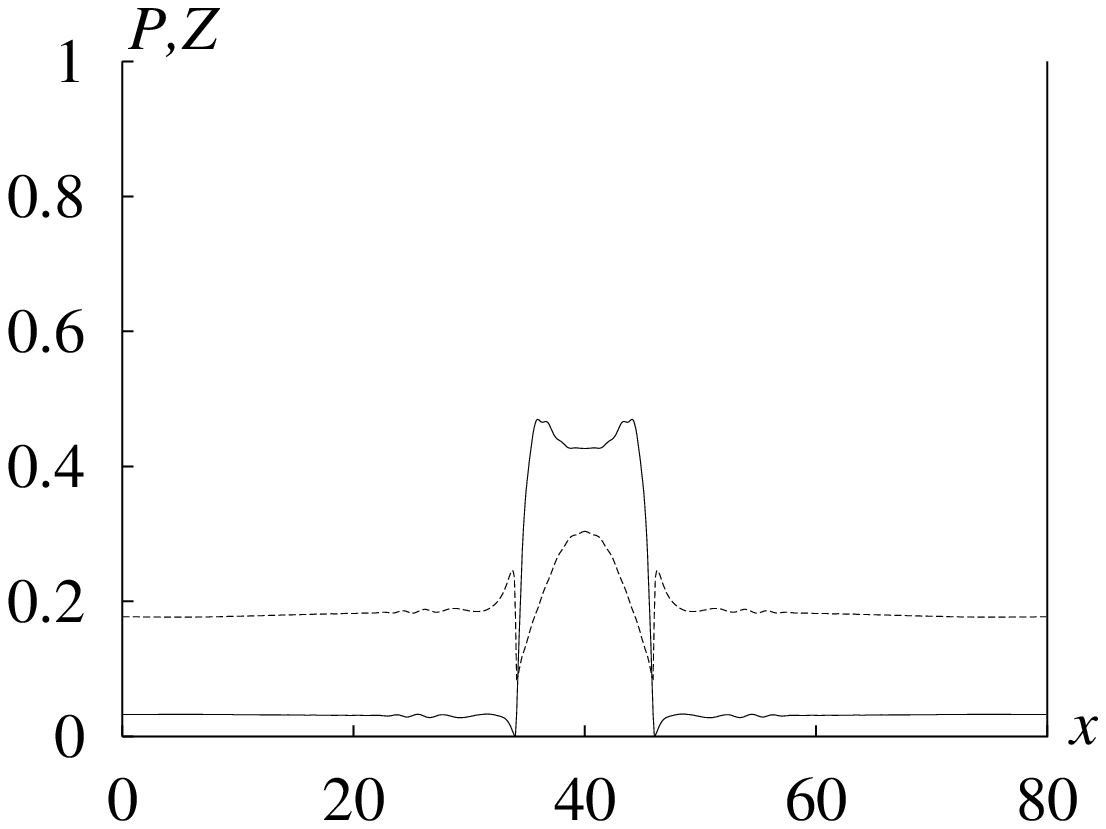} \\
(a)
&
(b)
&
(c)
&
(d)
\end{tabular}}
\centerline{\begin{tabular}{cccc}
\panel{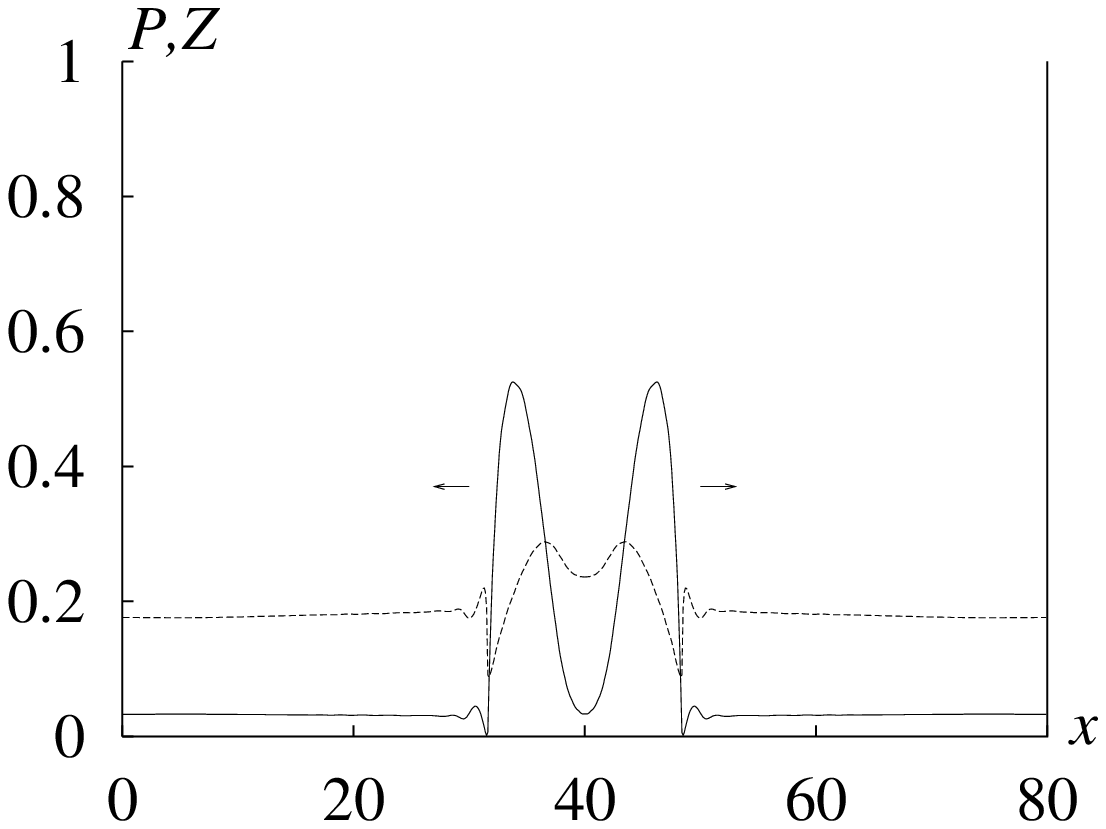} & \panel{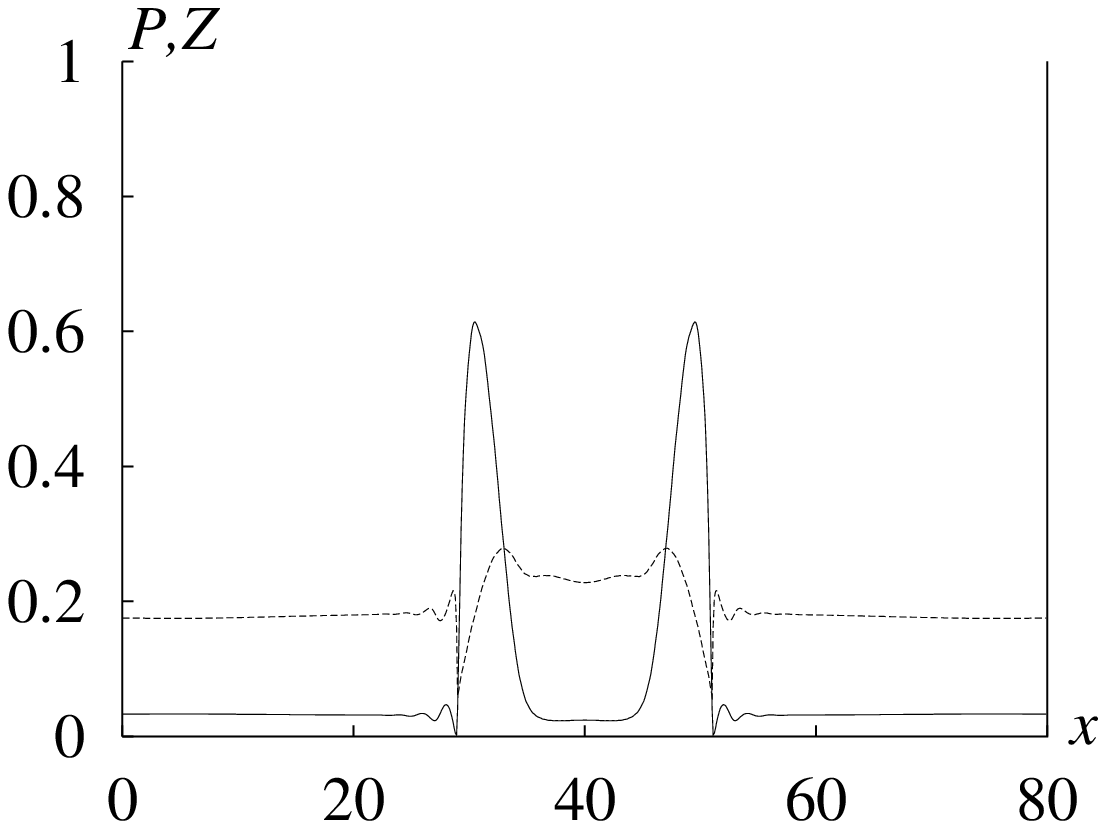} & \panel{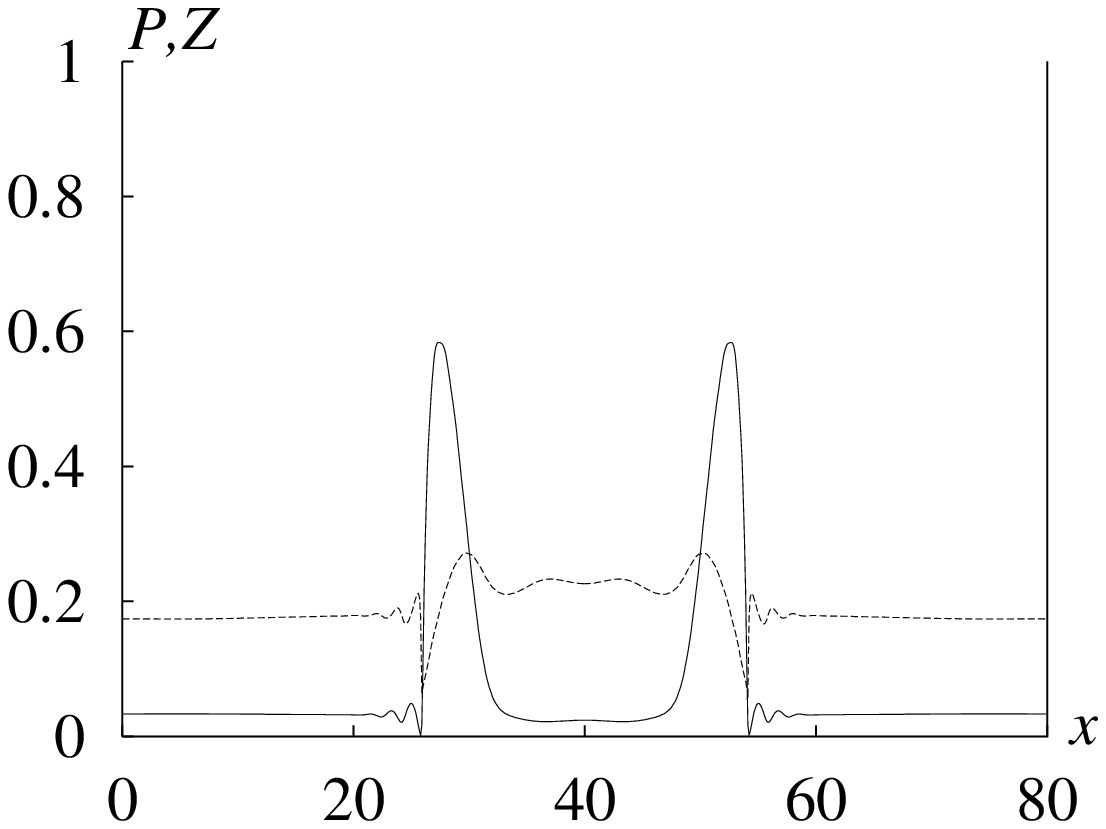} & \panel{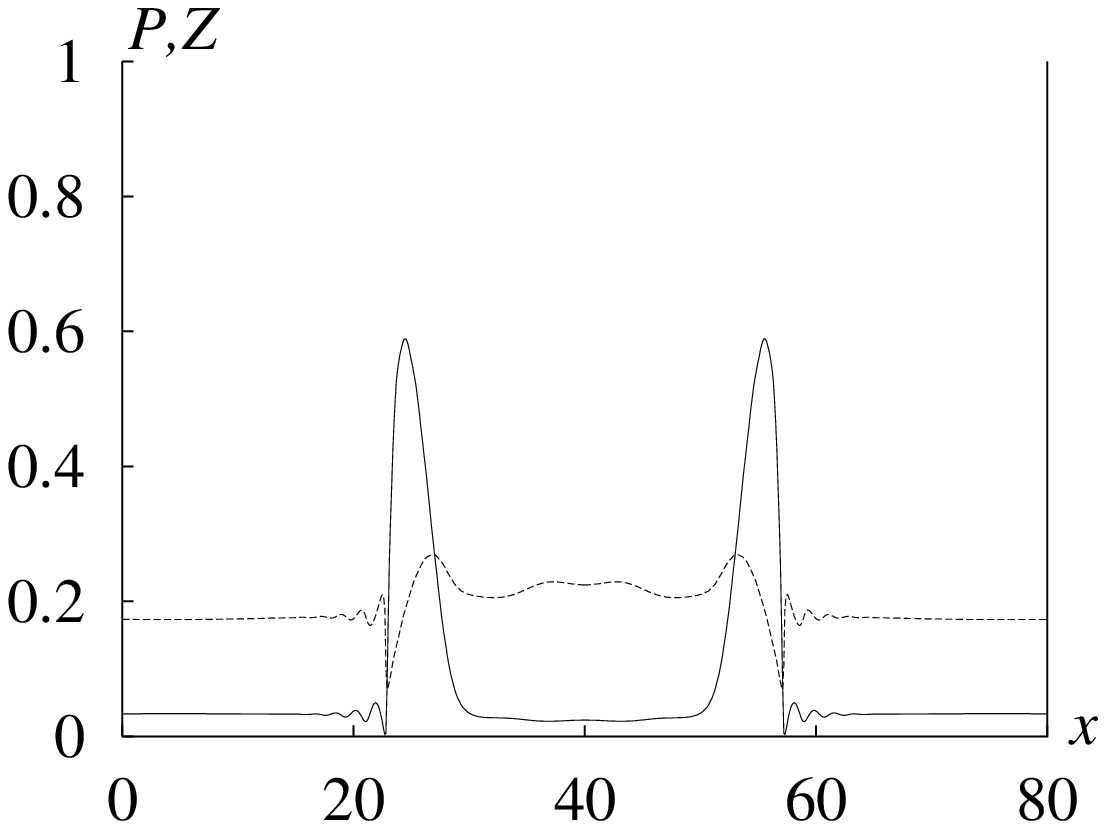}  \\
(e)
&
(f)
&
(g)
&
(h)
\end{tabular}}
\caption{ Soliton interaction.$h_-=5, h_+=1$ $D=0$. The time interval between the panels
is 5.
}
\label{solitDYN}
\end{figure*}

We consider the details of the soliton-like interaction of taxis waves
for the case $D=0$.
\Fig{solitDYN} shows a sequence of wave profiles during such an interaction.
As noted above, a feature of taxis waves is the low level of
predators ahead of the prey wave, as the predators are attracted backwards
by the prey density gradient. This backward gradient of predators encourages
the forward movement of prey (see \Fig{solitDYN}a).
The meeting of two prey waves creates a high peak of prey density (\Fig{solitDYN}a-c).
This higher local density of prey attracts predators,
which abandon the margins of the collision zone (\Fig{solitDYN}b,c).
The local growth of predators causes escape of the prey from the center of
the collision zone towards the margins abandoned by the
predators. These events invert the gradients of the populations and
re-create front structures on the margins of the collision zone
(\Fig{solitDYN}c,d), which then cause generation of two new, ``reflected''
taxis waves (\Fig{solitDYN}e-h), which subsequently restore their normal
amplitude. So, the phenomenon of reflection is driven by interaction of
both the pursuit and evasion taxis terms, forming a positive feedback loop.
The interplay and positive feedback between the two taxis terms can
also be elucidated by considering a simple linear analogue of equations \eq{RDT}:

\begin{equation}
\df{P}{t} = h_-\ddf{Z}{x} ,  \qquad
\df{Z}{t} = - h_+\ddf{P}{x} ,\label{RDTlin}
\end{equation}

which are obtained from \eq{RDT} by putting $f_{1,2}=D=0$ and removing
nonlinearity from the taxis terms. System \eq{RDTlin} is equivalent to
a Schr\"odinger equation for $u=h_+^{1/2}P+ih_-^{1/2}Z$. This is consistent
with both the oscillatory fronts of taxis waves and their
ability to reflect from each other. The role of nonlinearities appears
to be in selecting a unique amplitude and shape of propagating waves,
and restricting, compared to \eq{RDTlin}, the relative values of $h_{\pm}$
that allow reflection. Adding diffusion in \eq{RDTlin} destroys propagating waves,
but not necessarily in \eq{RDT} where its dissipative effect may be
compensated by the nonlinear kinetics.

% ---------------------------------------------------------------------

\begin{figure}[htbp]
\centerline{\includegraphics{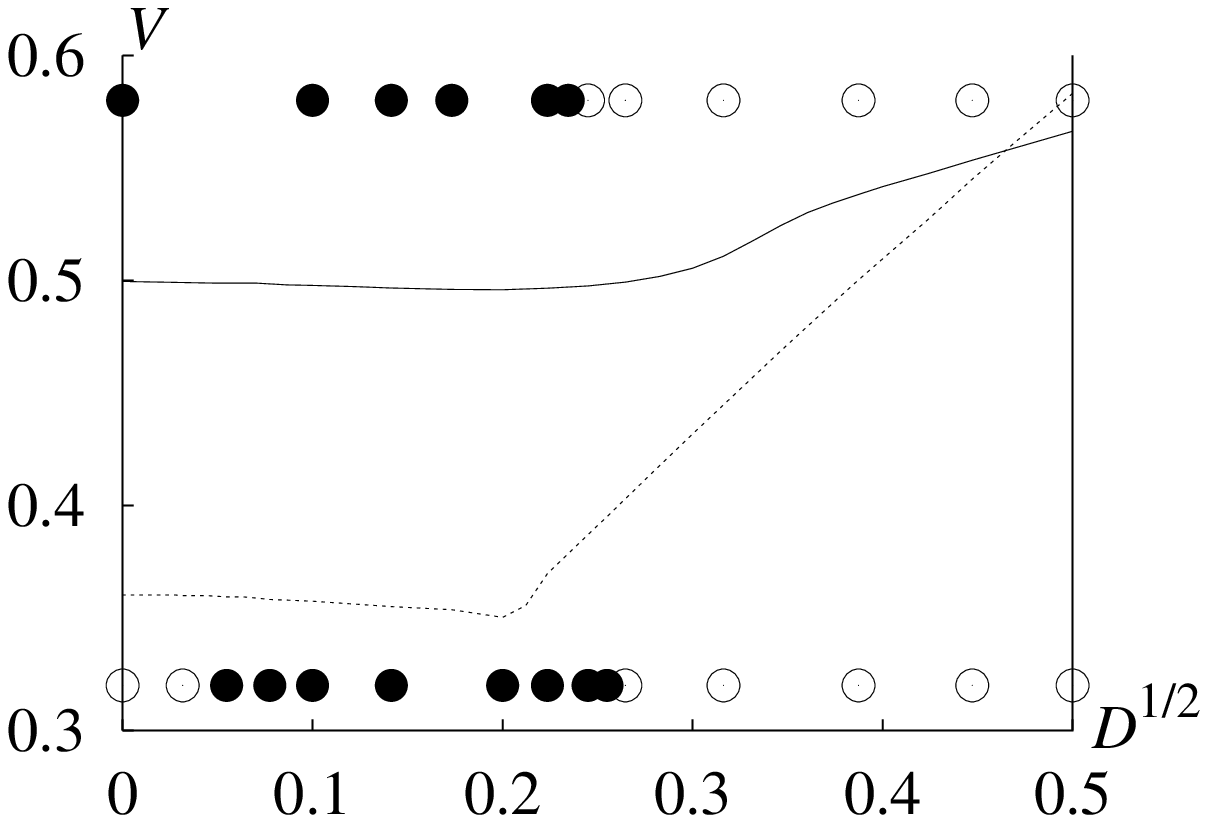}}
\caption{
Wave propagation velocity as function of the square root of the diffusion coefficient.
Solid line and the upper row of symbols: $\gamma=0.016$, $h_-=5$, $h_+=1$.
Dotted line and the lower row:  $\gamma=0.01$, $h_-=h_+=1$.
As before, filled circles denote existence of quasisoliton waves, and hollow circles
annihilating waves.
In reaction-diffusion systems, this dependence is always a straight line.
}
\label{VD}
\end{figure}

As we have already noted, some properties of taxis waves are essentially
different from those of solitary waves observed in excitable reaction-diffusion
systems.
This difference can be seen in \Fig{VD} which shows dependence
of the wave propagation velocity on the diffusion coefficient $D$.
This dependence is clearly different from $\propto D^{1/2}$ law
obeyed by reaction-diffusion waves.
There is a marked change of this dependence near the transition between
annihilating and reflecting waves, which is yet another evidence of
different mechanism of taxis waves, especially of quasisolitons.

%------------------------------------------------------V(D)

%----------------------------------------------------Bacteria
\begin{figure*}[htbp]
\setlength{\unitlength}{1mm} \newcommand{\panel}[1]{\begin{picture}(44,44)(0,0) % (0,-20)
\put(1,1){\mbox{\resizebox{43mm}{!}{\includegraphics{#1}}}}
\end{picture}}
\centerline{\begin{tabular}{cccc}
\panel{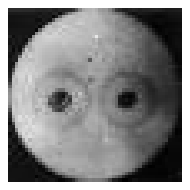} & \panel{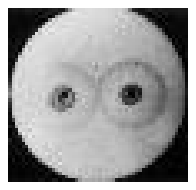} & \panel{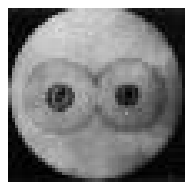} & \panel{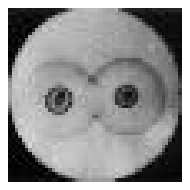} \\
(a) $t=0$
&
(b) $t=20 min$
&
(c) $t=28 min$
&
(d) $t=33 min$
\end{tabular}}
\caption{ Soliton-like interaction of bacterial population waves in
a nutrient medium in a Petrie dish (diameter 9 sm) \cite{Tsyg93}: 
(a) Bacterial waves, formed by \textit{E.coli}, propagate from two
inoculation points. This moment is chosen as zero time. 
(b) The bacterial waves collide. 
(c),(d) The bacterial waves penetrate through/reflect from each other.
}
\label{bact}
\end{figure*}
% ---------------------------------------------------------------------

\paragraph*{Conclusions.}
We have studied a spatially distributed predator-prey system of equations,
in which, in addition to or instead of diffusion terms, we
have included terms describing taxis of each species on the other's
gradient: predators pursuing prey, and prey escaping predators.
We have found that the taxis terms change the shape of the propagating
waves and increase the propagation speed, which is an evidence of a
different mechanism of propagation of these waves. In this change the
major role is played by the pursuit terms. Also, the taxis terms can
change the interaction between propagating waves, i.e. make them
penetrate/reflect, rather than annihilate. For this effect, both
pursuit and evasion terms are essential.
We have also identified regimes of ``wave-splitting'', that can be observed
both during normal propagation, and as a result of interaction of taxis waves.

%----------------------------------------

Quasisoliton interaction of taxis waves is not a purely mathematical exercise.
In \cite{Tsyg93}, soliton-like interaction of bacterial population taxis waves
were observed in vitro, where colliding waves continued to
propagate after collision without delay (see \fig{bact}).

%---------------------------------------------------------------------------
However, apart from immediate application to predator-prey or similar systems,
our results illustrate that inclusion of taxis terms can dramatically change the
behaviour of the system. Reaction-diffusion systems are the most studied
class of models describing waves and patterns in spatially distributed systems.
Meanwhile mechanisms of spatial interaction in real system vary widely beyond
simple diffusion; let us mention just the role of ``cross-diffusion'' in
the propagation of forest boundary \cite{xdiff}, 
in chemical and biological pattern formation \cite{Jorne74,Jorne77, Almir91}, 
turbulence-shear flow interaction in plasmas \cite{Cast02}, 
and cross-diffusion interaction between displacement and velocity 
in the Burridge-Knopoff model of tectonic slips \cite{Cart97}. 
We hope that our present results may stimulate more attention into properties 
of dissipative nonlinear PDE systems with spatial terms beyond simple 
diffusion term, as these constitute a new class of models with
significantly different properties.

This study was supported in part by EPSRC grant GR/S08664/01 (UK) and by
RFBR grant 03-01-00673 (Russia).
% --------------------------------------------------
\bibliography{qs2}
% ------------------------------------------------------------------
\end{document}